\documentclass[a4paper,11pt]{article}
\usepackage{jheppub} 
\usepackage{graphicx} 
\usepackage{graphics}
\usepackage{amsmath, amssymb, amsthm} 
\numberwithin{equation}{section}
\usepackage{mathtools}
\usepackage{caption}
\usepackage{subcaption}
\usepackage{hyperref}
\usepackage{float}
\usepackage{braket}
\usepackage{multicol}
\usepackage{subfiles}
\usepackage{siunitx} 
\usepackage{placeins}
\usepackage{url}
\usepackage{comment}
\usepackage{setspace}
\usepackage{bbold}
\usepackage{bm}
\usepackage{overpic}

\usepackage{tabularray}

\title{\boldmath Covariant eigenmode overlap formalism for gravitational wave signals in electromagnetic cavities}

\author[a]{Jordan Gu\'e,}
\author[b]{Tom Krokotsch,}
\author[b,c]{Gudrid Moortgat-Pick}
\affiliation[a]{Institut de F\'{i}sica d’Altes Energies (IFAE), The Barcelona Institute of Science and Technology, Campus UAB, 08193 Bellaterra (Barcelona), Spain}
\affiliation[b]{Universit\"at Hamburg,  Luruper Chaussee 149, 22761 Hamburg, Germany}
\affiliation[c]{Deutsches Elektronen-Synchrotron DESY,
 Notkestraße 85, 22607 Hamburg, Germany}

\emailAdd{tom.krokotsch@desy.de}
\emailAdd{jgue@ifae.es}

\abstract{We develop a coordinate invariant formalism which describes the mechanical and electromagnetic interaction of gravitational waves (GWs) with a wide class of resonant detectors. We solve the GW-modified equations of electrodynamics and elasticity with dynamic boundary conditions using an eigenmode expansion. Furthermore, we take damping effects and electromagnetic back-action on mechanical systems covariantly into account. The resulting coupling coefficients are particularly useful for high-frequency gravitational wave experiments using microwave cavities and allow a straightforward numerical implementation for arbitrary detector geometries.}

\begin{document}
\maketitle
\flushbottom

\section{Introduction}
\subsection{Overview}
Since their first detection 10 years ago by the LIGO/VIRGO collaboration \cite{LIGOScientific:2016aoc}, gravitational waves (GW) have revolutionized the way we can study the universe. Since the universe is highly transparent to GWs, they are a unique probe of both astrophysical and cosmological events, in particular before the emission of the cosmic microwave background. Many different projects have been proposed to cover the GW spectrum, which would offer a complete view of the gravitational universe. Among them are PTA/NANOGrav/MeerKAT at nHz \cite{Manchester13,Kramer13,Demorest13,Lee16,Joshi22,Miles23}, binary resonances at $\mu$Hz \cite{Foster25}, LISA, Taiji, TianQin at mHz \cite{Colpi24,Li24,Ruan20}, Atom Interferometry at $0.1$ Hz \cite{Badurina19, Abe21}, LIGO/VIRGO/KAGRA and GEO-600 in the $1-10^3$ Hz band \cite{Abbott23}, and many proposed experiments, like interferometers, resonant mass detectors, electromagnetic oscillators or photon generation experiments, which will target high-frequency GWs (HFGWs) above 10 kHz \cite{Aggarwal25}. 

In this range, the expected signals are mostly coming from Beyond the Standard Model (BSM) physics, such as Primordial Black Holes (PBH) mergers, Exotic Compact Objects (ECO) mergers, black hole superradiance, inflation or cosmological phase transitions (see \cite{Aggarwal25} for a review of the expected sources). Since many of those sources are stochastic, the entire characteristic frequency spectrum of each signal needs to be mapped to disentangle it efficiently from the noise of the detector. Since the number of known astrophysical sources delivering signals in the high-frequency range is limited, such experiments will be clean probes of new physics without suffering from an astrophysical stochastic foreground.

Calculating the interaction of GWs with HFGW detectors presents several new challenges as compared with low-frequency detectors, and is the main subject of this paper. The key difference is that HFGWs span the frequency range from kHz, where acoustic resonances determine the mechanical response of elastic detectors, to GHz and above, where electromagnetic resonances in microwave or optical cavities occur. This transition region between kHz and GHz frequencies, does not allow for many approximations, and requires a detailed analysis, which we provide here.

The main example we will consider for this purpose are microwave cavities, which offer a promising technology to search for HFGWs in the entire region between kHz and GHz frequencies. The setup involves creating an electromagnetic (EM) background, either in the form of a static magnetic field or an EM cavity mode, similarly as in axion haloscopes \cite{ADMX:2020ote,berlin2020axion}, and waiting for the GW to populate a given signal mode of the cavity. Both the perturbation of the background field directly (the `inverse Gertsenshtein effect'), and the interaction of the background field with the GW-perturbed mechanical structure can create EM signals. 

Calculating the signal power induced by a GW requires choosing a coordinate system on linearized curved space. A common choice for GW detectors are \emph{proper detector} (PD) coordinates, which can be thought of as laboratory coordinates in which distances are measured by rulers which are rigid and not affected by the GW for distances much shorter than the GW wavelength. Alternatively, \emph{transverse-traceless} (TT) coordinates are popular, where the coordinate system itself is freely falling and deforming along with the GW. Consequently, using different coordinate gauges leads to different physical effects on the EM fields and mechanical structure.
A previous work computed the expected power deposited in the cavity by a GW, but did not take the GW perturbation of the boundary in PD coordinates into account \cite{Berlin22}. However, subsequent works argued for using TT coordinates, where the perturbation of the cavity walls can be neglected when the GW wavelength is comparable to the size of the cavity \cite{Ahn2024, Ratzinger2024}. 

In this work, we develop a formalism which can describe the GW interaction with cavities using any coordinate system. More precisely, we present a way to compute the gauge-invariant electromagnetic fields in the local inertial frame of an observer as used in \cite{Ratzinger2024} by decomposing the resulting fields into electromagnetic eigenmodes of the unperturbed cavity. The benefit of this formalism is that the solution of Maxwell's equations and the equations of elasticity in curved spacetime in the presence of dynamic boundary conditions, can be described by overlap coefficients with the eigenmodes of the unperturbed cavity, which are typically easy to obtain. This allows a straightforward numerical implementation for arbitrary cavity shapes. Importantly, we can use the formalism at any frequency in every frame and include background EM fields of any complexity. Therefore, we can develop a physical picture of how mechanical and electromagnetic GW interactions create gauge independent results when added together. 

Our decomposition improves on previous treatments \cite{berlin_mago20_2023, Löwenberg2023} by being valid at non-resonant frequencies as well. Furthermore, we include electromagnetic and mechanical losses while preserving covariance and take back-action from the background fields on the mechanical structure into account. We find that damping cannot be neglected when investigating the so-called free falling limit of HFGW detectors where their displacement due to GWs is neglected in TT coordinates. Importantly, we argue that elastic solids can never become freely falling at their boundary in the same sense as a cloud of disconnected particles.

Overall, we present the most complete description of GWs interacting with cavities to date. We believe our results could be used to re-evaluate or extend the frequency range  of sensitivity calculations for e.g. axion haloscopes based on microwave cavities \cite{ADMX:2020ote, Valero:2024ncz}, microwave cavities loaded with radio-frequency power \cite{Fischer_2025, SHANHE, Fermilab_heterodyne, Slac_heterodyne}, plasma haloscopes \cite{plasma_haloscopes}, optical cavities \cite{Atonga:2025utf} and phonon detectors \cite{Kahn:2023mrj}.

The paper is organized as follows. In Sec.~\ref{sec:covariant_EM}, we compute the EM eigenmode coefficients by using overlap coefficients, and taking into account perturbed boundary conditions for the fields, which appear as surface currents. These terms depend on the mechanical dynamics of the cavity walls. In Sec.~\ref{sec:elasticy_theory}, we express the displacement of the elastic walls of the cavity using overlap coefficients as well. In this section, we neglect the EM pressure from the EM fields inside the cavity that affects the elasticity equations. This back-action effect is considered in Sec.~\ref{sec:back_action}, such that coupled equations of motion for the displacement and EM mode coefficients are obtained. In Sec.~\ref{sec:signal}, we compute the EM signal power generated by the GW and measured by an antenna in different frequency regimes and different experimental setups. Finally, in Sec.~\ref{sec:discussion}, we discuss the relevance of our results and their application in other experimental setups, such as LC circuits.

\subsection{Theoretical background and conventions}

We use the mostly plus convention for the Minkowski metric $\eta_{\mu\nu}=\text{diag}(-1,+1,+1,+1)_{\mu\nu}$ and the Levi-Civita symbol with $\bar{\epsilon}_{0123}=\bar{\epsilon}_{123}=+1$. Greek indices denote spacetime components $\mu, \nu,\dots=0, 1, 2, 3$ and Latin indices denote only the spatial parts $i, j,\dots=1, 2, 3$. Bold letters denote a vector or matrix built from the spatial components $\bm{\mathcal{A}}=(\mathcal{A}^i)_{i\in\{1,2,3\}}$, $\bm{\mathcal{B}}=(\mathcal{B}^{ij})_{i,j\in\{1,2,3\}}$. We use a natural unit system for electromagnetism where $c=\epsilon_0=1$. The electromagnetic field tensor $F^{\mu\nu}$ is chosen so that $F^{0i}=F_{i0}=E^i$ and $F^{ij}=\epsilon^{ijk}B_k$, in which case Maxwell's equations in linearized curved space with metric $g_{\mu\nu}=\eta_{\mu\nu}+h_{\mu\nu}+\mathcal{O}(h^2)$ become
\begin{subequations}\label{eq:GWMaxwell_before_perturbation}
\begin{align}
    &\partial_\nu F^{\mu\nu}+\frac12F^{\mu\nu}\partial_\nu h=J^\mu \label{eq:Maxwell_inhomogeneous}\,,\\ \label{eq:Maxwell_homogeneous}
    &\partial_\mu F_{\nu\rho}+\partial_\rho F_{\mu\nu}+\partial_\nu F_{\rho\mu}=0\,,
\end{align}
\end{subequations}
which follow from $\nabla_\nu F^{\mu\nu}=J^\mu$ and $\nabla_\mu F_{\nu\rho}+\nabla_\rho F_{\mu\nu}+\nabla_\nu F_{\rho\mu}=0$, and where we used $h\equiv h_\mu^{\,\mu}$.

The transformation for a monochromatic GW with wave vector $k^\mu=(\omega, \bm{k})^\mu$ between PD ($x_\mu^\text{PD}$) and TT ($x_\mu^\text{TT}\equiv \tilde{x}_\mu$) coordinates is given by \cite{Ratzinger2024}
\begin{align}\label{eq:TT_PD_coord_transfo}
    x_\mu^\text{PD}&=\tilde{x}_\mu+(\xi^\text{TT}_\text{PD})_\mu\\&\equiv\tilde{x}_\mu+h_{\mu\nu}^\text{TT}\tilde{x}^\nu\left(\frac12+i\bm{k}\cdot\tilde{\bm{x}}F(\bm{k}\cdot\tilde{\bm{x}})\right)-\frac{i}{2}k_\mu \,\tilde{\bm{x}}\cdot(\bm{h}^\text{TT}\cdot\tilde{\bm{x}})\,F(\bm{k}\cdot\tilde{\bm{x}})\nonumber
\end{align}
with $F(x)=(e^{-ix}+ix-1)/x^2=-\sum_{n=0}^\infty(-ix)^n/(n+2)!$. The GW strain then transforms according to $h_{\mu\nu}^\text{PD}=h_{\mu\nu}^\text{TT}-\partial_\mu(\xi^\text{TT}_\text{PD})_\nu-\partial_\nu(\xi^\text{TT}_\text{PD})_\mu$\,. A closed-form expression for the transformation can also be derived for non-monochromatic waves \cite{Fischer_waveforms}.

We introduce a coordinate system defined by a set of orthonormal tetrads $\{e_{\underline{\alpha}}\}$ along an observer's world line $x^\mu(\tau)$ with $\eta_{\alpha\beta}=g_{\mu\nu}e_{\underline{\alpha}}^\mu e_{\underline{\beta}}^\nu$ and $e_{\underline{0}}^\mu=u^\mu$, where $u^\mu(\tau)$ is the observer's four-velocity and $\tau$ the proper time. Indices numbering the tetrads are underlined to distinguish them from coordinate indices. This allows the definition of the electric field in the tetrad basis \cite{Ratzinger2024,Misner1973}
\begin{equation}\label{eq:observed_E_field}
    E^\text{tetrad}_a=F_{\mu\nu}e^\mu_{\underline{a}}u^\nu
\end{equation}
and the magnetic field
\begin{equation}\label{eq:cov_B_field}
    B^\text{tetrad}_a=\frac12\epsilon_{abc}F_{\mu\nu}e^\mu_{\underline{b}}e^\nu_{\underline{c}}\,.
\end{equation}
Whenever the tetrad is chosen to follow the worldline of a readout device such as an antenna $x^\mu=x^\mu_\text{ant}$, we refer to the electric field in the tetrad frame as \emph{observed} electric field $\bm{E}^\text{obs}$.

We take effects from the GW into account using a perturbation method as in \cite{Ratzinger2024} by decomposing quantities like $F_{\mu\nu}=\bar{F}_{\mu\nu}+\delta F_{\mu\nu}$ into a background value in flat space, which we denote with a bar, and its $\mathcal{O}(h)$ perturbations due to the GW, which we denote with a $\delta$ (see App.~\ref{sec:notes_on_perturbation_scheme} for more details) and neglect all $\mathcal{O}(h^2)$ quantities.

The inhomogeneous GW-Maxwell equations \eqref{eq:Maxwell_inhomogeneous} for the EM field decomposition then become
\begin{subequations}
\begin{align}
\partial_\nu \bar F^{\mu\nu}&=\bar J^\mu\,,\\\label{eq:GWMaxwell}
    \partial_\nu \delta F^{\mu\nu}&=J_\text{eff}^\mu+\delta J^\mu \,,
\end{align}
\end{subequations}
with the effective current density 
\begin{equation}\label{eq:Jeff}
    J_\text{eff}^\mu=-\frac12(\partial_\nu h)\bar F^{\mu\nu}-\partial_\nu\left(h^\nu_{\;\rho} \bar F^{\rho\mu}-h^\mu_{\;\rho} \bar F^{\rho\nu}\right)\,.
\end{equation}
The homogeneous equations Eqs. \eqref{eq:Maxwell_homogeneous} hold for $\delta F_{\mu\nu}$ and $\bar{F}_{\mu\nu}$ separately and receive no additional $\mathcal{O}(h)$ corrections.

For one background tetrad of the observer's coordinate system, we can use $\bar{u}^\mu=\bar{e}_{\underline{0}}^\mu=\delta_0^\mu$. In that case, a proper time derivative equals a coordinate time derivative up to $\mathcal{O}(h)$ terms and we write $\partial_\tau\delta X=\partial_t\delta X=\partial_0\delta X=\delta\dot{X}$.

For the Fourier transform of a time-dependent function $\mathcal{A}(t)$, we use the convention 
\begin{equation}
    \mathcal{A}(t)=\int \frac{d\omega}{2\pi} \tilde{\mathcal{A}}(\omega)e^{i\omega t}\,,\quad \tilde{\mathcal{A}}(\omega)=\int dt\mathcal{A}(t)e^{-i\omega t}\,,
\end{equation}
where $\omega=2\pi f$ is the angular Fourier frequency. For complex-valued monochromatic waves, we adopt the sign convention $h(\bm{x},t)\propto e^{+i(\omega_g t-\bm{k}_g\cdot\bm{x})}$, where $\omega_g$ and $\bm{k}_g=\omega_g\hat{\bm{k}}_g$ denote the frequency and wavevector of a monochromatic GW. 

As shorthand notation for the root-mean-square integral of a vector field $\bm{\mathcal{A}}(\bm{x})$, we occasionally use
\begin{equation}
    \langle\bm{\mathcal{A}}\rangle\coloneqq\sqrt{\frac1V\int dV|\bm{\mathcal{A}}(\bm{x})|^2}\,,
\end{equation}
where $V$ is the domain in which $\bm{\mathcal{A}}$ is defined.

The power spectral density (PSD) $S_f(\omega)$ of a (stochastic) function $f(t)$ is defined with the convention
\begin{subequations}
    \begin{align}\label{eq:PSD_definition}
    &S_f(\omega)\delta(\omega-\omega')=\langle \tilde{f}^*(\omega)\tilde{f}(\omega')\rangle_\text{ens}\,,\\
    &\langle f(t)^2\rangle_t=\frac{1}{(2\pi)^2}\int_{-\infty}^{\infty}d\omega\,S_f(\omega)\,,\label{eq:PSD_time_avg}
\end{align}
\end{subequations}

where $\langle\cdot\rangle_\text{ens}$ is an ensemble average and $\langle\cdot\rangle_t$ is a time average. The PSD of $f(t)=e^{i\omega_0t}$ is $S_f(\omega)=(2\pi)^2\delta(\omega-\omega_0)$.

Surface normal vectors are always chosen to point \emph{out} of the enclosed volume. 

\section{GW perturbation of electromagnetic fields with boundary conditions}\label{sec:covariant_EM}
\begin{figure}[ht]
\centering
\includegraphics[width=.99\linewidth]{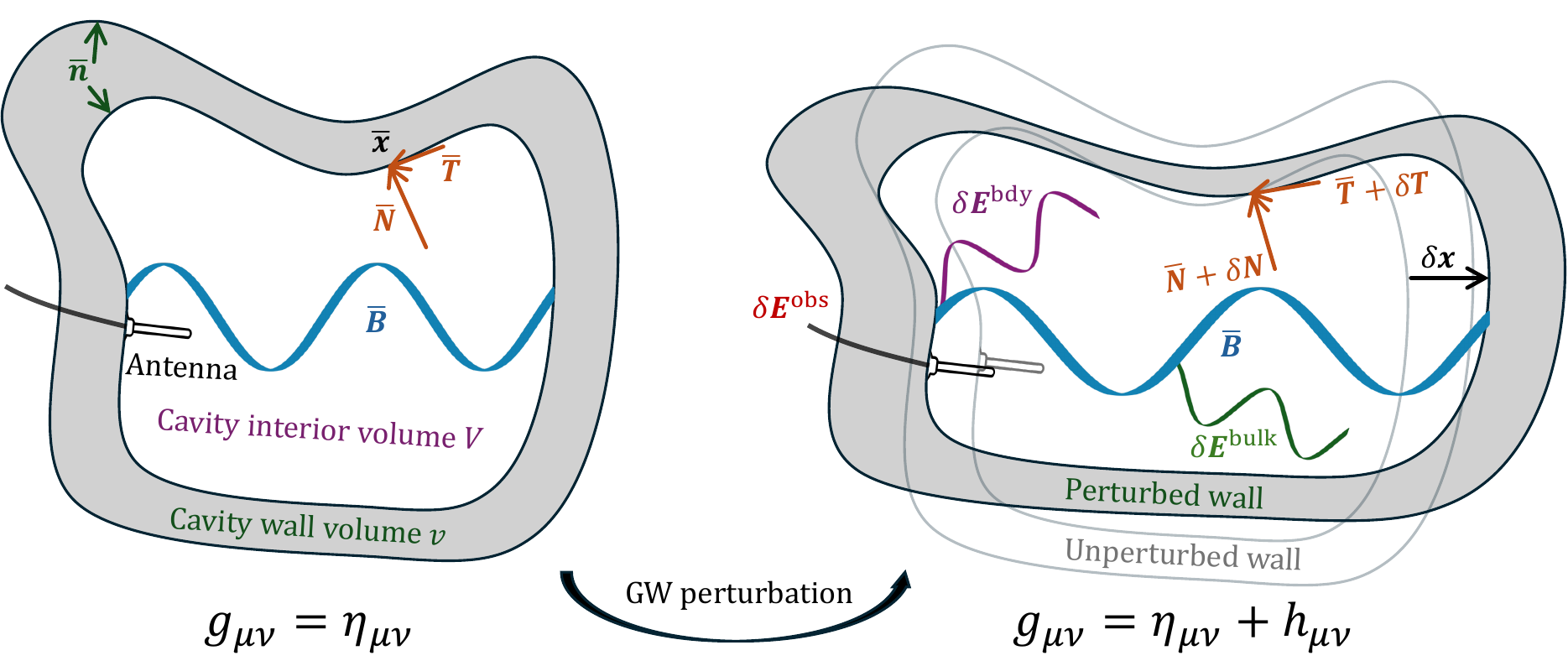}

\caption{Illustration of the setup described in this work. Electromagnetic fields such as $\bar{\bm{B}}$ with cavity boundary conditions are being monitored by an observer like an antenna. If a GW perturbs the metric of flat space $g_{\mu\nu}=\eta_{\mu\nu}+h_{\mu\nu}$, both the background field, as well as the mechanical structure get perturbed. The displacement of the boundary $\delta\bm{x}$ and rotation of the surface normal $\bar{\bm{N}}$ and tangent $\bar{\bm{T}}$ causes electric fields $\delta\bm{E}^\text{bdy}$ to be emitted at the boundary. Furthermore, the perturbation of the background field causes electric fields $\delta\bm{E}^\text{bulk}$ to be emitted throughout the cavity volume. The total signal $\delta\bm{E}^\text{obs}$ measured by the antenna is coordinate invariant and the quantity we calculate in this work.}
\label{fig:cavity_illustration}
\end{figure}

In this section, we provide a solution to the GW-Maxwell equations \eqref{eq:GWMaxwell} in a cavity whose boundary is moving as illustrated in Fig. \ref{fig:cavity_illustration}. In order to take the dynamic boundary condition into account, we first derive an effective surface current at the boundary of the unperturbed cavity which can act as a source of EM fields in the cavity.
A perfectly electrically conducting material can only support electric fields parallel to the surface normal vector $\bm{N}$. Consequently, all tangential components of the electric fields in the local inertial frame of a point on the surface must vanish \cite{Rawson-Harris1972, Ratzinger2024}
\begin{equation}\label{eq:boundary_condition}
      F_{\mu\nu}(x)u^\nu(\tau)T_{\underline{1},\underline{2}}^\mu(\tau)=0\,,
\end{equation}
where $T_{\underline{1},\underline{2}}^\mu$ are tangential vectors to the cavity wall forming a basis $\{u^\mu, T^\mu_{\underline{1}},T^\mu_{\underline{2}},N^\mu\}$ and all expressions are evaluated on the cavity boundary whose movement is described with the displacement field $x^\mu=\bar{x}^\mu+\delta x^\mu$ and the associated velocity $u^\mu$. 

As derived in App.~\ref{ap:EM_boundary_condition}, the GW perturbation of Eq.~\eqref{eq:boundary_condition} implies a surface \emph{tangential} electric field component at the boundary of the \emph{unperturbed} cavity 
\begin{equation}\label{eq:boundary_condition_perturbation}
   \bar{\bm{N}}\times\delta\bm{E}|_{\partial V}=\bar{\bm{N}}\times\left[\partial_0(\bar{\bm{B}}\times\delta\bm{x})-\nabla(\bar{\bm{E}}\cdot\delta\bm{x})\right]_{\partial V}\eqqcolon\bar{\bm{N}}\times\bm{\mathcal{V}}|_{\partial V}\,,
\end{equation}
where we have defined the boundary source vector for the tangential electric field $\bm{\mathcal{V}}$ in the second equality. The right-hand side of Eq. \eqref{eq:boundary_condition_perturbation} can be interpreted as a surface current density, which acts as a source term for $\delta\bm{E}$. The electric field perturbation $\delta\bm{E}$ can then be determined from Maxwell's equations \eqref{eq:GWMaxwell}, while $\bm{\mathcal{V}}$ follows from the mechanical response of the cavity to a GW.

Any \emph{real} conductor will allow tangential electric fields to penetrate the material by a small amount. Describing the losses in the walls with a surface impedance $Z_s$, we can modify the boundary condition of a perfect electric conductor \eqref{eq:boundary_condition} to an impedance (or Leontovich) boundary condition \cite{jackson_classical_1999}
\begin{subequations}\label{eq:boundary_condition_perturbation_damping}
    \begin{align}
     &F_{\mu\nu}(x)u^\mu(\tau)T_{\underline{1}}^\nu(\tau)=Z_sF_{\mu\nu }T_{\underline{1}}^\mu N^\nu\,,\\
     &F_{\mu\nu}(x)u^\mu(\tau)T_{\underline{2}}^\nu(\tau)=-Z_sF_{\mu\nu }T_{\underline{2}}^\mu N^\nu\,,
\end{align}
\end{subequations}
which is the generalization of the condition in flat space $\bm{N}\times\bm{E}=\bm{N}\times(Z_s\bm{B}\times\bm{N})$. Since damping effects are only significant near electromagnetic resonances in the cavity where $\delta F_{\mu\nu}\gg h\bar{F}_{\mu\nu}$, we can simplify the perturbation of the right-hand side in Eq. \eqref{eq:boundary_condition_perturbation_damping} to find
\begin{equation}
    \bar{\bm{N}}\times\delta\bm{E}\big|_{\partial V}=  \bar{\bm{N}}\times(\bm{\mathcal{V}}+Z_s\delta\bm{B}\times\bar{\bm{N}})\big|_{\partial V}+\mathcal{O}(Z_s h\bar{B})\,,
\end{equation}
which is the boundary condition we will use in the following to describe damped oscillations in the cavity.

For the magnetic flux to be conserved through the surface, we require \cite{Ratzinger2024}
\begin{align}\label{eq:boundary_condition_B}
  F_{\mu\nu}T^\mu_{\underline{1}}T^\mu_{\underline{2}}|_\text{in} &= F_{\mu\nu}T^\mu_{\underline{1}}T^\mu_{\underline{2}}|_\text{out} \, ,
\end{align}
where \textit{in/out} means that the field is respectively evaluated slightly inside and outside the conductor. Perturbing the above quantities, we find the surface \textit{normal} component of the magnetic field perturbation at the unperturbed boundary
\begin{subequations}
\begin{align}\label{eq:boundary_condition_perturbation_B}
    \bar{\bm{N}} \cdot \delta \bm{B}|_{\partial V} &=\bar{\bm{N}} \cdot  (\bm{\mathcal{W}}^\text{out}-\bm{\mathcal{W}}^\text{in})|_{\partial V} \equiv \bar{\bm{N}} \cdot  \bm{\mathcal{W}}|_{\partial V} \, ,
\end{align}
where 
\begin{align}
    \bm{\mathcal{W}}^\text{out/in} &= -\left[\delta x^\rho \partial_\rho \bar{\bm{B}}+\bar{\bm{B}}(\nabla\cdot\delta\bm{x})-(\bar{\bm{B}}\cdot\nabla)\delta \bm{x}\right]_\text{out/in}\, ,
\end{align}
\end{subequations}
and we used that oscillating EM fields in the conductor are suppressed $\delta \bm{B}|_\text{in}\ll\delta \bm{B}|_\text{out}\equiv\delta \bm{B}$.
Again, we find an effective surface current which can excite electromagnetic fields within the cavity in addition to further volume currents entering Maxwell's equations. Note that typical experimental setups either have $\bar{\bm{B}}|_\text{in}=\bar{\bm{B}}|_\text{out}$ or $\bar{\bm{B}}|_\text{in}\approx 0$ so that many terms in Eq. \eqref{eq:boundary_condition_perturbation_B} cancel.

\subsection{Eigenmode decomposition}\label{sec:EM_eigenmode_decomposition}
 The electromagnetic eigenmodes of a cavity form a complete basis (under the $L^2$ inner product) on the space of all electromagnetic fields that fulfill the same boundary conditions as the eigenmode. Since the eigenmodes of a cavity with dynamic boundary conditions are impractical to obtain, we attempt to expand the EM field perturbations $\delta \bm{E}$, $\delta \bm{B}$ in terms of the eigenmodes of the \emph{unperturbed} cavity $\{\bm{E}_n\}$, $\{\bm{B}_n\}$, where $n\in\mathbb{N}=\mathbf{I}\cup\mathbf{S}$ indexes all eigenmodes, which we separate into \emph{solenoidal} i.e. $\nabla\cdot\bm{E}_s=\nabla\cdot\bm{B}_s=0$ eigenmodes where $s\in\mathbf{S}$ and \emph{irrotational} i.e. $\nabla\times\bm{E}_i=0$ eigenmodes $i\in\mathbf{I}$, which have $\bm{B}_i=0$, as magnetic fields are solenoidal by nature. Solenoidal modes can be obtained from the equations
\begin{subequations}\label{eq:eigenmode_maxwell}
\begin{align}
    \nabla\times\bm{E}_s&=-i\omega_s\bm{B}_s\,, \\
\nabla\times\bm{B}_s&=i\omega_s\bm{E}_s \,.
\end{align}
\end{subequations}
The irrotational modes are obtained from the equation $\nabla^2\phi_i=-\lambda_i\phi_i$ so that $\bm{E}_i=-\nabla\phi_i$. 
Both types of modes satisfy the unperturbed boundary conditions
\begin{subequations}\label{eq:unperturbed_eigenmode_boundary_condition}
    \begin{align}
    \bm{E}_n\times\bar{\bm{N}}\big|_{\partial V}&=0 \, ,\\
    \bm{B}_n\cdot\bar{\bm{N}}\big|_{\partial V}&=0\,.
\end{align}
\end{subequations}
For irrotational modes, this implies $\nabla\phi_i\times\bar{\bm{N}}=0$ i.e. the potential $\phi_i$ is constant on the surface and can be chosen to vanish $\phi_i|_{\partial V}=0$.
We normalize the eigenmodes so that 
\begin{equation}
    \int dV\bm{E}_l^*\cdot\bm{E}_k=\int dV\bm{B}_l^*\cdot\bm{B}_k=\delta_{lk}V\,.\
\end{equation}
However, any expansion of $\delta \bm{E}$ in terms of unperturbed eigenmodes will inevitably violate the boundary conditions in Eq. \eqref{eq:boundary_condition_perturbation} or Eq. \eqref{eq:boundary_condition_perturbation_damping} as it automatically satisfies Eq. \eqref{eq:unperturbed_eigenmode_boundary_condition}. Therefore, we instead expand the EM field perturbations $\delta \bm{E}$, $\delta \bm{B}$ as
\begin{subequations}
\begin{align}\label{eq:eigenmode_expansion}
\delta\bm{E}(\bm{x},t)&=\sum_n \big(e_n(t)-f_n(t)\big)\bm{E}_n(\bm{x})+\bm{F}(\bm{x},t)\,,\\
\delta\bm{B}(\bm{x},t)&=\sum_n \big(b_n(t)-g_n(t)\big)\bm{B}_n(\bm{x})+\bm{G}(\bm{x},t)\,,
\end{align}
with the overlap coefficients
\begin{align}
    e_n(t)&=\frac{1}{V}\int dV\,\delta\bm{E}(\bm{x},t)\cdot\bm{E}_n^{*}(\bm{x})\,,\\
    b_n(t)&=\frac{1}{V}\int dV\,\delta\bm{B}(\bm{x},t)\cdot\bm{B}_n^{*}(\bm{x})\,,
\end{align}
and
\begin{align}
    f_n(t)&=\frac{1}{V}\int dV\,\bm{F}(\bm{x},t)\cdot\bm{E}_n^{*}(\bm{x})\,,\\
    g_n(t)&=\frac{1}{V}\int dV\,\bm{G}(\bm{x},t)\cdot\bm{B}_n^{*}(\bm{x})\,,
\end{align}
\end{subequations}
where the vector fields $\bm{F}(\bm{x}), \bm{G}(\bm{x})$ are introduced to ensure that boundary conditions are always valid, i.e we have
\begin{subequations}
\begin{align}
       \bar{\bm{N}}\times\bm{F}\big|_{\partial V} &=\bar{\bm{N}}\times \bm{\mathcal{V}}\big|_{\partial V} \label{eq:F_V_eq}\,, \\
       \bar{\bm{N}}\cdot \bm{G}\big|_{\partial V} &=\bar{\bm{N}}\cdot\bm{\mathcal{W}}\big|_{\partial V} \, ,
\end{align}
\end{subequations}
and choose the remaining degrees of freedom so that $\bar{\bm{N}}\cdot\bm{F}\big|_{\partial V}=0=\bar{\bm{N}}\times \bm{G}\big|_{\partial V}$. Then, at the boundary $\bm{F}, \bm{G}$ are respectively orthogonal to all the electric and magnetic components of the unperturbed eigenmodes. This means, we have constructed a quantity $\delta\bm{E}-\bm{F}$ which fulfills the same boundary conditions as the unperturbed eigenmodes $\bm{E}_n$ and then used the completeness of our eigenmode basis to obtain Eq. \eqref{eq:eigenmode_expansion}. The remaining challenge is now to use Maxwell's equations to derive equations of motion for $e_n$ and $b_n$ and use them to expand $\delta \bm{E}$ and $\delta\bm{B}$. 
As shown in App.~\ref{ap:EM}, we find 
\begin{subequations}\label{eq:perturbation_coupling_eom}
\begin{align}
    & \ddot{e}_s+\frac{\omega_s}{Q_s}\dot{e}_s+\omega_s^2e_s=-\partial_tj_s^\text{bulk}-\frac{\omega_s}{Q_s}j_s^\text{bulk}-i\omega_sj_s^\text{bdy},\label{eq:perturbation_coupling_eom_1}\\
    &\ddot{b}_s+\frac{\omega_s}{Q_s}\dot{b}_s+\omega_s^2b_s=-\partial_tj_s^\text{bdy}-i\omega_sj_s^\text{bulk}\,, \\
    &\dot{e}_i=-j_i^\text{bulk}\,,
\end{align}
where we have defined a source term due to the perturbation of the boundary
\begin{align}\label{eq:boundary_current}
    j^\text{bdy}_n\coloneqq -\frac{1}{V}\int_{\partial V} d\bm{A}\cdot\left(\bm{B}_n^{*}\times\bm{\mathcal{V}}\right)\,,
\end{align}
and a source term due to volume currents
\begin{align}\label{eq:bulk_current}
    j_n^\text{bulk}\coloneqq\frac{1}{V}\int dV\bm{E}_n^*\cdot\left(\bm{J}_\text{eff}+\delta\bm{J}\right)\,,
\end{align}
\end{subequations}
with $d\bm{A} = dA\bar{\bm{N}}$, and the magnitude of the damping terms is described by the quality factors $Q_s$. Above, we have also included the perturbation $\delta\bm{J}$ of the current $\bar{\bm{J}}$ which sources the background field $\bar{F}^{\mu\nu}$ for cases where it is not spatially separated from the cavity volume. In the Fourier domain we find the solution
\begin{subequations}\label{eq:EM_mode_coeff_no_F}
\begin{align}\label{eq:EM_mode_coefficient_e_S}
    \tilde e_s(\omega)  &= -i\frac{\omega_s\,\tilde{j}_s^\text{bdy}(\omega)+\omega\,\tilde{j}_s^\text{bulk}(\omega)}{\omega_s^2-\omega^2+\frac{i\omega \omega_s}{Q_s}}\,, \\\label{eq:EM_mode_coefficient_b_S}
    \tilde b_s(\omega)  &= -i\frac{\omega\,\tilde{j}_s^\text{bdy}(\omega)+\omega_s\,\tilde{j}_s^\text{bulk}(\omega)}{\omega_s^2-\omega^2+\frac{i\omega \omega_s}{Q_s}} \,, \\
    \tilde e_i(\omega)  &= \frac{i}{\omega }\tilde{j}_i^\text{bulk}(\omega) \,, 
\end{align}
\end{subequations}
where we have neglected the term $\frac{\omega_s}{Q_s}j_s^\text{bulk}$ in Eq. \eqref{eq:perturbation_coupling_eom_1}, as it is much smaller than $\dot{j}_s^\text{bulk}$ for most relevant frequencies.
Since the lifting functions $\bm{F}, \bm{G}$ can be fully expanded in eigenmodes in the interior $int(V)=V\backslash\partial V$ of the cavity volume, we have $\sum_n f_n\bm{E}_n=\bm{F}$, and the perturbed electric field $\delta \bm{E}$ becomes
\begin{subequations}\label{eq:delta_E_no_F}
\begin{align}\label{eq:delta_E_bulk}
    \delta\tilde{\bm{E}}(\bm{x} \in int(V),\omega) &=\sum_n \tilde e_n(\omega)\bm{E}_n(\bm{x}) \,, \\
    \delta\tilde{{\bm{E}}}_\perp(\bm{x} \in \partial V,\omega)     &=\sum_n \tilde e_n(\omega)\bm{E}_{n,\perp}(\bm{x}) \,, \\
    \delta\tilde{\bm{E}}_\parallel(\bm{x} \in \partial V,\omega)     &=\tilde{\bm{F}} = \tilde{\bm{\mathcal{V}}}_\parallel\, ,\label{eq:delta_E_bounday_parallel}
\end{align}
\end{subequations}
which is discussed in more detail in App.~\ref{ap:EM}. Analogous arguments hold for $\delta \bm{B}$.

\subsection{Coordinate invariance}\label{sec:coordinate_invariance_EM}
The electric field perturbation calculated in the previous section depends on the coordinate system used on linearized curved space. However, the signal observed in an actual experiment can not depend on the arbitrary choice of a coordinate system. The discrepancy arises since the apparatus used to observe the electric field can be perturbed by the GW as well, which can add additional signal contributions. Thus, only the total EM fields locally observed in the reference frame of an observer are truly gauge invariant. Following \cite{Ratzinger2024}, we calculate these coordinate independent fields by using the local coordinate frame spanned by an orthonormal tetrad attached to the worldline of the observer, such as an antenna. The observed fields in the tetrad frame are given in Eqs.~\eqref{eq:observed_E_field} and \eqref{eq:cov_B_field}. The observed electric field can be expanded in our perturbation formalism as
\begin{subequations}\label{eq:delta_E_obs_def_with_E_ant}
\begin{equation}
    \delta  \bm{E}^\text{obs}=\delta\bm{E}+\delta\bm{E}^\text{ant} \,, 
\end{equation}
where $\delta\bm{E}$ is given by our solution in Eqs.~\eqref{eq:delta_E_no_F} and the contribution due to the observer's motion is given by
\begin{equation}\label{eq:delta_E_ant_definition}
    \delta E^\text{ant}_i=\delta x^\rho(\partial_\rho \bar{E}_i)+\bar{E}_j\delta e^j_i+\bar{E}_a\delta u^0+\epsilon_{ijk}\bar{B}_k\delta u^j\,.
\end{equation}
\end{subequations}
The perturbation of the tetrad components along the observer's worldline is given by \cite{Misner1973, Ratzinger2024}
\begin{subequations}
\begin{align}\label{eq:tetrad_time_evo_full}
    \delta \dot{e}_{\underline{\alpha}}^\mu+\Gamma^\mu_{0\nu}\bar{e}^\nu_{\underline{\alpha}}=-(\delta\Omega^\text{FW}+\delta\Omega^\text{SR})_{\;\nu}^{\mu}\bar{e}^\nu_{\underline{\alpha}}\,,
\end{align}
where the rotation tensor perturbation for a Fermi-Walker transport is given by
\begin{align}
   (\Omega^\text{FW})_{\;\nu}^{\mu}= (\delta\Omega^\text{FW})_{\;\nu}^{\mu}=\delta a^\mu\bar{u}_\nu-\delta a_\nu\bar{u}^\mu\,,
\end{align}
and the spatial rotation tensor perturbation due to an elastic displacement of the antenna \cite{Belgacem24}
\begin{align}
     (\Omega^\text{SR})_{\;j}^{i}=(\delta\Omega^\text{SR})_{\;j}^{i}=\frac12\left(\partial^i\delta u_j-\partial_j\delta u^i+\partial^ih_{j0}-\partial_j h^i_{\:\:0}\right)\,, \quad(\delta\Omega^\text{SR})^0_{\;\mu}=0\,,
\end{align}
\end{subequations}
which follows from the covariant form of the vorticity tensor $\Omega^\text{SR}_{\mu\nu}=\frac12P_\mu^{\:\:\rho} P_{\nu}^{\:\:\sigma}(\nabla_\rho u_\sigma-\nabla_\sigma u_\rho)$ where $P_{\mu\nu}=g_{\mu\nu}+u_\mu u_\nu$ is a projection operator which ensures $\Omega^\text{SR}_{\mu\nu}u^\nu=0$ \cite{Ellis71}.
Therefore, we find
\begin{equation}\label{eq:delta_eij_time_evo}
    \delta \dot{e}^i_{\underline{j}}=-\frac12\left(\dot{h}^i_{\,k}+\partial^i\delta u_k-\partial_k\delta u^i\right)\bar{e}^k_{\underline{j}}\,.
\end{equation}

This enables us to demonstrate explicitly that our eigenmode formalism yields the same observed electric field in all coordinate systems, by considering the simple example of a static background magnetic field. In that case the total observed electric field perturbation in Eq. \eqref{eq:delta_E_obs_def_with_E_ant} is given by
\begin{equation}\label{eq:observed_E_static_B}
    \delta  \bm{E}^\text{obs}=\delta\bm{E}-\bar{\bm{B}}\times\delta\bm{u}^\text{ant}\,,
\end{equation}
where $\delta \bm{u}^\text{ant}$ describes the velocity of the observer (usually an antenna) and is not necessarily related to the movement of the wall $\delta\bm{u}$ used in the previous section.
Under an $\mathcal{O}(h)$ coordinate transformation $x'^\mu=x^\mu+\xi^\mu$, the GW strain components change as 
\begin{equation}\label{eq:gauge_trafo_h}
    h_{\mu\nu}'=h_{\mu\nu}-\partial_\mu\xi_\nu-\partial_\nu\xi_\mu\,,
\end{equation} which leads to the transformation of the effective current Eq. \eqref{eq:Jeff} for static magnetic fields
\begin{equation}\label{eq:gauge_transfo_j}
    \bm{J}'_\text{eff}=\bm{J}_\text{eff}-\bm{\bar{B}}\times\ddot{\bm{\xi}}-\nabla\times\left[\nabla\times\left(\bm{\bar{B}}\times\bm{\xi}\right)\right]\,.
\end{equation}
Importantly, $\bm{J}_\text{eff}$ does not transform like a vector field. The electric boundary source vector in Eq. \eqref{eq:boundary_current} becomes 
$\bm{\mathcal{V}}=\bm{\bar{B}}\times\delta\bm{u}$ and transforms as 
$\bm{\mathcal{V}}'=\bm{\mathcal{V}}+\bm{\bar{B}}\times\bm{\dot{\xi}}$. Furthermore, the real current perturbation is invariant $\delta\bm{J}'=\delta\bm{J}$ because there is no background $\bar{\bm{J}}=0$.
Using these transformations, as shown in App.~\ref{ap:EM}, we find that all the expansion coefficients transform as
\begin{align}\label{eq:mode_e_transformation}
    \tilde e'_n (\omega) &=\tilde e_n(\omega) +\frac{i\omega}{V}\int dV\bm{E}^{*}_n\cdot\left(\bm{\bar{B}}\times\tilde{\bm{\xi}}\right)\, .
\end{align}
Thus, we find the transformation of the field from Eq.~\eqref{eq:delta_E_no_F}
\begin{equation}\label{eq:delta_E_transformation}
    \delta\tilde{\bm{E}}'(\bm{x} \in int(V))=\delta\tilde{\bm{E}}+\frac{i\omega}{V}\sum_n\bm{E}_n\int dV\bm{E}^{*}_n\cdot\left(\bm{\bar{B}}\times\tilde{\bm{\xi}}\right)= \delta \tilde{\bm{E}}+i\omega\bm{\bar{B}}\times\tilde{\bm{\xi}}\,,
\end{equation}
where we have used that our eigenmode decomposition forms a complete basis for all physical fields in the interior of the perturbed cavity. The tangential component of the field at the surface is given by $\bm{F} = \bm{\mathcal{V}}_\parallel$ in Eq. \eqref{eq:delta_E_bounday_parallel} which transforms in the same way as above. In other words, $\delta \bm{E}$ transforms the same way inside the cavity and at its surface.
Finally, we can evaluate the transformation of Eq.~\eqref{eq:observed_E_static_B} by using the transformation of the antenna contribution 
$(\delta \bm{E}^\text{ant})'=\delta \bm{E}^\text{ant}+\bm{\bar{B}}\times\bm{\dot{\xi}}$
\begin{equation}
    \delta\tilde{\bm{E}}^\text{obs}=\delta\tilde{\bm{E}}'-\bar{\bm{B}}\times(\delta{\tilde{\bm{u}}^{\text{ant}}})'=\delta\tilde{\bm{E}}-\bar{\bm{B}}\times\delta\tilde{\bm{u}}^\text{ant}\,,
\end{equation}
and can confirm the expected coordinate invariance.
Note that the transformation in Eq.~\eqref{eq:delta_E_transformation} is consistent with the coordinate transformation of the field strength tensor directly
\begin{equation}
    \delta F_{i0}'=\delta F_{i0}+\frac{\partial x^j}{\partial x'^ i}\frac{\partial x^k}{\partial x'^0}\bar{F}_{jk}=\delta F_{i0}-\dot{\xi}^k\bar{F}_{ik}+\mathcal{O}(h^2)\,,
\end{equation}
which confirms that the eigenmode overlap formalism preserves coordinate invariance, and we can calculate our coupling coefficients in whatever frame is most convenient.

While true gauge invariance is only guaranteed for the observed field, we still find approximate gauge invariance for the mode coefficients near resonances where both $e'_s$ and $e_s$ in Eq. \eqref{eq:mode_e_transformation} are enhanced by a quality factor $Q_s$ and the relative difference is thus only $\mathcal{O}(Q_s^{-1})$, as can be seen from Eqs. \eqref{eq:perturbation_coupling_eom}.

\section{GW perturbation of elastic solids with boundary conditions}\label{sec:elasticy_theory}

Beside the electromagnetic response of a cavity to a GW, the second important physical response is the mechanical deformation of the cavity walls due to an incoming GW, as it determines the boundary condition for the EM fields in Eq.~\eqref{eq:boundary_condition_perturbation} and creates the surface current Eq. \eqref{eq:boundary_current}. The covariant equations and boundary conditions for the displacement field $\delta \bm{x}$ of an elastic solid are given by \cite{Hudelist_2023}
\begin{subequations}
\begin{align}\label{eq:mech_elasticity_equation}
    \rho\delta\ddot{x}^i-\partial_j\delta\sigma^{ij}&=F_g^i+\partial_j\sigma_h^{ij}\,,\\
    \delta\sigma^{ij}\bar{n}_j\big|_{\partial v_\text{f}}&=-\sigma_h^{ij}\bar{n}_j\big|_{\partial v_\text{f}}\,,\label{eq:mech_boundary_condition}\\
    \delta x^i\big|_{\partial v_\text{c}}&=\delta x^i_\text{ext}\big|_{\partial v_\text{c}}\,,\label{eq:mech_boundary_condition_Dirichlet}
\end{align}
where we have separated the stress tensor in flat space
\begin{align}\label{eq:stress_tensor_flat}
    \delta\sigma^{ij} = \lambda\delta^{ij}\nabla\cdot\delta\bm{x}+\mu(\partial^i\delta x^j+\partial^j\delta x^i) + \delta \sigma^{ij}_\text{damp.}\, ,
\end{align}
from its perturbation due to a GW
\begin{align}\label{eq:stress_tensor_h}
    \sigma_h^{ij}=\frac\lambda2\delta^{ij}h^k_k+\mu h^{ij}+\delta \sigma^{ij}_\text{$h$, damp.}\,,
\end{align}
where $\lambda$ and $\mu$ are the scalar Lamé parameters, $\bar{\bm{n}}$ is the normal vector field of the unperturbed wall boundaries, $\rho$ is the mass density of the cavity walls and $\delta \sigma^{ij}_\text{damp.}$ and $\delta \sigma^{ij}_\text{$h$, damp.}$ account for internal energy dissipation (through the viscous parameters of the body $\eta,\zeta$, see App.~\ref{ap:elasticity}). Furthermore, we have defined the GW tidal force density\footnote{In contrast to the expression in the long-wavelength limit in PD coordinates $F_g^i=\frac{\rho}{2}\ddot{h}_{ij}^\text{TT}\bar{x}^j$, we find $\bm{k}_g\cdot\bm{F}_g\neq0$  when $\bm{k}_g\cdot\bm{x}\gtrsim1$, where $\bm{k}_g$ is the wave vector of the GW, i.e. the GW force is no longer purely transverse.}
\begin{align}
    F_g^i=-\frac{\rho}2(2\dot{h}_0^{\,i}-\partial^ih_{00})\,.
\end{align}
\end{subequations}
In Eqs. \eqref{eq:mech_boundary_condition} and \eqref{eq:mech_boundary_condition_Dirichlet}, we have included the possibility of mixed boundary conditions where $\partial v_\text{f}$ is the part of the boundary which is subject to the Neumann boundary condition \eqref{eq:mech_boundary_condition} of a free oscillation, and where $\partial v_\text{c}$ is the part of the surface which is constrained by an external mechanical system due to a Dirichlet boundary condition \eqref{eq:mech_boundary_condition_Dirichlet}, i.e where the movement is dictated by an externally applied displacement $\delta \bm{x}_\text{ext}$. For example, if parts of the wall are held fixed in PD coordinates, we have a clamped boundary condition in PD $\delta x^i_\text{PD}\big|_{\partial v_\text{c}}=0$,\footnote{Physically, this is only ever possible until some GW frequency, above which the supporting structure starts vibrating with a comparable amplitude as the cavity walls. This threshold frequency is approximately linear in the speed of sound in the material $v_s \sim \sqrt{\lambda/\rho} \sim \sqrt{\mu/\rho}$. In this work we set $\partial v_c$ to zero and use $\partial v_\text{f}\equiv\partial v$.} but not in TT. 

\subsection{Eigenmode decomposition}\label{sec:mech_eigenmode_decomposition}
Just like in the electromagnetic case, we can solve the equations by using the eigenmodes of the unperturbed cavity.
In certain configurations, the GW does not induce a perturbation of the boundary condition Eq.~\eqref{eq:mech_boundary_condition} at lowest order, such that a perturbative expansion of $\delta\bm{x}$ in eigenmodes can be exact. In PD coordinates, this is the case in the long wavelength limit $\omega \bar{x} \ll 1$, as pointed out in e.g. \cite{Ratzinger2024, Belgacem24}. However, this description becomes incomplete in the more general situation where $\sigma^{ij}_h$ cannot be neglected and the boundary conditions become dynamical. Therefore, we construct a quantity again that fulfills the unperturbed boundary conditions by definition and expand the solution to Eq.~\eqref{eq:mech_elasticity_equation} as
\begin{subequations}
\begin{align}\label{eq:mech_eigenmode_expansion}
    \delta\bm{x}(\bm{x}, t)=\sum_m (q_m(t)-y_m(t))\bm{U}_m(\bm{x})+\bm{y}(\bm{x}, t)\,,
\end{align}
where the modes with $m\in\mathbb{N}$ are made up of \emph{solenoidal} i.e. $\nabla\cdot\bm{U}^S_m=0$ and \emph{irrotational} i.e. $\nabla\times\bm{U}_m^I=0$ contributions with the same eigenfrequency. In the following, we take the eigenmodes to be real, $\bm{U}_m \in \mathbb{R}^3$. The expansion coefficients are given by the overlap integrals 
\begin{align}
q_m(t)=\frac{1}{M}\int dv\,\rho\,\delta\bm{x}(\bm{x}, t)\cdot\bm{U}_m(\bm{x})\,,
\end{align}
and
\begin{align}
y_m(t)=\frac{1}{M}\int dv\,\rho\,\bm{y}(\bm{x}, t)\cdot\bm{U}_m(\bm{x})\,.
\end{align}
\end{subequations}
The mechanical eigenmodes are obtained from
\begin{subequations}\label{eq:mech_eigenmode_elasticity_eq}
\begin{align}
\rho\omega_m^2U_m^{i}+\partial_j\bar{\sigma}^{ij}_m&=0 \, \label{eq:mech_eigenmode_equation}\,,\\
\bar{\sigma}^{ij}_m\bar{n}_j\big|_{\partial v_\text{f}}&=0\,,\label{eq:mech_eigenmode_boundary}\\
U_m^i\big|_{\partial v_\text{c}}&=0\,,
\end{align}
where 
\begin{align}
    \bar{\sigma}^{ij}_m &= \lambda\delta^{ij}\nabla\cdot\bm{U}_m+\mu(\partial^i U^j_m+\partial^j U^i_m)\, , 
\end{align}
\end{subequations}
and the function $\bm{y}$ is enforcing the boundary condition
\begin{align}\label{eq:BC_y}
&\left(\lambda\delta^{ij}\nabla\cdot\bm{y}+\mu(\partial^i y^j+\partial^j y^i)+\eta\partial^j  \dot{y}^i +\left(\frac{\eta}{3}+\zeta\right)\delta^{ij}\partial_k \dot{y}^k\right)\bar{n}_j\big|_{\partial V} =-\sigma_h^{ij}\bar{n}_j\big|_{\partial V}\,.
\end{align}
In Eqs.~\eqref{eq:mech_eigenmode_elasticity_eq}, we choose to construct the eigenmodes as the solutions of the undamped equations. While damping adds a contribution to those equations, including the boundary condition Eq.~\eqref{eq:mech_boundary_condition}, this effect is absorbed into a surface quality factor (see App.~\ref{ap:elasticity}), as for EM modes.
For the modes, we choose the normalization
\begin{align}
    \int dv\,\rho\,\bm{U}_k \cdot \bm{U}_l&=\delta_{kl}\int dv\,\rho\eqqcolon M \delta_{mn}\,,
\end{align}
where $M$ denotes the total mass of the cavity walls.

As derived in App.~\ref{ap:elasticity}, the equations of motion for the mode coefficients read
\begin{subequations}
\begin{align}\label{eq:mech_EOM_q}    \ddot{q}_m+\frac{\omega_m}{Q_m}\dot{q}_m+\omega^2_mq_m=\frac{1}{M}\left(f_m^\text{bulk}+f_m^\text{bdy}\right)\,,
\end{align}
where we have separated forces acting within the wall
\begin{align}\label{eq:f_bulk}
    f_m^\text{bulk}=\int dv\,\bm{U}_m\cdot\left(\bm{F}_g+\nabla\bm{\sigma}_h\right) \, ,
\end{align}
from a surface pressure due to the modified boundary condition
\begin{align}\label{eq:f_boundary}
    f_m^\text{bdy}=-\int_{\partial v_\text{f}} d \bm{a} \cdot (\bm{\sigma}_h \cdot \bm{U}_m)-\int_{\partial v_\text{c}} d \bm{a} \cdot ( \bm{\bar{\sigma}}_m\cdot\delta\bm{x}_\text{ext}) \, ,
\end{align}
\end{subequations}
and where the quality factors differ between irrotational and solenoidal modes (see App.~\ref{ap:elasticity}). In Fourier space, the solution to this equation is
\begin{align}\label{eq:mech_coeff_general}
    \tilde q_m(\omega) &= \frac1 M\frac{\tilde{f}_m^\text{bulk}(\omega)+\tilde{f}_m^\text{bdy}(\omega)}{\omega^2_m - \omega^2 +i\frac{\omega \omega_m}{Q_m}}\, .
\end{align}
Similarly as in the EM case, the lifting function $\bm{y}$ is only relevant at the boundary, as dictated by Eqs.~\eqref{eq:BC_y}. Since the eigenmodes are only unable to fully expand derivatives of $\bm{y}$ at the boundary, their contribution cancels entirely in the displacement 
\begin{subequations}\label{eq:delta_x_no_y}
\begin{align}\label{eq:delta_x_solution_expansion}
    \delta\tilde{\bm{x}}(\bm{x},\omega) =\sum_m \tilde q_m(\omega)\bm{U}_m(\bm{x}) \, .
\end{align}
However, this does not hold for spatial derivatives of the displacement and we need to use
\begin{equation}\label{eq:nabla_delta_x_with_y}
    \partial_i\delta\tilde{x}^j(\bm{x} \in \partial v,\omega)     =\sum_m \tilde q_m(\omega)\partial_iU_{m}^j(\bm{x})+\partial_i\tilde{\mathcal{Y}}(\bm{y})^j \, ,
\end{equation}
at the boundary, where we defined
\begin{align}
    \tilde{\bm{\mathcal{Y}}}(\bm{y}) = \tilde{\bm{y}}-\frac{1}{M}\sum_m\bm{U}_m\int \, dv \, \rho \,\bm{U}_m \cdot \tilde{\bm{y}}  \, .
\end{align}
\end{subequations}
In the bulk, the contribution from $\bm{y}$ to $ \partial_i\delta\tilde{x}^j$ cancels as for $\delta\tilde{x}^j$.
An example for this is described in App.~\ref{sec:toy_example}, where Fig. \ref{fig:toy_example_neumann} demonstrates the convergence of $\delta\bm{x}$ to an exact solution without using $\bm{y}$, while the lifting function is crucial for describing derivatives of $\delta\bm{x}$. Therefore, it is usually not necessary to solve Eq.~\eqref{eq:BC_y} for $\bm{y}$, since only derivatives of $\bm{y}$ remain when using $\delta\bm{x}$ in derived quantities.
Note that, since $\bm{\mathcal{Y}}$ is non-resonant, its contribution is negligible to $\mathcal{O}(Q_m^{-1})$ near a mechanical resonance.

The expression for the mode coefficients Eq.~\eqref{eq:mech_coeff_general} can be simplified depending on the frequency regime of the experiment and on the frame in which the calculation of the displacement is performed. When expressed in the PD frame at long GW wavelengths $\omega_g L \ll 1$ and using Eq. \eqref{eq:TT_PD_coord_transfo}, the GW force becomes the well known expression $F_g^{\text{PD},i}\approx\frac{\rho}{2}\,\partial^ih_{00}^{\text{PD}}\approx-\rho\omega^2 h^{\mathrm{TT},i}_{\,j}\bar{x}^j/2$, and typically dominates over the bulk source term $\nabla\bm{\sigma_h}$ and the boundary term \eqref{eq:f_boundary}, as they are suppressed by the factor $\mu/(\rho\omega \bar{x})\sim v_s^2/\omega \bar{x}\sim10^{-6}(v_s/1\,\text{km}\,\text{s}^{-1})^2(2\pi\,\text{kHz}/\omega)(1\,\text{m}/\bar{x})$, where $v_s$ is the speed of sound in the material. Therefore, we recover the known results from e.g. \cite{Ratzinger2024, Belgacem24} in this limit, where the boundary condition is approximately not perturbed by the GW. The expansion coefficients in frequency space then simplify to
\begin{align}\label{eq:mech_EOM_q_PD_low_freq}
    &\tilde q^\mathrm{PD}_m (\omega) =\frac{-\omega^2\int dv\,\rho\,\bm{U}_m \cdot \tilde{\bm{h}}^\mathrm{TT} \cdot \bm{x}}{2M\left(\omega^2_m - \omega^2 +i\frac{\omega \omega_m}{Q_m}\right)} \, ,
\end{align}
which is the solution of the well known equation of motion of the mode coefficients at long wavelength in the PD frame, see e.g. \cite{berlin_mago20_2023, Belgacem24}. As expected, there is no boundary term anymore and the GW perturbation acts as a tidal force density.\footnote{This is because the GW wavelength is much larger than the cavity size such that the background can be considered as flat and all the effects of the GW can be treated as an external Newtonian force.} Since the boundary condition is not perturbed at lowest order, one can also set $\bm{y}^\mathrm{PD}=0$ in this regime. Another important regime  is reached when $\omega \ll \omega_{m0}$, where $\omega_{m0}$ is the lowest resonant frequency, and thus $\tilde{q}^\text{PD}_m \ll \tilde{h}^\text{TT} L$, which is the so-called \emph{rigid limit}.

In TT coordinates, $\bm{F}_g = \nabla \bm{\sigma}_h=0$, i.e the GW only affects the boundary condition \eqref{eq:mech_boundary_condition}. In this case, the expansion coefficients simplify to
\begin{align}\label{eq:mech_EOM_q_TT}
\tilde q^\mathrm{TT}_m(\omega)  &= -\frac{\mu \int_{\partial v_\text{f}} d \bm{a} \cdot (\tilde{\bm{h}}^\mathrm{TT} \cdot \bm{U}_m)}{M\left(\omega^2_m - \omega^2 +i\frac{\omega \omega_m}{Q_m}\right)}\,,
\end{align}
where we considered only free boundary conditions. As expected, the coefficients are only proportional to a boundary term. Interestingly, in contrast to Eq. \eqref{eq:mech_EOM_q_PD_low_freq} the displacement in TT coordinates directly depends on the material parameters, which could be seen as conflicting with the equivalence principle (EP). However, for an extended solid supported by elastic forces, the EP only applies to the center of mass motion of the solid and not to tidal deformations of the structure. Here, we compute the displacement of the material \textit{with respect to} the center of mass, such that the result is consistent with the EP.

\subsection{Elastic versus pure free falling limit}\label{sec:free_fall_limit}
A common assumption is that GW detectors excited at frequencies much larger than the lowest mechanical resonances behave like a cloud of disconnected particles and are not mechanically deformed in TT coordinates \cite{Domcke_Dielectric}. This is usually called a \emph{free falling limit}. However, we will see that elastic solids never enter free fall in the same sense as a group of disconnected particles and the damping mechanism of the mechanical oscillator can not be neglected when defining a free falling regime. Therefore, we will need to distinguish between \emph{elastic} and \emph{pure} free fall.

First, let us consider the behavior of the coefficients $q_m$ for increasing GW frequency $\omega_g$. For $\omega_g=\omega_m$, Eq. \eqref{eq:mech_EOM_q_TT} scales as 
\begin{equation}\label{eq:q_TT_FF_scaling}
    \tilde q^\mathrm{TT}_m(\omega_m) \sim \tilde h^\mathrm{TT} L Q_m \frac{L}{t_{\text{wall}}}\left(\frac{v_s}{\omega_m L}\right)^{2+\alpha}\,,
\end{equation} 
where $L$ is the typical length scale of the cavity, $t_{\text{wall}}$ the thickness of the walls and $\alpha$ parametrizes how the dimensionless overlap integral scales with $v_s/\omega_m L$.\footnote{A simple estimate of the dimensionless surface integral in a cubic cavity leads to a scaling $\frac{1}{L^2h}\int d\bm{a}\cdot{(\bm{h}^\text{TT}\cdot\bm{U}_m)}\sim (v_s/\omega_m L)^2$ when $\omega_m \rightarrow \infty$.} Note that for high frequencies, the resonant peaks are so close to each other in frequency space that any GW frequency will be on resonance. We can measure the smallness of the displacement by comparing with the displacement induced in a non-resonant object made of strongly interacting particles, where $\delta x^\mathrm{TT} \sim h^\mathrm{TT} L$, i.e. the rigid limit in TT coordinates. Therefore, when $\omega_m L/v_s \gg (Q_m L/t_{\text{wall}})^{1/(2+\alpha)}$, the coefficients $q_m$ are suppressed. We also expect the lifting function to decrease $\bm{y}\propto\omega_g^{-1}$, however it does not enter $\delta\bm{x}$. Therefore, this is the condition on $\omega_g$ we denote \emph{elastic freely falling (FF) limit}. As shown in App. \ref{ap:elasticity}, the mechanical quality factor decreases linearly with frequency such that we can parametrize it as $Q_m = Q_{m0} \omega_{m0}/ \omega_m$, where $Q_{m0}$ corresponds to the quality factor of the first resonance at the frequency $\omega_{m0}$. Parametrically, $\omega_{m0}\sim v_s/L$, such that, using Eq. \eqref{eq:q_TT_FF_scaling}, the condition for elastic free fall $q_m^\text{TT}\ll h^\text{TT}L$ is equivalent to 
\begin{equation}\label{eq:free_falling_limit}
    \left(\frac{\omega_g L}{v_s}\right)^{3+\alpha} \gg Q_{m0}\frac{L}{t_{\text{wall}}}\,.
\end{equation} 
For typical normal conducting cavities suspended in superfluid helium, we expect $Q_{m0} \gg 1$ and a ratio of cavity length over width $L/t_{\text{wall}} \gg 1$ such that the naive condition $\omega L/v_s \gg 1$ is not accurate anymore and the system will enter elastic free fall at different frequencies.

However, the same argument does not apply to spatial derivatives of $\delta\bm{x}$ as can be seen in the $\mathcal{\bm{Y}}$ contribution in Eq. \eqref{eq:nabla_delta_x_with_y}. This can also be understood independent of our formalism, since the boundary condition \eqref{eq:mech_boundary_condition} in TT coordinates requires $\delta\bm{\sigma}$ to not scale with frequency (unless the GW direction is aligned with the surface normal). Therefore, free elastic solids do not allow $\nabla\delta\bm{x}^\text{TT}|_{\partial v}\sim\delta\bm{\sigma}|_{\partial v}\to0$ as $\omega_g\to\infty$ even though $\delta\bm{x}^\text{TT}\to0$ does happen (see App.~\ref{sec:toy_example} for an example). This is in contrast to a \emph{non-elastic solid} (i.e. $\mu=\lambda=0$), which has the solution $\delta\bm{x}^\text{TT}=\nabla\delta\bm{x}^\text{TT}=0$ and all terms involving the displacement field vanish. An example of such a detector is an optical cavity built from infinitesimally thin freely falling mirrors. This \emph{pure free fall} is therefore not obtained as a high-frequency limit of elastic free fall. In the two examples of detectors we discuss later, we will find that elastic and pure free fall lead to the same signal power on resonance, since no spatial derivatives of the displacement field appear in the coupling parameters. However, we use perfect-conducting boundary conditions for the background EM fields to derive this equivalency, and including losses could lead to $\mathcal{O}(Q_n^{-1})$ terms violating it. 
Further, the distinction between elastic and pure free fall mainly applies to symmetric components of the Jacobian $\nabla\delta\bm{x}$, as only symmetric combinations of the Jacobian are constrained by the boundary condition \eqref{eq:mech_boundary_condition}. In particular, vorticity $\bm{V}_{\delta x}=\frac12(\nabla\delta\bm{x}-\nabla\delta\bm{x}^T)$, which ultimately affects the perturbation of an observer's tetrad from Eq. \eqref{eq:delta_eij_time_evo}, is not constrained by \eqref{eq:mech_boundary_condition}. Therefore, it is not guaranteed that vorticity also stays constant at high frequencies, where $\bm{V}_{\delta x}\to \bm{V}_y-\sum_my_m\bm{V}_{U_m}\,$.\footnote{In a two-dimensional toy example, we found that expanding the vorticity of an elastic lifting function $\bm{y}$ in up to 100 mechanical eigenmodes obtained from simulation did not converge to $\bm{V}_{y}$ at the boundary of an elastic solid.}  Importantly, this possible distinction is never resonantly enhanced and often negligible in resonant experiments, like electromagnetic cavities. However, a detailed investigation of cases where elastic and pure free fall need to be distinguished is beyond the scope of this work.

Note that we expect the equations of linear elasticity to break down above some cutoff frequency, when the GW wavelength becomes similar to the lattice constant of a metal, which happens typically around $10^{18}$ Hz. However, we leave an analysis of the free falling limit in this regime to future work.

\subsection{Coordinate invariance}
We can verify that our result in Eq.~\eqref{eq:mech_EOM_q} is also consistent with the gauge transformation $x^\mu\to x^\mu+\xi^\mu$. Using Eq. \eqref{eq:gauge_trafo_h}, we find $\bm{F}_g'=\bm{F}_g+\rho\ddot{\bm{\xi}}$ and 
\begin{equation}
    (\sigma_h^{ij})'=\sigma_h^{ij}-\delta^{ij}\left(\lambda+\left(\zeta-\frac{2\eta}{3}\right)\partial_t\right)\partial_k\xi^k-\left(\mu+\eta \partial_t\right)(\partial^i\xi^j+\partial^j\xi^i)\,.
\end{equation} 
Using the same relations that led to Eq.~\eqref{eq:mech_EOM_q}, we find 
\begin{equation}
    \tilde q'_m(\omega) =\tilde q_m(\omega) +\frac{1}{M}\int dv\,\rho\,\bm{U}_m\cdot\bm{\xi}\,. 
\end{equation}
 Since $\{\bm{U}_m\}$ is a complete basis of physical displacement fields \textit{inside} the cavity walls, we find from Eq.~\eqref{eq:mech_eigenmode_expansion} that $\delta\bm{x}'=\delta\bm{x}+\bm{\xi}$, which is consistent with the gauge transformation for a position vector. 

\section{Electromagnetic back-action on elastic solids}\label{sec:back_action}

In Sec.~\ref{sec:covariant_EM}, we computed the EM mode coefficients, which depend on the perturbation of the boundary conditions through $\bm{\mathcal{V}}$ in Eq. \eqref{eq:boundary_current}. In Sec.~\ref{sec:elasticy_theory}, we computed the displacement field $\delta\bm{x}$ inside the cavity walls in the presence of a GW, from
Eqs.~\eqref{eq:mech_elasticity_equation} and \eqref{eq:mech_boundary_condition}. Those equations describe a linear elastic medium, but they neglect the electromagnetic fields in which the cavities typically operate. We now take into account the EM pressure from this background affecting the elasticity equations. This includes a deformation due to a monochromatic background EM field and dynamic forces due to the simultaneous presence of the background- and GW induced fields. The former is the sum of a static deformation and a dynamical one oscillating at $2\omega_0$. We will assume that the static deformations are already accounted for in the EM and mechanical eigenmodes, while the dynamical one will act as a source of noise. In this work, we will focus on the GW-induced signal and not the noise of the apparatus, and therefore not evaluate this effect. 

\subsection{Equations of motion}
The stress-energy tensor of the EM fields is given by
\begin{equation}
T_\text{EM}^{\mu\nu}=g_{\rho\sigma}F^{\mu\rho}F^{\nu\sigma}-\frac14 g^{\mu\nu}F^{\rho\sigma}F_{\rho\sigma} \, ,
\end{equation}
which causes mechanical stress according to the Maxwell stress tensor $\sigma_\text{EM}^{ij}=-T^{ij}_\text{EM}$ \cite{lan84}. Therefore, the equations of linear elasticity need to be modified to 
\begin{subequations}
\begin{align}\label{eq:elasticity_and_EM_equations}
    \rho\delta\ddot{x}^i-\partial_j\delta\sigma^{ij}=F_g^i+F_L^i+\partial_j\sigma_h^{ij}\,,
\end{align}
where we split $\sigma_\text{EM}^{ij}=\bar{\sigma}_\text{EM}^{ij}+\delta\sigma^{ij}_\text{EM}$ and where $F_L^\mu=-\partial_\nu \delta T^{\mu\nu}_\text{EM}-\Gamma_{\nu\rho}^\mu \bar{T}_\text{EM}^{\rho\nu}-\Gamma_{\nu\rho}^\nu \bar{T}_\text{EM}^{\mu\rho}$ is the Lorentz force, and
\begin{align}\label{eq:EM_stress_tensor_perturbation}
    \delta\sigma^{ij}_\text{EM}=\,-&\bar{F}^{i\rho}\delta F^j_{\:\:\rho}-\delta F^{i\rho}\bar{F}^{j}_{\:\:\rho}+\frac{1}{2}\eta^{ij}\bar{F}^{\rho\sigma}\delta F_{\rho\sigma}\\\nonumber
    &+h^{\rho\sigma}\bar{F}^{i}_{\:\:\sigma}\bar{F}^{j}_{\:\:\rho}+h^{i\rho}\bar{F}^{j\sigma}\bar{F}_{\rho\sigma}+h^{j\rho}\bar{F}^{i\sigma}\bar{F}_{\rho\sigma}-\frac{1}{4}h^{ij}\bar{F}^{\rho\sigma}\bar{F}_{\rho\sigma}-\frac{1}{2}\eta^{ij}h^{\sigma\delta}\bar{F}^\rho_{\:\:\delta}\bar{F}_{\rho\sigma} \, .
\end{align}
\end{subequations}
Using this, the free boundary conditions at an interface between two media becomes
\begin{equation}\label{eq:elasticity_and_EM_boundary_condition}
\bar{n}_j(\delta\sigma^{ij}+\sigma_h^{ij}+\delta\sigma_\text{EM}^{ij})\big|_{\partial v^\text{in}}=\bar{n}_j\delta\sigma_\text{EM}^{ij}\big|_{\partial v^\text{out}}\,,
\end{equation}
where we have assumed that outside of the elastic solid is a vacuum with $\mu=\lambda=0$, but have included the possibility of EM fields within and outside the cavity wall. In the following, we do not explicitly consider the deformations of the boundary of the unperturbed cavity $\partial v$ due to the background EM fields (also known as Lorentz force detuning \cite{Padamsee98}), i.e we neglect $\bar{\sigma}^{ij}_\mathrm{EM} \bar{n}_j$ in Eq.~\eqref{eq:mech_eigenmode_boundary} and assume this effect is already absorbed in the definition of the eigenmodes.  In principle, in Eq.~\eqref{eq:elasticity_and_EM_boundary_condition}, an additional term of the form $\delta n_j(\bar{\sigma}_\text{EM}^{ij}|_{\partial v^\text{in}}-\bar{\sigma}_\text{EM}^{ij}|_{\partial v^\text{out}})$ should arise, where $\delta n$ is the perturbation of the wall normal vector due to the GW. However, this term is negligible since $\bar{\sigma}_\text{EM}\sim 10^6\,\text{Pa}\,(\bar{B}/1\,\text{T})^2\ll \sigma_h/h\sim\mu\sim 10^{10}\,\text{Pa}$ for typical material parameters.
Including these terms and assuming only the free boundary condition Eq.~\eqref{eq:mech_eigenmode_boundary}, the equations of motions for the mode expansion coefficients become
\begin{subequations}
\begin{align}\label{eq:mech_coeff_EoM_back_action}
    \ddot{q}_m+\frac{\omega_m}{Q_m}\dot{q}_m+\omega^2_mq_m&=\frac{1}{M}\left(f_m^\text{bulk}+f_m^\text{bdy}+f_m^L+f_m^\text{EM}\right)\,,
\end{align}
where we have defined the overlaps
\begin{align}\label{eq:back_action_Lorentz_force}
    f_m^L&=\int dv\,\bm{U}_m\cdot\bm{F}_L\,,\\\label{eq:back_action_EM_force}
    f_m^\text{EM}&=\int_{\partial v} d\bm{a}\cdot\left(\delta\bm{\sigma}_\text{EM}^\text{out}-\delta\bm{\sigma}_\text{EM}^\text{in}\right)\cdot\bm{U}_m\,,
\end{align}
\end{subequations}
and $\delta\bm{\sigma}_\text{EM}^\text{out/in}=\lim_{\epsilon\to0}\delta\bm{\sigma}_\text{EM}(\bar{\bm{x}}\pm\epsilon\bar{\bm{n}})$ for $\bar{\bm{x}}\in\partial v$.\footnote{In reality, $\delta\bm{\sigma}_\text{EM}^\text{in}$ is not necessarily evaluated exactly at the inner boundary. For example, oscillating components with wavelength $\lambda$ which decay inside the conductor as $e^{-x/\lambda_p}$, where $\lambda_p$ is the penetration depth, are effectively evaluated where their amplitude becomes negligible compared to its incident amplitude, as long as $\lambda \gg \lambda_p$.}

We now compute the equations of motion for the EM field coefficient taking this effect into account.  Here, we only present the coupled equations of motion assuming solenoidal modes, because, as we shall see throughout the rest of the paper, back action effects are only relevant near EM resonances. As we will focus on the electric field signal in this work (see Sec.~\ref{sec:signal}) and since the electric coefficient of the irrotational mode does not depend on the displacement, see Eq.~\eqref{eq:EM_mode_coeff_no_F}, back action will have no effect on irrotational signal contributions. 

In order to allow a static or monochromatic oscillating background field, we write 
\begin{subequations}\label{eq:B0_E0_definition}
\begin{align}
    &\bar{\bm{B}}(t,\bm{x})=\text{Re}[\,e^{i\omega_0t}\bm{B}_0(\bm{x})]\,, \\ &\bar{\bm{E}}(t,\bm{x})=\text{Re}[\,e^{i\omega_0t}\bm{E}_0(\bm{x})]\,,
\end{align}
\end{subequations}
where $\bm{B}_0$ and $\bm{E}_0$ can be complex, to account for a possible phase shift between electric and magnetic fields. For example, they could be given by an EM eigenmode of the cavity itself. From Eqs.~\eqref{eq:EM_mode_coefficient_e_S}, \eqref{eq:boundary_condition_perturbation} and \eqref{eq:mech_coeff_EoM_back_action}, this leads to the coupled equations $\tilde e_s(\omega)\supset \tilde j_s^\text{bdy}\supset A \tilde q_m(\omega-\omega_0)+B\tilde q_m(\omega+\omega_0)$ and $\tilde q_m(\omega)\supset \tilde f_m^\text{EM}+\tilde f_m^L \supset C \tilde e_n(\omega-\omega_0)+ D\tilde e_n(\omega+\omega_0)$, where $A,B,C,D$ are coupling coefficients. This means that the mechanical and EM oscillations are up- and down converted by the background oscillation. However, the part of $\tilde j^\text{bdy}_s$ where $\tilde q_m$ is up-converted and the part of $\tilde f^\text{EM}_m+\tilde f_m^L$ where $\tilde e_n$ and $\tilde b_n$ are down-converted are expected to dominate, because they generate a feedback loop. Therefore, we decompose
\begin{equation}
    \tilde f_m^\text{EM}(\omega)+\tilde f_m^L(\omega)=\sum_n \left(f_m^{b}\tilde b_n(\omega+\omega_0)+f_m^{e}\tilde e_n(\omega+\omega_0)\right)+\mathcal{O}(\bar{F}^2h)\,,
\end{equation}
where $f_m^{b}$ and $f_m^{e}$ are time-independent coefficients collecting all the terms proportional to $\tilde b_n$ and $\tilde e_n$ respectively and are evaluated with $\bar{\bm{B}}\approx\frac12e^{-i\omega_0 t}\bm{B}_0^*$ and $\bar{\bm{E}}\approx\frac12e^{-i\omega_0 t}\bm{E}_0^*$. Furthermore, we have neglected terms in $\tilde f_m^\text{EM}(\omega)$ and $\tilde f_m^L(\omega)$ of the form $\bar{F}^2h$, since they are not enhanced by EM resonances like the other terms including $\delta F^{\mu\nu}$, and are generally suppressed.\footnote{If the background is oscillating, the ratio of the $hF^2$ contribution of the volume integral of the Lorentz force with the volume integral of the GW tidal force is schematically given by $\left(\int \, dv \, \bm{U}_m \cdot \bm{F}^h_L\right)/\left(\int \, dv \, \bm{U}_m \cdot \bm{F}^h_g\right) \sim (\langle \bar E\rangle ^2/\rho) \times (\lambda_p/t_{\text{wall}}) \times ((\omega_0+\omega)/(\omega^2 L) \ll 1$, for all the frequencies of interest here where $t_{\text{wall}}$ is the thickness of the walls. Similarly, the ratio $\int \, d\bm{a} \cdot \delta\bm{\sigma}^h_\mathrm{EM} \cdot \bm{U}_m /\int \, dv \, \bm{U}_m \cdot \bm{F}^h_g \sim (\langle \bar E\rangle ^2/\rho)  \times (1/(\omega^2 L t_{\text{wall}}) \ll 1$. For all the frequencies of interest in this paper and considering the typical EM amplitude that can be pumped into the cavity (see e.g. \cite{berlin_mago20_2023}), this ratio is $\ll 1$. In the case of a static background, similar conclusions can be drawn.} Similarly, we decompose
\begin{equation}\label{eq:jn_bdy_decomposition}
    \tilde j_n^\text{bdy}(\omega)=\sum_mj_n^m \tilde q_m(\omega-\omega_0)+j_n^\text{bdy, $h$} \tilde h(\omega-\omega_0)\,,
\end{equation}
where $j_n^m$ and $j_n^\text{bdy, $h$}$ collect all terms proportional to $\tilde q_m$ and $\tilde h$ respectively and are evaluated with $\bar{\bm{B}}\approx\frac12e^{i\omega_0 t}\bm{B}_0$ and $\bar{\bm{E}}\approx\frac12e^{i\omega_0 t}\bm{E}_0$. The decomposition Eq.~\eqref{eq:jn_bdy_decomposition} is necessary since in the freely falling limit, the boundary condition can still retain a frequency independent perturbation as $q_m^\text{TT}\to0$ (see Sec.~\ref{sec:signal}).

The coupled equations of motion in the Fourier domain then become
\begin{subequations}\label{eq:EoM_back_action}
\begin{align}
 \tilde b_s(\omega+\omega_0)&=-i\frac{\omega_sj_s^\text{bulk, $h$}\tilde h(\omega)+(\omega+\omega_0)\left(j_s^\text{bdy, $h$} \tilde h(\omega) +\sum_m j_s^m\,\tilde q_m(\omega)\right)}{\omega_s^2-(\omega+\omega_0)^2+\frac{i(\omega+\omega_0)\omega_s}{Q_s}}\,,\\
    \tilde e_s(\omega+\omega_0)&=-i\frac{(\omega+\omega_0)j_s^\text{bulk, $h$}\tilde h(\omega)+\omega_s\left(j_s^\text{bdy, $h$} \tilde h(\omega) +\sum_m j_s^m\,\tilde q_m(\omega)\right)}{\omega_s^2-(\omega+\omega_0)^2+\frac{i(\omega+\omega_0)\omega_s}{Q_s}}\,, \\
    \tilde q_m(\omega)&=\frac1M\frac{(f_m^\text{bulk, $h$}+f_m^\text{bdy, $h$})\tilde h(\omega)+\sum_n\left(f_m^{b}\tilde b_s(\omega+\omega_0)+f_m^{e}\tilde e_s(\omega+\omega_0)\right)}{\omega^2_m-\omega^2+\frac{i\omega\omega_m}{Q_m}}\label{eq:EoM_back_action_q}\, .
\end{align}
\end{subequations}
Above, we used that $\tilde j_n^\text{bulk}, \tilde f_m^\text{bulk}, \tilde f_m^\text{bdy}$ have only explicit dependence on $\tilde h$.

In order to solve this system of coupled equations, we will only consider solutions in specific GW frequency regimes in Sec.~\ref{sec:signal}. 
Let us nevertheless discuss the relevance of the new terms, which are encoded in the couplings $f^e_m, f^b_n$. 
In particular, back action effects are only negligible if 
\begin{align}\label{eq:ratio_back_action}
    \left|\frac{f^{b,e}_m j^m_s\,\omega_s/M}{\left(-\omega^2+\frac{i\omega\omega_m}{Q_m}+\omega^2_m\right)\left(-(\omega+\omega_0)^2+\frac{i(\omega+\omega_0)\omega_s}{Q_s}+\omega^2_s\right)}\right|\ll1\,,
    \end{align}
for all combinations of $f^{b}_m j^m_s$ and $f^{e}_m j^m_s$.
In order to assess the importance of back-action, we need to distinguish between the different experimental schemes.

\subsection{Back-action in heterodyne setup}

A \emph{heterodyne} detection scheme consists of a background field given by an eigenmode of a superconducting cavity excited on resonance at frequency $\omega_0$ so that a GW up-converts some power to another eigenmode \cite{ballantini_microwave_2005}. In this case, the fields $\bar{F}_{\mu\nu}$ and $\delta F_{\mu\nu}$ (appearing in $\delta \bm{\sigma}_\mathrm{EM}$ and $\bm{F}_L$) only exist within the cavity wall up to a small distance away from the surface, which is given by the penetration depth $\lambda_p\propto 1/\sqrt{\omega}$ \cite{Pozar12}.

At frequencies above the elastic free fall limit, even assuming superconducting cavities, i.e with $Q_n \sim 10^{10}$ (and other typical experimental parameters for pump amplitude, walls' density and walls' thickness, see e.g. \cite{berlin_mago20_2023} or Sec.~\ref{sec:signal}), the criterion in \eqref{eq:ratio_back_action} is fulfilled even on resonance, such that back action is negligible. At lower frequencies, when the GW frequency matches both mechanical and EM resonances, the left-hand side of \eqref{eq:ratio_back_action} becomes
\begin{equation}\label{eq:back_action_relevance_heterodyne}
    \left|\frac{Q_nQ_m\langle\bar{B}\rangle^2V^{1/3}}{\omega_m^2M}\right| \approx 10^7 \,\frac{V^{1/3}}{0.1\,\text{m}}\left(\frac{\langle \bar{B}\rangle }{0.1\,\text{T}}\right)^2\frac{10\,\text{kg}}{M}\frac{Q_n}{10^{10}}\frac{Q_m}{10^4}\left(\frac{10\,\text{kHz}}{\omega_m/2\pi}\right)^2 \, ,
\end{equation}
up to dimensionless coupling constants which are $\lesssim \mathcal{O}(1)$.
Thus we see that back action is only negligible if the mechanical/EM quality factors or the pump amplitude are much lower. For example, if $Q_n=Q_m=10^3$ and $\langle \bar{B} \rangle=10^6$ V/m, the quantity in Eq.~\eqref{eq:back_action_relevance_heterodyne} becomes small compared to $1$, such that back action is negligible.

In the regime where back action is relevant, one can show that the contribution  of the surface integral in $f^{e,b}_m$ is much larger than the contribution from the volume integral from the Lorentz force. This means that significant back action will only arise from $f_m^\text{EM}$ in Eq. \eqref{eq:back_action_EM_force}.

Furthermore, $\delta\bm{\sigma}_\text{EM}=0$ on the outer surface of the cavity, as all EM fields are contained in the cavity. In addition, as mentioned above, $\delta\bm{\sigma}^\mathrm{in}_\text{EM}$ is evaluated well inside the walls (deeper than the penetration depth) such that its amplitude is negligible compared to the \textit{out} component and therefore we will neglect it. Finally, the relevant contribution from the perturbation of the EM stress energy tensor at the inner surface of the cavity is given by (dropping the \textit{out} superscript)
\begin{align}\label{eq:heterodyne_back_action_coupling_full}
    &f_m^\text{EM}=\int_{\partial v} d\bm{a}\cdot\delta\bm{\sigma}_\text{EM}\cdot\bm{U}_m=-\int _{\partial V}d\bm{A}\cdot\delta\bm{\sigma}_\text{EM}\cdot\bm{U}_m\, \\
    &=\int_{\partial V} d\bm{A}\cdot\left(\bm{U}_m\,(\bar{\bm{B}}\cdot\delta\bm{B}+\bar{\bm{E}}\cdot\delta\bm{E})-\delta\bm{E}(\bm{U}_m\cdot\bar{\bm{E}})-\delta\bm{B}(\bm{U}_m\cdot\bar{\bm{B}})- \bar{\bm{E}}(\bm{U}_m\cdot\delta\bm{E})\right)  \nonumber\, ,
\end{align}
up to terms of the form $\int d \bm{A} \cdot \bm{U}_m  h\bar{F}^2$ which are negligible, as discussed above.
Using Eq.~\eqref{eq:delta_E_no_F}, we find
\begin{align}\label{eq:heterodyne_back_action_coupling}
    f_m^\text{EM}=&\sum_s\int_{\partial V} d\bm{A}\cdot\left[\bm{U}_m\,(b_s\bar{\bm{B}}\cdot\bm{B}_s-e_s\bar{\bm{E}}\cdot\bm{E}_s)\right]\nonumber \\&-\int_{\partial V}dA\left[(\bar{\bm{N}} \times \bm{\mathcal{V}})\cdot(\bm{U}_m \times \bar{\bm{E}})+ (\bar{\bm{N}}\cdot\bm{\mathcal{W}})\,(\bm{U}_m \cdot \bar{\bm{B}})\right]\, ,
\end{align}
where we considered that the signal mode is solenoidal. 
The additional terms involving $\bm{\mathcal{V}},\bm{\mathcal{W}}$ in Eq. \eqref{eq:heterodyne_back_action_coupling} are actually negligible as we show now. $\bm{\mathcal{V}},\bm{\mathcal{W}}$ have contributions $\propto h(t)$ and $\propto \delta \bm{x}$, however we can neglect the $\propto h(t)$ contributions as they are non resonant, as discussed above. Then we can make the decomposition $\delta \bm{\sigma}_\mathrm{EM} \approx b_s(t)\delta \bm{\sigma}^b_\mathrm{EM} +e_s(t)\delta \bm{\sigma}^e_\mathrm{EM} + q_m(t)\delta \bm{\sigma}^q_\mathrm{EM}$, where the last term is sourcing the contributions $\propto\bm{\mathcal{V}},\bm{\mathcal{W}}$ in Eq. \eqref{eq:heterodyne_back_action_coupling}. The terms $\propto q_m(t)$ modify Eq.~\eqref{eq:mech_coeff_EoM_back_action} by introducing an additional term $(\int d\bm{a} \cdot \delta \bm{\sigma}^q_\mathrm{EM} \cdot \bm{U}_m) q_m$ on the LHS, effectively leading to a frequency shift $\omega^2_m \rightarrow \omega^2_m + \int d\bm{a} \cdot \delta \bm{\sigma}^q_\mathrm{EM} \cdot \bm{U}_m$. However, this additional term is $\ll \omega^2_m$ for all frequencies discussed in this paper. Therefore, we can neglect this contribution, such that back action is simply given by 
\begin{align}
    f_m^\text{EM}=\sum_s\int_{\partial V} d\bm{A}\cdot\bm{U}_m\,(b_s\bm{B}_s\cdot \bar{\bm{B}}-e_s\bm{E}_s\cdot\bar{\bm{E}})\, .
\end{align}
Note that on resonance, i.e when $\omega_s = \omega+\omega_0$, we find that $b_s=e_s$ such that the above expression simplifies to the back action terms as discussed in \cite{ballantini_microwave_2005, Löwenberg2023}.

\subsection{Back-action in setup with magnetostatic background}\label{sec:elasticity_B0}

In the case of static magnetic background fields, we need to consider how the background magnetic field enters the cavity volume. A normal conducting material will allow the static magnetic field to permeate the cavity wall almost undisturbed, which means $\bar{F}_{\mu\nu}^\text{in}=\bar{F}_{\mu\nu}^\text{out}$ on both boundaries of the wall. However, a jump in the fields $\delta F_{\mu\nu}$ at the inner boundary causes a contribution to $q_m$ in the boundary integral Eq. \eqref{eq:back_action_EM_force}. Furthermore, the Lorentz force contains terms $\propto h\bar{F}^2$ which act throughout the entire volume of the wall. However, this contribution is subdominant near resonances. If the cavity wall is superconducting, the magnetic field will be strongly expelled from the wall, which leads to  $\bar{F}_{\mu\nu}^\text{in}\neq\bar{F}_{\mu\nu}^\text{out}$ and several possible back-action terms, depending on the specific experimental setup. In addition, we have $f^e_m=0$, as there is no electric background. 

However, importantly, in most cases of magnetostatic background experiments, the GW cannot excite both EM and mechanical resonances at the same time, since microwave resonances usually lie in the free falling limit discussed in Sec.~\ref{sec:free_fall_limit}. In this case, the criterion \eqref{eq:ratio_back_action} is fulfilled and back action is negligible. For example, in the typical frequency regime of operation of cavities employing static magnetic field backgrounds, $\omega \gg \omega_m$ such that $j^m_n\rightarrow 0$, as the displacement becomes independent of the mode coefficients. In this case, the mode coefficients become
\begin{subequations}\label{eq:EM_coeff_B0_FF}
\begin{align}
\tilde e_s(\omega \gg \omega_m)&=-i\frac{(\omega+\omega_0)j_s^\text{bulk, $h$}+\omega_sj_s^\text{bdy, $h$}}{\omega_s^2-\omega^2+\frac{i\omega \omega_s}{Q_s}}\tilde h(\omega)\,, \\
\tilde b_s(\omega\gg \omega_m)&=-i\frac{\omega_sj_s^\text{bulk, $h$}+(\omega+\omega_0)j_s^\text{bdy, $h$}}{\omega_s^2-\omega^2+\frac{i\omega \omega_s}{Q_s}}\tilde h(\omega) \, ,
\end{align}
\end{subequations}
i.e back action can be neglected. 

\section{GW-induced signal power}\label{sec:signal}
In the previous sections, we computed the EM field inside the cavity generated by the passage of a GW. Now, we will show how to find the signal power measured in an experiment in a gauge invariant way. Electromagnetic fields at microwave frequencies are commonly measured using antennas. Electromagnetic fields induce a current in the antenna, which can be directly read out. As the antenna itself can be perturbed by the GW with amplitude $\delta\bm{x}^\text{ant}$, we need to use the electric field in the local tetrad frame of the antenna to calculate the resulting signal. While this field is already a gauge invariant quantity according to Eq. \eqref{eq:observed_E_field}, it is not guaranteed that it will be the same field observed by a readout device. This is because cables carrying the signal current while still containing background fields, could receive additional $\mathcal{O}(h\bar{B})$ contributions due to the GW. We can only safely take the EM field at the antenna to be the observed signal (after including amplification and losses), when solely $\mathcal{O}(h\bar{B})$ and not $\mathcal{O}(\bar{B})$ fields travel through the readout system, as further perturbations are $\mathcal{O}(h^2)$.

The induced current in the antenna conductor is given by
\begin{equation}\label{eq:current_antenna}
    I=\frac{1}{Z_\text{eff}}\int d\bar{\bm{l}}\cdot\delta\bm{E}^{\text{obs}}\,,
\end{equation}
where $\delta\bm{E}^{\text{obs}}$ is given in Eq. \eqref{eq:delta_E_obs_def_with_E_ant} is the electric field in the local tetrad coordinates of the antenna with $\delta\bm{x}=\delta\bm{x}^\text{ant}\,$ in a tetrad aligned with the cartesian axes $\bar{e}^\mu_{\underline{\alpha}}=\delta^\mu_\alpha$,  $\bar{\bm{l}}$ is a curve following the unperturbed antenna and $Z_\text{eff}$ is an effective impedance containing the load of the antenna, reflections at the input, and back-reaction effects due to the surrounding cavity \cite{Balanis05}.\footnote{If one were interested in measuring the magnetic field, one could use a closed loop where the current is related to the flux $\int d\bar{\bm{A}}\cdot\delta\bm{B}^{\text{obs}}$.} As antennas are themselves conductors, they are a part of the boundary, and their surface area contributes to the integral in Eq. \eqref{eq:boundary_current}. However, the current in Eq. \eqref{eq:current_antenna} is due to the electric field incident on the antenna and not the resulting total field which fulfills the boundary condition \eqref{eq:boundary_condition_perturbation_damping}. Therefore, we use the electric field in the interior Eq. \eqref{eq:delta_E_bulk} rather than Eq. \eqref{eq:delta_E_bounday_parallel} in Eq. \eqref{eq:current_antenna}.

The coupling of the antenna to the individual cavity modes are usually experimentally characterized by the external quality factors \cite{Padamsee98}
\begin{equation}\label{eq:Qext}
    Q_n^\text{ext}=\frac{\omega_n \varepsilon_n}{P_n}=\frac{Z_\text{eff}}{2}\frac{\omega_nV}{\left\lvert\int d\bar{\bm{l}}\cdot\bm{E}_n\right\rvert^2} \, ,
\end{equation}
where $\varepsilon_n=V/2$ is the stored EM energy in the $n$th cavity eigenmode in our mode normalization.

In the following, it will be instructive to decompose the observed electric field 
\begin{equation}
    \delta\bm{E}^\text{obs}=\delta\bm{E}^\text{bulk}+\delta\bm{E}^\text{bdy}+\delta\bm{E}^\text{ant}\,,
\end{equation}
where $\delta\bm{E}^\text{bulk}$ collects all terms proportional to $j_n^\text{bulk}$, $\delta\bm{E}^\text{bdy}$ collects all terms proportional to $j_n^\text{bdy}$ and $\delta\bm{E}^\text{ant}$, from Eq. \eqref{eq:delta_E_ant_definition}, all terms due to the movement of the observer. 

We can characterize the resulting signal in terms of a power spectral density (see Eq. \eqref{eq:PSD_definition})
\begin{equation}
    S_\text{sig}(\omega)=Z_\text{eff}S_I(\omega)\approx \frac12\frac{\omega_n}{Q_n^\text{ext}}VS_{e_n}(\omega)\,,
\end{equation}
which can be used to obtain the average signal power received by the antenna using Eq. \eqref{eq:PSD_time_avg}
\begin{equation}
    P_\text{sig}=\frac{1}{(2\pi)^2}\int_{-\infty}^{\infty}d\omega\,S_\text{sig}(\omega)\,.
\end{equation}

\begin{figure}[ht!]
\centering
\begin{subfigure}{0.56\textwidth}
  \begin{overpic}[width=\textwidth]{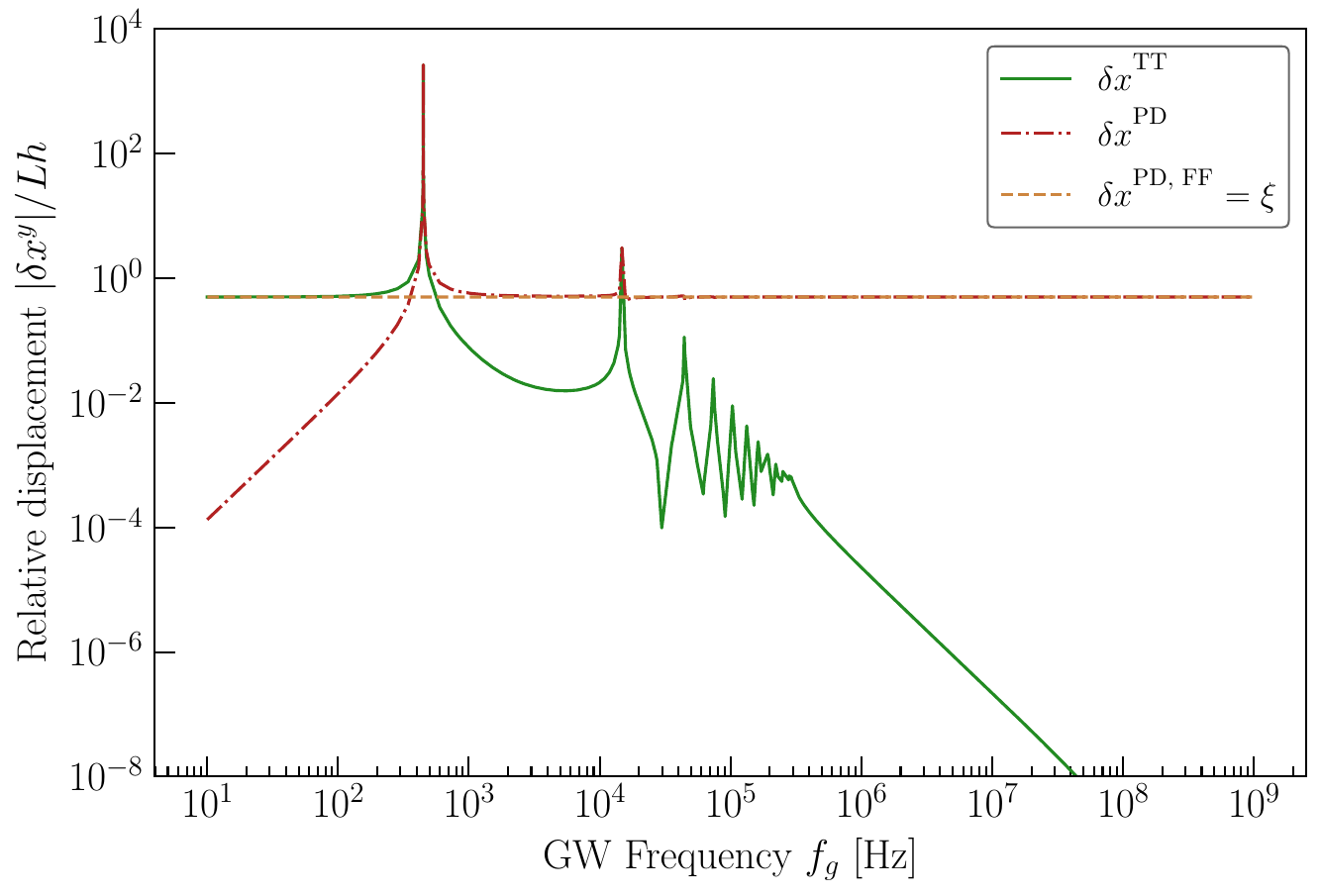}
    \put(15,17){\scriptsize $\leftarrow$ Rigid}
    \put(21, 12){\scriptsize $\delta x^\text{PD}\ll\xi$}
    \put(37,24){\scriptsize $\leftarrow$ Elastic $\rightarrow$}
    \put(64,37){\scriptsize Free fall $\rightarrow$}
    \put(64, 32){\scriptsize $\delta x^\text{TT}\ll\xi$}
  \end{overpic}
\end{subfigure}
~ 
\begin{subfigure}{0.4\textwidth}
\includegraphics[width=\textwidth]{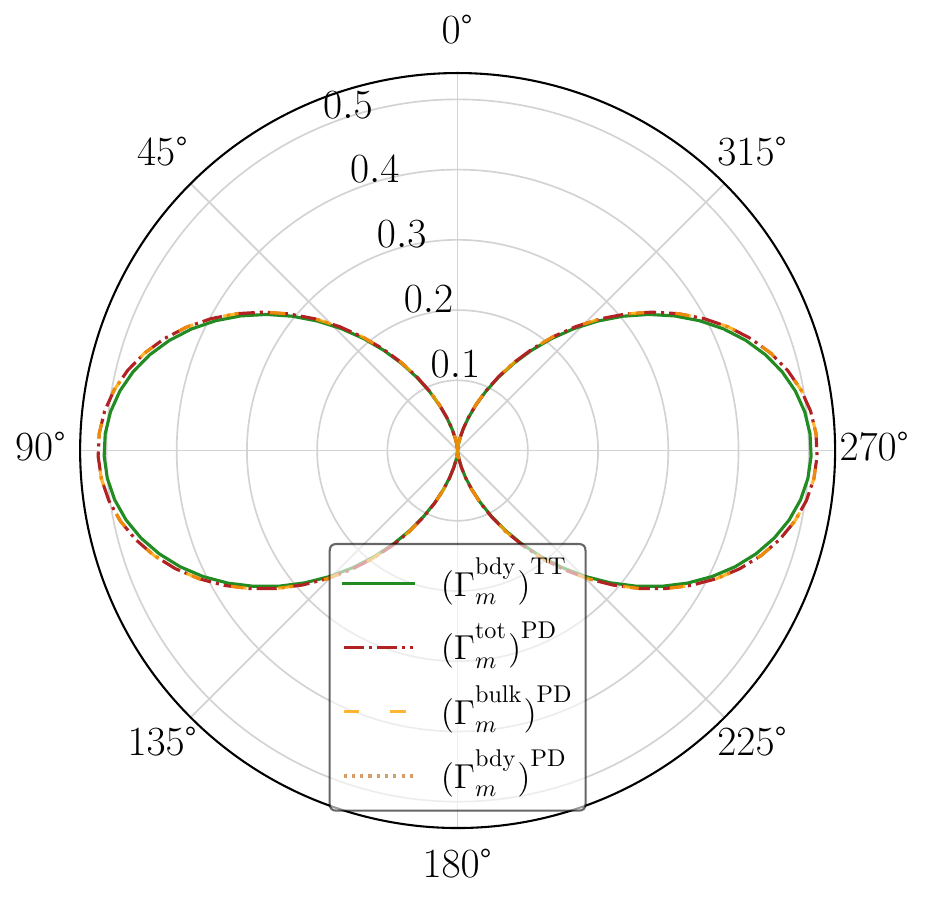}
\end{subfigure}

\caption{The mechanical displacement of the wall of a cylindrical microwave cavity as calculated using TT and PD coordinates evaluated near the inner radius of the cavity. The GW has a wave vector $\bm{k}_g=\omega_g\hat{\bm{x}}$ and is plus polarized with amplitude $h$. The cavity has an inner radius and length $R=L=1\,\text{m}$. Further details on the numerical calculation can be found in section \ref{sec:details_on_examples}. The polar plot on the right is showing the normalized mechanical response from Eqs. \eqref{eq:mech_coupling_coefficients} for different angles with the cylinder axis for the mechanical mode with the lowest resonant frequency. The total PD and TT curves overlap up to numerical accuracy.}
\label{fig:delta_x}
\end{figure}

Now, we will calculate the signal power both in magnetostatic and microwave backgrounds. After discussing the resulting spectrum, we will focus on the most important frequency ranges near electromagnetic resonances $|\omega-\omega_n|\lesssim \omega_n/Q_n$. In such case, $\delta\bm{E}^\text{obs}\approx\delta\bm{E}^\text{bulk}+\delta\bm{E}^\text{bdy}\approx e_n\bm{E}_n$, as the remaining terms are non-resonant. 
This allows us to neglect irrotational modes, whose coefficients are suppressed by $1/Q_n \ll 1$, and we will only consider the solenoidal EM modes, such that $n\in\mathbf{S}$ in the following.
Furthermore, we will consider GWs which can (approximately) be decomposed $h^\text{TT}_{ij}=\hat{h}_{ij}(\bm{x})h(t)$ so that $h(t)$ is the strain amplitude in TT coordinates.
The displacement of the cavity walls in Eq. \eqref{eq:delta_x_solution_expansion} with negligible back-action is then given by 
\begin{subequations}
\begin{align}\label{eq:mech_displacement_spectrum_expansion}
    \delta\tilde{\bm{x}}(\omega)=\frac{V^{1/3}}2\sum_m\bm{U}_m\frac{\omega^2}{\omega_m^2-\omega^2+i\frac{\omega\omega_m}{Q_m}}(\Gamma_m^\text{bdy}+\Gamma_m^\text{bulk})\tilde{h}(\omega)\,,
\end{align}
where we have defined the dimensionless GW-mechanical coupling coefficients
\begin{align}\label{eq:mech_coupling_coefficients}
    \Gamma_m^\text{bdy/bulk}=\frac{2\tilde{f}_m^\text{bdy/bulk}(\omega)}{\omega^2MV^{1/3}\tilde{h}(\omega)}\,.
\end{align}
\end{subequations}
Fig. \ref{fig:delta_x} illustrates the mechanical response for different GW frequencies and incidence angles using a simple model for the mechanical eigenmodes discussed in appendix \ref{sec:details_on_examples}. It demonstrates that the sum of the coupling coefficients in Eq. \eqref{eq:mech_coupling_coefficients} is frame independent on resonance and how the freely falling limit arises at higher frequencies, as $\delta \bm{x}^\mathrm{TT} \rightarrow 0$ for $\omega_g \rightarrow \infty$. Furthermore, we find the expected rigid limit $\delta\bm{x}^\text{PD}\to0$ for $\omega_g\to0$. Since we are not using a complete set of mechanical modes to create Fig. \ref{fig:delta_x}, inaccuracies of the displacement can occur away from resonances. In App.~\ref{sec:numerical_considerations}, we explain how to mitigate such problems.

\subsection{Detector with magnetostatic background}

In the case of static background magnetic fields, one has $\bm{\mathcal{V}}= \bm{\bar{B}}\times\delta\bm{u}$ with $|\bar{\bm{B}}|=\bar{B}$ and $\omega_0=0$ in Eqs.~\eqref{eq:EoM_back_action}. As discussed in Sec.~\ref{sec:back_action}, back-action effects are negligible in these setups, such that the mode coefficients are given in Eq.~\eqref{eq:EM_mode_coeff_no_F}. The gauge invariant signal PSD then becomes
\begin{subequations}
\begin{align}
    S_\text{sig}(\omega)=\frac12\frac{\omega_n}{Q_n^\text{ext}}V\frac{\omega^2\omega_n^2}{(\omega_n^2-\omega^2)^2+\left(\frac{\omega\omega_n}{Q_n}\right)^2}\left|\eta_n^\text{bdy}+\frac{\omega}{\omega_n}\eta_n^\text{bulk}\right|^2\bar{B}^2S_h(\omega)\,,
\end{align}
with the electromagnetic coupling coefficients
\begin{align}\label{eq:EM_couplings_static}
        \eta_n^\text{bdy/bulk}=\frac{\tilde{j}_n^\text{bdy/bulk}(\omega)}{\omega\bar{B}\tilde{h}(\omega)}\,.
\end{align} 
\end{subequations}
Typically, magnetostatic cavity experiments are designed to search for GWs with $f_g\gtrsim10^8$ Hz, which corresponds to the fundamental frequency of a $\mathcal{O}(1\,\text{m})$ cavity. 
In this case, the GW frequency is much larger than the fundamental mechanical eigenfrequency, such that in TT coordinates, $\delta \bm{x}^\mathrm{TT} \rightarrow 0$ (see Eq.~\eqref{eq:mech_EOM_q_TT} and the discussion after that). This is the freely falling limit and the coupling coefficients in Eq.~\eqref{eq:EM_couplings_static} become
\begin{subequations}\label{eq:eta_TT_magnetostatic}
\begin{align}
(\eta_n^\text{bdy})^\text{TT}&=0
\,,\\
(\eta_n^\text{bulk})^\text{TT}&= \frac{1}{\omega V\bar{B}\tilde{h}}\int dV\,\bm{E}^{*}_n\cdot\tilde{\bm{J}}^\text{TT}_\text{eff}  =\frac{i}{V\bar{B}}\int dV\,\bm{E}_n^*\cdot\left(\hat{\bm{h}}^\text{TT}\cdot(\bar{\bm{B}}\times\hat{\bm{k}}_g)\right)\label{eq:eta_bulk_TT_magneto}\, ,
\end{align}
\end{subequations}
where we have assumed a plane GW with direction $\hat{\bm{k}}_g$ in the last equality.
The coefficients in PD coordinates are
\begin{subequations}\label{eq:eta_PD_magnetostatic}
\begin{align}
(\eta_n^\text{bdy})^\text{PD}&=-\frac{i}{\bar{B}V\tilde{h}}\int d\bm{A}\cdot\left(\bm{B}^{*}_n\times(\bm{\bar{B}}\times\tilde{\bm{\xi}}_\text{PD}^\text{TT})\right)
\label{eq:eta_bdy_PD_magneto}\,,\\(\eta_n^\text{bulk})^\text{PD}&= \frac{1}{\omega V\bar{B}\tilde{h}}\int dV\,\bm{E}^{*}_n\cdot\tilde{\bm{J}}^\text{PD}_\text{eff}  \, ,
\end{align}
\end{subequations}
where for $(\eta_n^\text{bdy})^\text{PD}$ we used the coordinate transformation Eq.~\eqref{eq:TT_PD_coord_transfo}.  Note that we do not need to distinguish between elastic and pure free fall here, as we would obtain the same result assuming either. Then, the signal power for a monochromatic GW on resonance becomes
\begin{align}\label{eq:static_B_power_res}
P_\text{sig} &= \frac12\frac{\omega_n }{ Q_n^\mathrm{ext}}V Q_n^2 |\eta_\mathrm{n}^\text{bulk}+\eta_\mathrm{n}^\text{bdy}|^2\bar{B}^2 h^2 \, ,
\end{align}
which yields the same result in any coordinate system up to the $\mathcal{O}(1/Q_n)$ terms we neglected. Due to the inclusion of the boundary coupling, the scaling of the signal power with the GW frequency and cavity size is not the same as reported in e.g. \cite{Berlin22}, where the power was evaluated in the PD frame but the contribution from the wall's movement was neglected.

\begin{figure}[ht!]
\centering
\includegraphics[width=.9\linewidth]{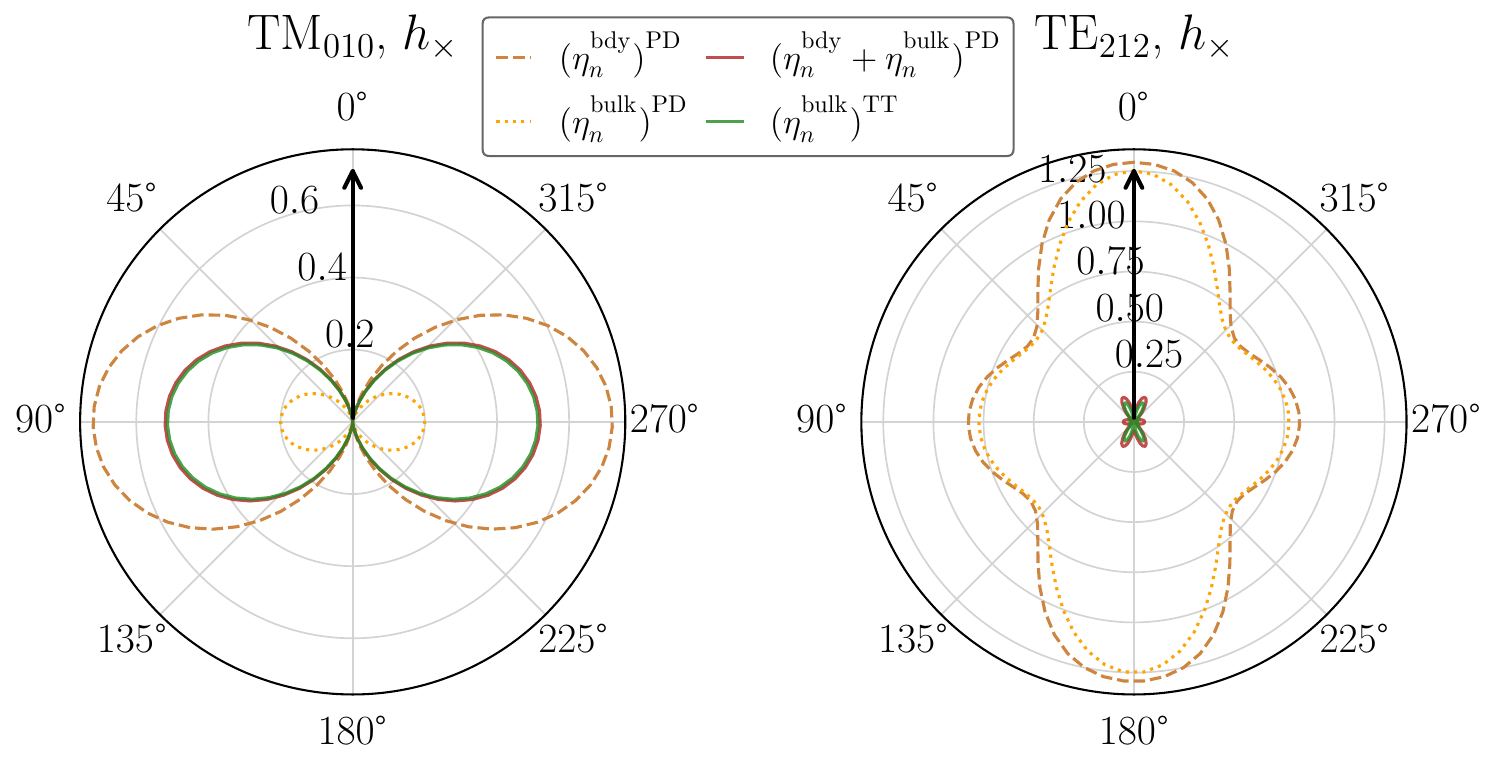}

\caption{An evaluation of the dimensionless geometric factors $\eta^\mathrm{bulk/bdy}_{n}$ both in PD and in TT frames respectively from Eqs.~\eqref{eq:eta_PD_magnetostatic} and \eqref{eq:eta_TT_magnetostatic}, for a static magnetic field in the direction shown by the black arrow for different angles of a GW around the cylinder axis. We assume that the GW frequency $\omega_g$ is the same as the resonant frequency of the eigenmode, such that we can consider freely falling cavity walls. We show the absolute value of the coupling and add the couplings for degenerate mode polarizations in quadrature and do not include $(\eta_n^\text{bdy})^\text{TT}=0$. The total PD and TT curves overlap up to numerical accuracy.}
\label{fig:test}
\end{figure}

In Fig.~\ref{fig:test}, we evaluate Eqs.~\eqref{eq:eta_TT_magnetostatic} and \eqref{eq:eta_PD_magnetostatic} assuming the GW frequency matches the eigenfrequency of two EM modes of a cylindrical cavity of equal length and radius.\footnote{In order to compare Fig. \ref{fig:test} with the results in \cite{Berlin22}, the normalization needs to be changed by dividing the coupling by a factor $V^{1/3}\omega_n$.} As expected, we find that the sum $\eta^\mathrm{bulk}_n+\eta^\mathrm{bdy}_n$ is equivalent in TT and PD coordinates, demonstrating their gauge invariance on resonance.\footnote{Note that these coefficients are not exactly gauge invariant, because they are only part of the expansion of the gauge dependent $\delta \bm{E}$. However, as mentioned above, on resonance, we have $\delta \bm{E} \approx\delta \bm{E}^\mathrm{obs}$.}

\begin{figure}[ht!]
\centering

  \centering
  \includegraphics[width=.7\linewidth]{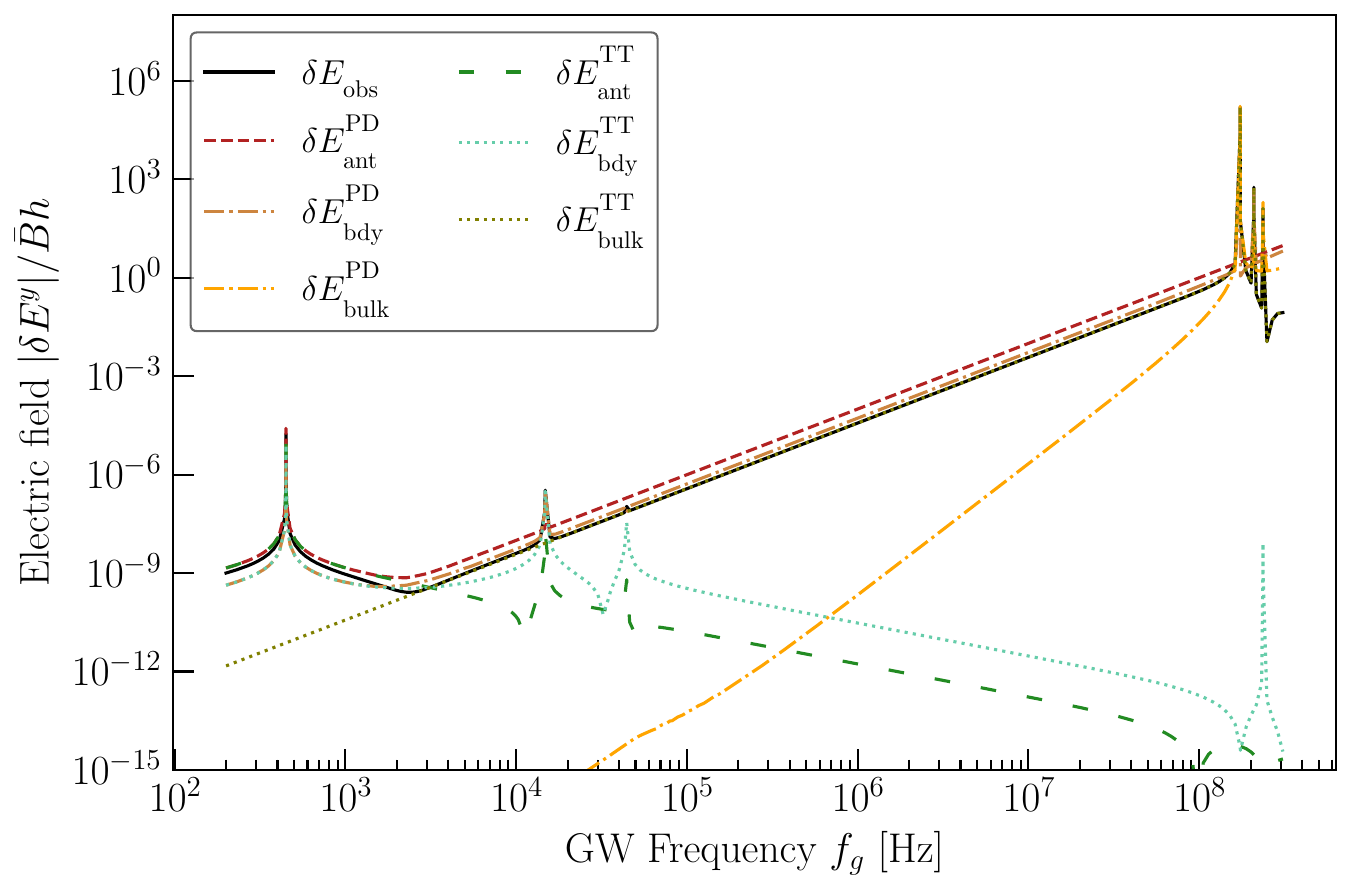}

\caption{Magnetostatic experiment. The observed electric field perturbation and its contributions in a cylindrical microwave cavity in a static magnetic field $\bar{\bm{B}}=\bar{B}\hat{\bm{z}}$ aligned with the cylinder axis. The GW has a wave vector $\bm{k}_g=\omega_g\hat{\bm{x}}$ and is plus polarized with amplitude $h$. The field is measured using a short pin antenna oriented along the $y$ axis, attached to the wall at $x=z=0$. The mechanical response is calculated as described in section \ref{sec:details_on_examples} including the lowest three mechanical modes and lowest four EM modes with non-zero coupling. The cylinder has equal length and radius of $1\,\text{m}$ and $Q_n=10^6$ for all modes.}
\label{fig:delta_E_static}
\end{figure}

\begin{figure}[ht!]
\centering

  \centering
  \includegraphics[width=.7\linewidth]{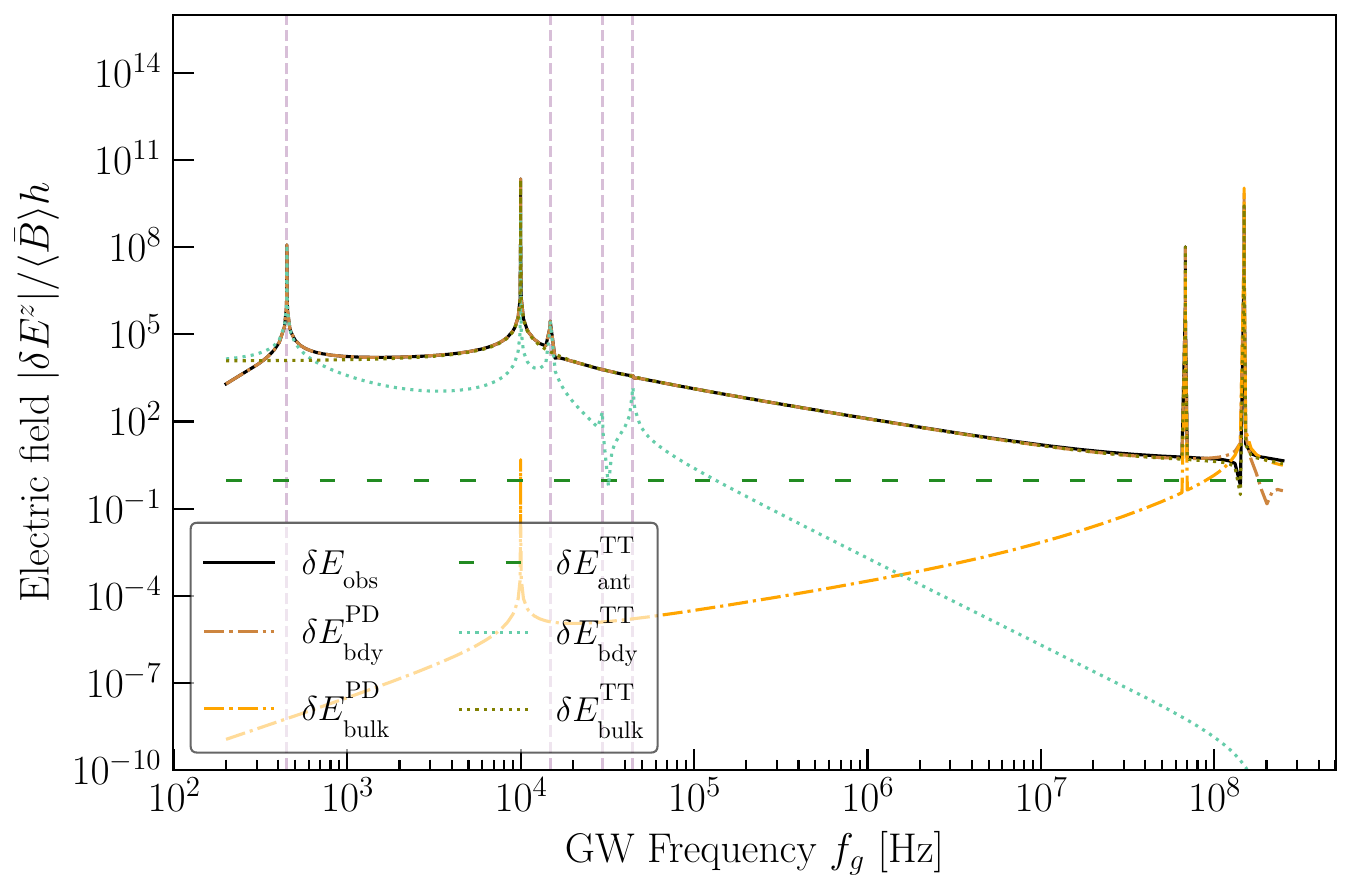}

\caption{Heterodyne experiment. The observed electric field perturbation and its contributions in two orthogonal coupled cylindrical microwave cavities loaded in phase with the TM$_{010}$ mode. The GW has a wave vector $\bm{k}_g=\omega_g\hat{\bm{x}}$ and is plus polarized with amplitude $h$. The field is measured using a short pin antenna oriented along the $z$ axis of one cavity at the center of mass of one cylinder where $\delta\bm{E}_\text{ant}^\text{PD}=0$. The antenna is assumed to be in the pure free fall. The response is calculated as described in section \ref{sec:details_on_examples}, including the out of phase oscillation of the lowest three mechanical modes and lowest four EM modes with non-zero coupling, including the out-of-phase oscillation of the TM$_{010}$ mode, taken to be $10\,\text{kHz}$ away from the symmetric oscillation. The cylinder has equal length and radius of $1\,\text{m}$ and $Q_n=10^{10}$ for all modes. Mechanical resonances are marked by vertical dashed lines.}
\label{fig:delta_E_heterodyne}
\end{figure}

A calculation of the sensitivity of magnetostatic cavity experiments outside EM resonances can be found in App.~\ref{ap:signals}. 

Fig.~\ref{fig:delta_E_static} illustrates an example for the gauge invariant electric field perturbation measured by an antenna in its tetrad frame in such an experiment at different frequencies.
Around $\text{kHz}$ frequencies, mechanical resonances enhance the signal, which is, however, suppressed by the off-resonant electromagnetic response. The mechanical resonances enter the signal power through the boundary current Eq.~\eqref{eq:boundary_current}. When combined with the expansion in Eq. \eqref{eq:mech_displacement_spectrum_expansion} the boundary coupling can also be expressed as
\begin{subequations}
\begin{align}\label{eq:eta_n_bdy}
    \eta_n^\text{bdy}=\frac{i}{V^{1/3}\tilde{h}}\sum_m\tilde{q}_mC_n^m=\frac{i}{2}\sum_m\frac{\omega^2}{\omega_m^2-\omega^2+i\frac{\omega\omega_m}{Q_m}}C_n^m(\Gamma_m^\text{bdy}+\Gamma_m^\text{bulk})\,,
\end{align}
using the coupling coefficient between one mechanical and one EM mode
\begin{align}
    C_n^m=-\frac{1}{V^{2/3}\bar{B}}\int_{\partial V} d\bm{A}\cdot\left(\bm{B}_n^*\times(\bar{\bm{B}}\times\bm{U}_m)\right)\,.
\end{align}
\end{subequations}
In this regime $\eta_n^\text{bdy}$ is not normalized to be $\mathcal{O}(1)$, since mechanical resonances can enhance the coupling.

As the antenna oscillates together with the wall, its movement contributes with a similar magnitude to the signal. Around frequencies $\gtrsim\text{GHz}$, electromagnetic resonances enhance the signal and the cavity enters the free falling regime. Therefore, the boundary is barely perturbed by the GW in TT coordinates anymore and the effective current dominates. In PD coordinates, the boundary and bulk terms have a comparable magnitude and interfere to give the same signal as in TT coordinates. Between these two regimes, where $\omega_g \gg \omega_m$, the signal scales roughly with $\omega^2_g$, dominated in PD by the boundary term $\tilde j^\mathrm{bdy,PD}_n$ and in TT by the bulk coupling $\tilde j^\mathrm{bulk,TT}_n$. The signal contribution proportional to the bulk term $\tilde j^\mathrm{bulk,PD}_n$, while being subdominant, scales with $\omega^3_g$ in this intermediate regime. 

As discussed in detail in App.~\ref{sec:numerical_considerations}, it is computationally challenging to compute $\delta\bm{E}^\text{bdy}$ in PD coordinates at frequencies in free fall, off EM resonances. Therefore, in Fig. \ref{fig:delta_E_static}, we compute $\delta\bm{E}^\text{obs}$ using TT coordinates and then obtain $(\delta \bm{E}^\text{bdy})^\text{PD}=\delta \bm{E}^\text{obs}-(\delta \bm{E}^\text{bulk})^\text{PD}-(\delta \bm{E}^\text{ant})^\text{PD}$, which ensures gauge invariance by construction and serves to illustrate the contributions to $\delta \bm{E}^\text{obs}$ in different frames and different frequencies.

\subsection{Heterodyne detector}

In order to evaluate the couplings in the case of oscillating electromagnetic backgrounds, we need to consider all terms in Eq.~\eqref{eq:boundary_condition_perturbation}. 
If we assume that the background fields are given by another cavity eigenmode, we can express the boundary source term in Eq.~\eqref{eq:boundary_current} as
\begin{align}\label{eq:mech_coupling_mago_approx}
    \tilde j_n^\text{bdy}(\omega)=&\frac{i}{2V}\int_{\partial V} d\bm{A}\cdot\delta\tilde{\bm{x}}(\omega-\omega_0)\,\left(\omega\bm{B}_n^*\cdot\bm{B}_0-\omega_n\bm{E}_n^*\cdot\bm{E}_0\right) \,,
\end{align}
where we used that $\bar{\bm{B}}=\frac12e^{i\omega_0t}\bm{B}_0$, and similarly for $\bar{\bm{E}}$ (see Eqs. \eqref{eq:B0_E0_definition} in Sec.~\ref{sec:back_action}). On resonance, when $\omega = \omega_n$, we recover the coupling integral as reported in \cite{ballantini_microwave_2005, meidlinger, berlin_mago20_2023, Löwenberg2023}. 
Using similar calculations as in the magnetostatic case, the resulting signal PSD near an electromagnetic resonance in a heterodyne setup is given by
\begin{subequations}
\begin{align}
    S_\text{sig}(\omega)=\frac12\frac{\omega_n}{Q_n^\text{ext}}V\frac{\omega^2\omega_n^2}{(\omega_n^2-\omega^2)^2+\left(\frac{\omega\omega_n}{Q_n}\right)^2}\left|\eta_n^\text{bdy}+\frac{\omega}{\omega_n}\eta_n^\text{bulk}\right|^2\langle\bar{B}\rangle^2 S_h(\omega-\omega_0) \, ,
    \end{align}
with the electromagnetic coupling coefficients
\begin{align}\label{eq:EM_couplings_heterodyne}
        \eta_n^\text{bdy/bulk}=\frac{\tilde{j}_n^\text{bdy/bulk}(\omega)}{\omega\langle\bar{B}\rangle\tilde{h}(\omega-\omega_0)}\,.
\end{align}
\end{subequations}

\subsubsection{Elastic regime}\label{sec:heterodyne_signal_power_elastic}

Around mechanical resonances, as mentioned in Sec.~\ref{sec:elasticy_theory}, PD coordinates are more convenient and Eq.~\eqref{eq:mech_EOM_q_PD_low_freq} can be used for the computation of the mechanical mode coefficients. As mentioned in Sec.~\ref{sec:elasticy_theory}, the benefit of PD coordinates at low frequency is that the boundary conditions are not perturbed such that $\bm{y}$ in Eqs.~\eqref{eq:delta_x_no_y} is negligible. Important applications for this calculation are setups like MAGO \cite{ballantini_microwave_2005, Fischer_2025}.

A useful way to write the boundary coupling in Eq. \eqref{eq:EM_couplings_heterodyne} is then
\begin{subequations}
\begin{align}\label{eq:eta_bdy_heterodyne_general_q}
        \eta_n^\text{bdy}(\omega)&=\frac{i}{2V^{1/3}\tilde{h}(\omega-\omega_0)}\sum_m\tilde{q}_m(\omega-\omega_0)C_n^m(\omega)\,,
\end{align}
where we introduced the coupling coefficient between one mechanical and one EM mode
\begin{align}
        C_n^m(\omega)=\frac{1}{V^{2/3}\langle\bar{B}\rangle}\int_{\partial V} d\bm{A}\cdot\bm{U}_m\left(\bm{B}_n^*\cdot\bm{B}_0-\frac{\omega_n}{\omega}\bm{E}_n^*\cdot\bm{E}_0\right)\,.
\end{align}
\end{subequations}
In cases where back-action is negligible this coupling can also be expressed as 
\begin{equation}
    \eta_n^\text{bdy}(\omega)=\frac{i}{4}\sum_m\frac{(\omega-\omega_0)^2}{\omega_m^2-(\omega-\omega_0)^2+i\frac{(\omega-\omega_0)\omega_m}{Q_m}}C_n^m(\omega)\left(\Gamma_m^\text{bdy}(\omega-\omega_0)+\Gamma_m^\text{bulk}(\omega-\omega_0)\right)\,.
\end{equation}

As an example, we will consider the case where a monochromatic GW excites both an EM and mechanical resonance, i.e $\omega - \omega_0 = \omega_m$ and $\omega=\omega_n$. As mentioned in Sec.~\ref{sec:back_action}, back actions effects become relevant in this regime. The coupled equations of motion in Fourier space are
\begin{subequations}\label{eq:EoM_heterodyne_low_freq}
\begin{align}
     e_n&=-Q_n\langle\bar{B}\rangle\left(\frac{i C_n^m}{2V^{1/3}}q_m+\eta_n^\text{bulk}h\right)\,, \\
    q_m&=\frac{Q_mV^{1/3}}{2i}\left(\Gamma_m^\text{GW}h+\frac{V^{1/3}\langle\bar{B}\rangle}{\omega_m^2M}(C^{m}_n)^*e_n\right)\, ,
\end{align}
\end{subequations}
where we have simplified $\Gamma^\text{GW}_m=\Gamma^\text{bulk}_m+\Gamma^\text{bdy}_m$, where we used that $j_n^\text{bdy}$ has no $\propto h$ component in the elastic regime, and that $e_n=b_n$ on resonance. The system of equations Eq.~\eqref{eq:EoM_heterodyne_low_freq} has the solution
\begin{equation}\label{eq:mode_coeff_e_BA}
     e_n=-\frac{1}{4}Q_n\langle\bar{B}\rangle h\frac{Q_m C_n^m\Gamma_m^\text{GW}+4\eta_n^\text{bulk}}{1+\frac{Q_nQ_m\langle\bar{B}\rangle^2V^{1/3}}{4\omega_m^2M}| C^m_n|^2}\,.
\end{equation}
In the alternative case that the parameters are chosen so that back-action is negligible (see Eq. \eqref{eq:back_action_relevance_heterodyne}), the boundary contribution dominates at low frequencies $\omega_g\ll\omega_n$ in PD coordinates and we can simplify
\begin{align}\label{eq:en_heterodyne_no_BA}
    e^\mathrm{PD}_n &= -\frac{1}{4}Q_nQ_m\langle\bar{B}\rangle\Gamma_m^\text{GW} C_n^mh  \, .
\end{align}
The power received by the antenna is then
\begin{subequations}
\begin{align}
    P_\text{sig}^\text{PD} &= \frac{h^2 Q^2_nQ^2_m V\langle \bar{B}\rangle^2 \omega_n}{32 Q_n^\mathrm{ext}}(\Gamma^\text{GW,PD}_m)^2 |C_n^m|^2 \, ,
\end{align}
where we can simplify the mechanical coupling coefficient in a long-wavelength approximation
\begin{align}
    \Gamma^\text{GW,PD}_m\approx  \frac{1}{MV^{1/3}}\int_{\partial v} dv\,\rho\, \bm{U}_m\cdot\hat{\bm{h}}^\text{TT}\cdot \bar{\bm{x}}\,.
\end{align}

\end{subequations}
In the case where the quality factors and/or the pump amplitude can be much larger, such as in superconducting cavities, back action arises and can have a significant effect. 
In particular, if the second term of the denominator in Eq.~\eqref{eq:mode_coeff_e_BA} gets significantly larger than $1$, the mode coefficient becomes
\begin{align}
    e_n^\text{PD} &\approx -\frac{h\omega^2_m M \left(\Gamma^\text{GW}_m\right)^\text{PD}}{V^{1/3}\langle \bar{B}\rangle (C^m_n)^*} \, ,
\end{align}
which is significantly lower than the amplitude without back action Eq.~\eqref{eq:en_heterodyne_no_BA} since the quality factors $Q_n$ and $Q_m$ do not enter anymore.
This suggests that in a heterodyne setup, in regimes where both mechanical and EM resonances are excited by the GW, it is not beneficial to increase the quality factors and pump amplitude as much as possible to reach the maximum signal power. This conclusion is also strengthened by the fact that noise can also grow with these parameters, as pointed out in \cite{berlin_mago20_2023}. 

\subsubsection{Free falling limit}

As for the magnetostatic background case, TT coordinates are most convenient to describe the system at high frequencies in free fall. In this case $(j^\text{bdy}_n)^\text{TT} \approx 0$ in Eq.~\eqref{eq:mech_coupling_mago_approx}, since $\delta\bm{x}^\text{TT}\to0$ and $j_n^\text{bulk}$ dominates. The final contribution to the signal is the electric field perturbation at the surface of the elastic freely falling antenna, which is given by (using Eqs.~\eqref{eq:delta_E_ant_definition},\eqref{eq:delta_eij_time_evo})
\begin{align}\label{eq:V_tt_FF_heterodyne}
    \delta \bm{E}^\text{ant,TT,FF}\approx \delta\bm{e}^\text{TT, FF}\cdot\bar{\bm{E}}\, .
\end{align} 
As discussed in Sec. \ref{sec:free_fall_limit}, the vorticity contributions in Eq. \eqref{eq:delta_eij_time_evo} are not guaranteed to vanish at high frequencies, if the antenna is described by a free elastic solid rather than a non-elastic solid in pure free fall.\footnote{While we approximate the antenna as a one-dimensional curve in Eq.~\eqref{eq:current_antenna}, effects from the elastic boundary condition can be understood better by considering the finite extent of the antenna, where Eq. \eqref{eq:current_antenna} is replaced by a volume overlap with the current density profile in the antenna. Differences between elastic and pure free fall are only expected up to a distance $\lambda_f \sim v_s/\omega_g$ from the boundary of the antenna inside the conductor. However, since the signal currents flow up to a skin depth $\lambda_p\sim 10^{-6}\,\text{m} \sqrt{\text{GHz}/(\omega_0+\omega_g)}$ \cite{Pozar12}, we only expect the contribution from elastic free fall to be significant if $\lambda_p\lesssim\lambda_f$. This is typically only the case for elastic free fall below $\sim$ GHz EM frequencies.}

In addition, as for the magnetostatic case, there is no back action effects at high frequencies. Considering that the GW is resonant with the signal mode i.e $\omega_0 + \omega_g = \omega_n$, the electric mode coefficient in TT coordinates takes the simple form 
\begin{align}
     e_n^\mathrm{TT} &= -\frac{Q_n\int dV\,\bm{E}^{*}_n\cdot\bm{J}^\text{TT}_\text{eff}}{\omega_nV} \, .
\end{align}
The resulting signal power in either frame is
\begin{align}
    P_n &= \frac{\omega_n Q^2_n \langle \bar B\rangle^2  h^2 V (\eta_n^\text{bdy}+\eta_n^\text{bulk})^2}{2 Q_n^\mathrm{ext}}  \, ,
\end{align}
where we can use the simplified version of the coupling coefficients in TT coordinates
\begin{subequations}\label{eq:eta_TT_FF_heterodyne}
\begin{align}
(\eta_n^\text{bdy})^\text{TT}&=0\, , \\
(\eta_n^\text{bulk})^\text{TT}&= \frac{1}{\omega_n V\langle\bar{B}\rangle\tilde{h}}\int dV\,\bm{E}^{*}_n\cdot\tilde{\bm{J}}^\text{TT}_\text{eff}  \, ,
\end{align}
\end{subequations}
or alternatively in PD coordinates
\begin{subequations}\label{eq:eta_PD_FF_heterodyne}
\begin{align}
(\eta_n^\text{bdy})^\text{PD}&=\frac{i}{2V\langle\bar{B}\rangle\tilde{h}}\int_{\partial V} d\bm{A}\cdot\tilde{\bm{\xi}}_\text{PD}^\text{TT}(\omega_n-\omega_0)\,\left(\bm{B}_n^*\cdot\bm{B}_0-\bm{E}_n^*\cdot\bm{E}_0\right) \,,\\(\eta_n^\text{bulk})^\text{PD}&= \frac{1}{\omega_nV\langle\bar{B}\rangle\tilde{h}}\int dV\,\bm{E}^{*}_n\cdot\tilde{\bm{J}}^\text{PD}_\text{eff}  \, .
\end{align}
\end{subequations}
Fig. \ref{fig:delta_E_heterodyne} shows the electric field perturbation in a heterodyne experiment for different frequencies. As in Fig. \ref{fig:delta_E_static}, mechanical resonances enhance the signal in the elastic regime and electromagnetic resonances typically enhance the signal at microwave frequencies. A key difference is that two EM modes can be tuned to be almost degenerate in frequency, allowing EM resonances at arbitrarily low frequencies. In this example, $\omega_n - \omega_0=\omega_m$ is never satisfied, i.e mechanical and EM resonances are never excited at the same time and back-action can be neglected. In the figure, we envision two weakly coupled cylindrical cavities placed behind each other so that the two cylinder axes are orthogonal. Thus all modes can resonate in- or out of phase in both cavities with a tunable frequency difference which we assume to be $10\,\text{kHz}$. Above the mechanical resonances, i.e when $\omega_m \ll \omega_g$ and when $\omega_g \gg \omega_n - \omega_0 = 2\pi \,(10^4\,\text{Hz})$, the boundary term is still dominant in PD while in TT, the bulk contribution leads to the highest signal and the gauge invariant signal scales as $1/\omega$. While being subdominant in this regime, the bulk contribution in PD scales as $\omega$. The antenna contribution is computed assuming the GW wavelength to be always much larger than the antenna size, such that the whole antenna can be considered as one single observer. Furthermore, we assume the antenna to be freely suspended, so that it can be considered as freely falling at all frequencies considered and we can use Eq.~\eqref{eq:V_tt_FF_heterodyne} in TT coordinates, and we neglect its internal structure.\footnote{In a more general setup, where the antenna is e.g. rigidly attached to the wall, one would have to solve the elasticity equations for the antenna as in Sec. \ref{sec:elasticy_theory} and explicitly compute the tetrad perturbation $\delta\bm{e}$. This is similar to what was assumed in Fig.~\ref{fig:delta_E_static} where the antenna contribution gets enhanced at mechanical resonances. As mentioned earlier, at frequencies above the freely falling limit, the possible contributions from vorticity are non resonant, and therefore, the signal will barely be impacted.} In this pure free falling approximation, Eq.~\eqref{eq:V_tt_FF_heterodyne} becomes 
\begin{align}
    \delta \bm{E}^\text{ant,TT,FF}\approx -\frac12\bm{h}^\text{TT}\cdot \bar{\bm{E}} \, .
\end{align}

We assume the antenna to be located at the center of mass of the cavity, which is the origin of our coordinate system, so that $\delta \bm{E}^\text{ant,PD}=0$. As for Fig. \ref{fig:delta_E_static}, we have computed $\delta\bm{E}^\text{bdy,PD}$ implicitly by obtaining $\delta\bm{E}^\text{obs}$ in TT coordinates and subtracting the bulk and antenna contributions in PD coordinates.

\begin{figure}[ht!]
\centering

  \includegraphics[width=.9\linewidth]{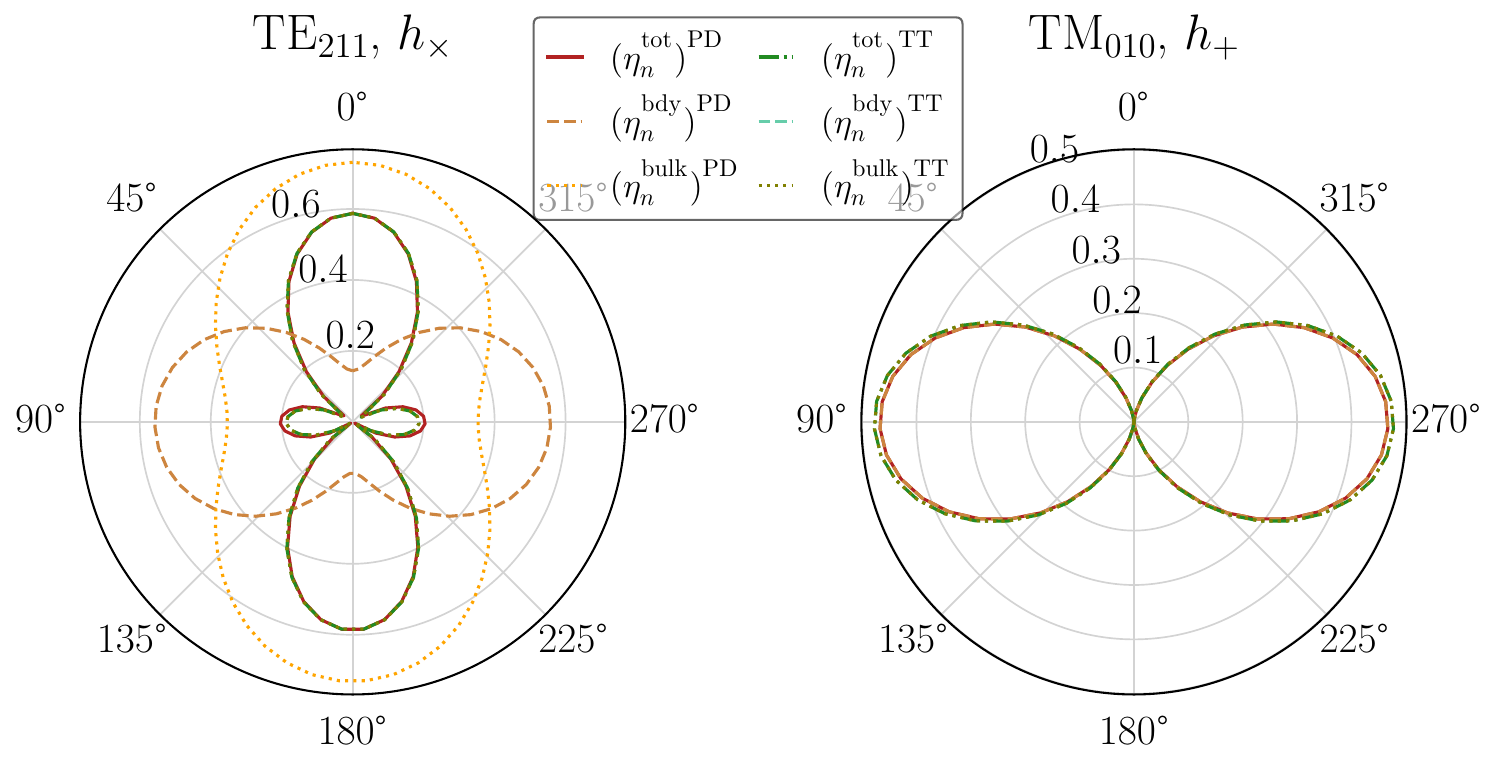}

\caption{The coupling coefficients in a heterodyne experiment in elastic free fall for a TM$_{010}$ mode in the background for different angles of a GW with the cylinder axis. The boundary and volume terms from Eqs. \eqref{eq:eta_TT_FF_heterodyne} and \eqref{eq:eta_PD_FF_heterodyne} and their sum $\eta_n^\text{tot}$ are evaluated in PD and TT coordinates. On the left, the signal mode is TE$_{211}$ and on the right, the signal mode is the same as the pump mode. We assume that the GW frequency $\omega_g$ is the same as the resonant frequency difference between background and signal mode. We show the absolute value of the coupling and add the couplings for degenerate mode polarizations in quadrature. The length and radius of the cylinder are $L=R=1\,\text{m}$. The total PD and TT curves overlap up to numerical accuracy.}
\label{fig:heterodyne_rotation}
\end{figure}

Fig. \ref{fig:heterodyne_rotation} shows the coupling coefficients in Eq. \eqref{eq:EM_couplings_heterodyne} for different angles of a resonant monochromatic GW in a heterodyne cavity experiment in elastic free fall. We show that the total coupling is gauge invariant for transitions between two different cavity modes when $\omega_g=\omega_n-\omega_0$, and also for transitions between the same cavity mode when $\omega_g\ll\omega_0$. The second case is e.g. important for setups where power in the pump mode is non-resonantly up-converted into the same mode or resonantly up-converted to an almost degenerate mode with the same geometry as in \cite{ballantini_microwave_2005} and shows the angular dependence of the resonance at $10$ kHz in Fig. \ref{fig:delta_E_heterodyne}.

\section{Discussion}\label{sec:discussion}

\begin{table}[t!]
    \centering
    \footnotesize
    \begin{tblr}{vlines, colspec={c| c c}}
    \hline
    & \SetCell[c=2]{} $j_n^\text{bdy}\big|_\mathrm{Magnetostatic}$ \\
    \hline
    Gauge & \SetCell[c=1]{} Elastic free fall  & \SetCell[c=1]{} Elastic regime \\
    \hline
    TT & 0 & $\frac{\mu Q_m}{\omega_m MV} \int_{\partial v} d \bm{a} \cdot (\bm{h}^\mathrm{TT} \cdot \bm{U}_m)\int_{\partial V} d\bm{A}\cdot (\bm{B}^*_n \times(\bar{\bm{B}}\times\bm{U}_m))$ \\
    \hline
    PD & $-\frac{i\omega_g}{V}\int_{\partial V} d\bm{A}\cdot (\bm{B}^*_n \times(\bar{\bm{B}}\times\bm{\xi}^\mathrm{TT}_\mathrm{PD}))$ & $\frac{\omega_m Q_m}{2MV}\int_{\partial v} dv\,\rho\,\bm{U}_m \cdot \bm{h}^\mathrm{TT} \cdot \bm{x}\int_{\partial V} d\bm{A}\cdot (\bm{B}^*_n \times(\bar{\bm{B}}\times\bm{U}_m))$ \\
    \hline
\end{tblr}
\caption{Simplified expression of the surface current source term Eq. \eqref{eq:boundary_current} in a magnetostatic setup, depending on the GW frequency and gauge. For simplicity, we assume that in the elastic regime, a single mechanical resonance is excited by the GW, i.e $\omega_g = \omega_m$ and we neglect non-resonant terms.}
\label{tab:boundary_source_v_magnetostatic}
\end{table}

\begin{table}[t!]
    \centering
    \footnotesize
    \begin{tblr}{vlines, colspec={c| c c}}
    \hline
    & \SetCell[c=2]{} $j_n^\text{bdy}\big|_\mathrm{Heterodyne}$ \\
    \hline
    Gauge & \SetCell[c=1]{} Elastic free fall  & \SetCell[c=1]{} Elastic regime \\
    \hline
    TT & $0$ & $\scriptscriptstyle\begin{aligned}
        -\frac{\mu Q_m\omega_n}{2\omega^2_mMV}&\int_{\partial v} d \bm{a} \cdot (\bm{h}^\mathrm{TT} \cdot \bm{U}_m)\\&\times\int_{\partial V} d\bm{A}\cdot\bm{U}_m\,\left(\bm{B}_n^*\cdot\bm{B}_0-\bm{E}_n^*\cdot\bm{E}_0\right)
    \end{aligned}$\\
    \hline
    PD & $\displaystyle\frac{i\omega_n}{2V}\int_{\partial V} d\bm{A} \cdot\bm{\xi}^\mathrm{TT}_\mathrm{PD}\,\left(\bm{B}_n^{*}\cdot\bm{B}_0-\bm{E}_n^{*}\cdot\bm{E}_0\right)$ & $\begin{aligned}-\frac{Q_m\omega_n}{4MV}&\int_{\partial v} dv\,\rho\,\bm{U}_m \cdot \bm{h}^\mathrm{TT} \cdot \bm{x}\\&\times\int_{\partial V} d\bm{A}\cdot\bm{U}_m\,\left(\bm{B}_n^*\cdot\bm{B}_0-\bm{E}_n^*\cdot\bm{E}_0\right)\end{aligned}$ \\
    \hline
\end{tblr}
\caption{Simplified expression of the surface current source term Eq. \eqref{eq:boundary_current} in a heterodyne setup, depending on the GW frequency and gauge. For simplicity, we assume that in the elastic regime, a single mechanical resonance is excited by the GW, i.e $\omega_g = \omega_m$, and in both frequency regimes, an EM resonance is excited by the GW, i.e $\omega_g+\omega_0 = \omega_n$. We neglect non-resonant terms and assume experimental parameters like the quality factors and EM pump power, such that back action is negligible and the signal power is maximized (see section \ref{sec:heterodyne_signal_power_elastic}).}
\label{tab:boundary_source_v_heterodyne}
\end{table}

In this paper, we develop a covariant formalism to calculate the effects of a GW interacting with a microwave cavity by using overlap coefficients with eigenmodes of the system on flat space. The benefit of this formalism is that the solution of the GW-Maxwell equations in the presence of dynamic boundary conditions becomes straightforward for any cavity geometries and EM background field configurations. The perturbation of the electromagnetic boundary conditions enters as surface currents when deriving the electromagnetic mode coefficients of the cavity. In Tabs.~\ref{tab:boundary_source_v_magnetostatic},\ref{tab:boundary_source_v_heterodyne}, we summarize the expression of such boundary integrals depending on the experimental setup (magnetostatic background or heterodyne setup) in the general case of an elastic detector and at high frequencies in the free fall limit. 

To account for the perturbations of the boundary conditions, we perform a quasi-eigenmode expansion of the electromagnetic and displacement fields, where the expansion is supplemented by a lifting function. As we demonstrate throughout the paper, those lifting functions are required when describing the field close to the boundary. Therefore, our formalism improves standard eigenmode perturbation theory.

In this paper, we also clarify the notion of free fall for bodies interacting with GWs. In particular, we distinguish between \emph{pure} and \emph{elastic} free fall where the former describes effectively point-like objects, i.e bodies without boundaries and the latter describes macroscopic objects whose boundary perturbation by a GW does not vanish at infinitely large frequencies. As we discuss throughout the paper, this distinction is important as the corresponding response to a GW can differ at high frequencies.

Other experiments searching for HFGWs like lumped-element circuits \cite{Brouwer22, Salemi21, ADMX_SLIC} can also benefit from the formalism presented here. In these experiments, a pickup loop measures the magnetic flux generated by the interaction between a passing GW and a toroidal magnetic field, i.e $\Phi \propto \int d\bm{A} \cdot \delta \bm{B}^\mathrm{obs}$. Therefore, the mechanical response of the antenna can enter the resulting flux as well, depending on the GW frequency and placement of the pickup loop in the background field.

Furthermore, as pointed out in \cite{Domcke25}, an additional contribution to the signal should arise due to the movement of the apparatus generating the magnetic field, an effect known as the magnetic Weber bar (MWB). While this effect should in principle contribute to the signal in cavity experiments, the cavity walls effectively shield any periodic EM perturbation from the exterior at the frequencies of interest of this paper, such that the MWB effect on the observed field measured by the antenna is negligible. 

Although we focused on the example of microwave cavities in this work, our calculations apply to electromagnetic cavities at any frequency range. Thus, they could be used to analyze optical resonators as well, as long as suitable boundary conditions are used.

As we prove throughout the paper, in specific frequency regimes and frames, some contributions to the signal become negligible, justifying approximations made in part of the existing literature. Important cases include the free-falling limit in TT coordinates, the long wavelength limit in PD coordinates, and the rigid limit in the PD frame. Our covariant formalism rules out contradictory calculations in the literature where such approximations were not used correctly.
Overall, the presented formalism can be used to extend signal calculations to new frequencies, especially in the intermediate range between acoustic and EM resonances for a wide class of GW detectors without relying on approximations. 

\acknowledgments
The authors acknowledge Dorian Amaral, Diego Blas and Saarik Kalia for helpful discussions, and Sebastian Ellis and Michel Paulsen for useful comments on this manuscript. We are particularly grateful to Asher Berlin and Wolfram Ratzinger for important comments on the derivation of the surface normal perturbation in Sec.~\ref{sec:covariant_EM}. JG is funded by Grant No. CNS$2023-143767$, funded by MICIU/AEI/$10.13039$/ $501100011033$ and by European Union NextGenerationEU/PRTR. TK acknowledges support by the Deutsche Forschungsgemeinschaft (DFG, German Research Foundation) under Germany's Excellence Strategy - EXC 2121 `Quantum Universe' - 390833306. The authors are grateful to the Mainz Institute for Theoretical Physics
(MITP) of the DFG Cluster of Excellence PRISMA+ for its hospitality and support during early stages of this work.

\appendix

\section{Details of the perturbation scheme}\label{sec:notes_on_perturbation_scheme}
An important subtlety arises when considering the raising and lowering of indices in the perturbation scheme. Consider a vector field $V$, whose components we decompose as usual
\begin{equation}
V^\mu=\bar{V}^\mu+\delta V^\mu\,,
\end{equation}
where all $\mathcal{O}(h)$ quantities are contained in $\delta V$ and all $\mathcal{O}(h^0)$ quantities in $\bar{V}$. The covariant vector corresponding to $V^\mu$ is given by 
\begin{equation}
    V_\mu=g_{\mu\nu}V^\nu=\eta_{\mu\nu}\bar{V}^\nu+\eta_{\mu\nu}\delta V^\nu+h_{\mu\nu}\bar{V}^\nu\,.
\end{equation}
Consequently, we need to make a choice on how to define the covariant quantities $\bar{V}_\mu$ and $\delta V_\mu$. As being the only quantity not $\mathcal{O}(h)$, we define $\bar{V}_\mu\equiv\eta_{\mu\nu}\bar{V}^\nu$. 
One option is to then define $\delta V_\mu\equiv\eta_{\mu\nu}\delta V^\nu+h_{\mu\nu}\bar{V}^\nu$ so that $V_\mu=\bar{V}_\mu+\delta V_\mu$ and $\delta V_\mu$ transforms like a covariant vector $\delta V_\mu'=\delta V_\mu-\bar{V}_\nu\partial_\mu\xi^\nu-\xi^\nu\partial_\nu \bar{V}_\mu$\,. The disadvantage is that great care must be taken when raising and lowering indices of the perturbed quantities since $\mathcal{O}(h)$ terms need to be added.

Another option is to define $\delta V_\mu\equiv\eta_{\mu\nu}\delta V^\nu$. The benefit is that the indices of $\bar{V}$ and $\delta V$ are now easily raised and lowered with the Minkowski metric. The disadvantage is that now $V_\mu\neq\bar{V}_\mu+\delta V_\mu$ and $\delta V_\mu$ does not transform as a covariant vector, like its index would suggest. Instead it transforms like a \emph{contravariant} vector
$\delta V'_\mu=\eta_{\mu\nu}\delta V'^\nu=\delta V_\mu+\bar{V}^\nu\partial_\nu \xi_\mu-\xi^\nu \partial_\nu \bar{V}_\mu$. The reverse applies to vectors to which the perturbation method has been applied as covariant vectors $W_\mu=\bar{W}_\mu+\delta W_\mu$. Then, $\delta W^\mu$ will transform as \emph{covariant} vector, no matter if the index is raised or lowered. In order to avoid mistakes, one has to keep track under which index position the perturbation method has been applied in the first place, and transform vectors accordingly, even if their indices have been changed later on. The same is true for higher order tensors. This second method is the one used in \cite{Ratzinger2024} (see their appendix B) and we are also using it in this work. 

Due to the importance of the index positions, we summarize the perturbation scheme of all relevant quantities below
\begin{subequations}
\begin{align}
    x^\mu&=\bar{x}^\mu+\delta x^\mu \, , \\
    u^\mu&=\dot{x}^\mu=\bar{u}^\mu+\delta u^\mu \, , \\
    g_{\mu\nu}&=\eta_{\mu\nu}+h_{\mu\nu} \, , \\
    F_{\mu\nu}&=\bar{F}_{\mu\nu}+\delta F_{\mu\nu}\, , \\
    T^{\mu\nu}&=\bar{T}^{\mu\nu}+\delta T^{\mu\nu} \, .
\end{align}
\end{subequations}

\section{Derivation of the electric boundary condition}\label{ap:EM_boundary_condition}

In this appendix, we explicitly derive the electric boundary condition in Eq.~\eqref{eq:boundary_condition_perturbation}.
Expanding to linear order in the GW strain, Eq.~\eqref{eq:boundary_condition} becomes
\begin{equation}\label{eq:EM_bdy_condition_perturbation}
    \bar{T}^i_{\underline{1},\underline{2}}(\delta F_{i0}(\bar{x})+\delta x^\nu\partial_\nu\bar{F}_{i0}(\bar{x})+\bar{F}_{i\nu}(\bar{x})\delta u^\nu)+\delta T_{\underline{1},\underline{2}}^i\bar{F}_{i0}(\bar{x})|_{\partial V}=0\,,
\end{equation}
where $V$ denotes the unperturbed volume of the cavity, where we have used $\bar{u}^\mu=\delta_0^\mu$ and the fact that $\bar{T}^0_{\underline{1},\underline{2}}=0$. The tangent perturbation $\delta T_{\underline{1},\underline{2}}$ is still taken to be attached to the perturbed worldline $x^\mu(\tau)$ of the surface and not to $\bar{x}^\mu(\tau)$. Next, we would like to rephrase this condition in terms of the normal vector $N^\mu$ by using the orthogonality condition $g_{\mu\nu}N^\mu T_{\underline{1},\underline{2}}^\nu=0$ to find
\begin{equation}\label{eq:N_T_overlap}
    \delta T_{\underline{1},\underline{2}}^\mu\bar{N}_\mu=-\delta N_\mu \bar{T}_{\underline{1},\underline{2}}^\mu-h_{\mu\nu}\bar{N}^\mu \bar{T}_{\underline{1},\underline{2}}^\nu\,.
\end{equation}
Using the (perfect conducting) boundary condition  $\bar{\bm{E}}=(\bar{\bm{E}}\cdot\bar{\bm{N}})\bar{\bm{N}}$ for the background field, we can therefore combine the two constraints in Eq. \eqref{eq:EM_bdy_condition_perturbation} and find
\begin{equation}  \label{eq:EM_bdy_condition_perturbation_2}
\bar{\bm{N}}\times\delta\bm{E}|_{\partial V}=\bar{\bm{N}}\times\left[\bar{\bm{B}}\times\delta\bm{u}-(\delta\bm{x}\cdot\nabla)\bar{\bm{E}}+\left(\delta \bm{N}+\bm{h}\cdot\bar{\bm{N}}\right)(\bar{\bm{E}}\cdot\bar{\bm{N}})\right]_{\partial V}\,.
\end{equation}

In order to find the perturbation of the normal $\delta\bm{N}$, we first define the tangential vectors by using the coordinates $y_1, y_2$ on the two-dimensional surface of the cavity $x^\mu=x^\mu(\tau, y_1,y_2)$, which are chosen so that $T_{\underline{1},\underline{2}}^\mu=\partial x^\mu/\partial y_{1,2}$ are orthonormal. By defining $\bar{T}_{\underline{1},\underline{2}}^\mu=\partial \bar{x}^\mu/\partial y_{1,2}$ and using the chain rule, we find the pertubation of the tangentials to be $\delta T_{\underline{1},\underline{2}}^\mu = \bar{T}_{\underline{1},\underline{2}}^\nu\partial_\nu\delta x^\mu$. By using this fact in Eq. \eqref{eq:N_T_overlap} and using the normalization condition $g_{\mu\nu}N^\mu N^\nu=1$, we know all scalar products of $\delta \bm{N}$ with our unperturbed spatial basis $\{\bar{\bm{N}},\bar{\bm{T}}_{\underline{1}},\bar{\bm{T}}_{\underline{2}}\}$. Therefore, we can construct the normal perturbation\footnote{Note that the perturbation of the normal vector cannot  be computed with e.g. the Fermi-Walker (FW) transport law. This is because the surface normal must remain orthogonal to the conductor's surface in the presence of a GW, which a FW-transported tetrad does not take into account. This is in contrast to the orientation of a tetrad carried by an observer, which evolves according to a FW transport, as we discuss in Sec.~\ref{sec:coordinate_invariance_EM}.}
\begin{align}
    \delta\bm{N}&=(\delta\bm{N}\cdot\bar{\bm{T}}_{\underline{1}})\bar{\bm{T}}_{\underline{1}}+(\delta\bm{N}\cdot\bar{\bm{T}}_{\underline{2}})\bar{\bm{T}}_{\underline{2}}+(\delta\bm{N}\cdot\bar{\bm{N}})\bar{\bm{N}}\nonumber\\
    &=-\left((\nabla\delta\bm{x})^T+\bm{h}\right)\cdot\bar{\bm{N}}+\left[\bar{\bm{N}}^T\cdot\left((\nabla\delta\bm{x})^T+\frac12\bm{h}\right)\cdot\bar{\bm{N}}\right]\bar{\bm{N}}\,,
\end{align}
where we have used the completeness relation $\delta^{ij}=\bar{T}_{\underline{1}}^i\bar{T}_{\underline{1}}^j+\bar{T}_{\underline{2}}^i\bar{T}_{\underline{2}}^j+\bar{N}^i\bar{N}^j$ and introduced the Jacobian matrix $(\nabla\delta\bm{x})_{ij}=\partial_j\delta x_i$. Plugging this result into Eq. \eqref{eq:EM_bdy_condition_perturbation_2} and using the homogeneous Maxwell equations \eqref{eq:Maxwell_homogeneous}, we can simplify the tangential electric field perturbation at the boundary and find the result in Eq. \eqref{eq:boundary_condition_perturbation}.

\section{Elasticity theory with damping}\label{ap:elasticity}
In order to take into account dissipation effects in the equations of elasticity, one can include viscosity in the stress-energy tensor of the body as \cite{Misner1973} 
\begin{subequations}
\begin{align}\label{eq:Tab_full}
T^{\mu\nu}= \rho u^\mu  u^\nu - \sigma^{\mu\nu}- 2\eta \chi^{\mu\nu}-\zeta P^{\mu\nu}\nabla_\rho u^\rho  \, ,
\end{align} 
where $\eta, \zeta$ are respectively the shear and bulk viscosities, $P_{\mu\nu}=g_{\mu\nu}+u_\mu u_\nu$ is a projection operator, 
and 
\begin{align}
    \chi_{\mu\nu}= \frac{1}{2}\left(P^{\rho}_{\:\:\nu}\nabla_\rho u_\mu +P^{\rho}_{\:\:\mu}\nabla_\rho u_\nu \right)-\frac{1}{3}P_{\mu\nu}\nabla_\rho u^\rho  \, , 
\end{align}
\end{subequations}
is the shear tensor of the body. We neglect heat conduction, but this can be added in a similar manner. Taking the divergence of Eq.~\eqref{eq:Tab_full} and linearizing with $\sigma^{\mu\nu}$ constructed as in Ref. \cite{Hudelist_2023}, we find  
Eq.~\eqref{eq:mech_elasticity_equation} with $\delta\sigma^{ij}$ as defined in Eq.~\eqref{eq:stress_tensor_flat} with
\begin{equation}
    \delta\sigma^{ij}_\text{damp.}=\eta\partial^j \delta \dot{x}^i +\left(\frac{\eta}{3}+\zeta\right)\delta^{ij}\partial_k \delta \dot{x}^k\, ,
\end{equation}
and $\sigma^{ij}_h$ as defined in Eq.~\eqref{eq:stress_tensor_h} with
\begin{equation}
    \delta\sigma^{ij}_\text{$h$, damp.}=\eta \dot{h}^{ij}+\frac12\left(\zeta-\frac{2\eta}{3}\right)\delta^{ij}\dot{h}^k_k\,,
\end{equation}
where we used $\delta \dot{x}^0 = h_{00}/2$, from $g_{\mu\nu}u^\mu u^\nu=-1$ and $u^\mu=(1,\vec 0)+\mathcal{O}(h)$. The equation of motion obtained is equivalent to the Kelvin-Voigt model, see e.g. \cite{Lobo98}. 

In order to derive equations of motion for the coefficients $q_m(t)$, we multiply the equations of elasticity \eqref{eq:mech_elasticity_equation} with an eigenmode distribution $\bm{U}_m$ and integrate over the volume $v$ of the cavity wall, which immediately leads to 
\begin{equation}
    \ddot{q}_m-\frac1M\int dv\,U_m^i\partial_j\delta\sigma_i^{\;j}=\frac1Mf_m^\text{bulk}\,.
\end{equation}
Integrating by parts and using the boundary conditions for $\delta\sigma^{ij}$ and $\sigma_m^{ij}$ we find that this is equivalent to 
\begin{subequations}
\begin{align}\label{eq:iowfeIEJWF}
     \ddot{q}_m+\omega_m^2q_m+\frac{\omega_m}{Q_m^v}\dot{q}_m+\frac1M\int da\,n_j\psi^{ij}_m\delta\dot{x}_i=\frac1M(f_m^\text{bulk}+f_m^\text{bdy})\,,
\end{align}
where we have defined 
\begin{align}
    \psi_m^{ij}=\eta\partial^j U_m^i +\left(\frac{\eta}{3}+\zeta\right)\delta^{ij}\partial_k U_m^k\,,
\end{align}
\end{subequations}
and used the fact that 
\begin{align}\label{eq:aoediaeo}
     \partial_j \psi_{m}^{kj}&= \frac{3\zeta + 4 \eta}{3\lambda + 6\mu}\partial_j ({\sigma}^I_m)^{kj}+\frac{\eta}{\mu}\partial_j ({\sigma}_m^S)^{kj}=-\rho\frac{\omega_m}{Q_m^v}U_m^k \, .
\end{align}

In Eq. \eqref{eq:aoediaeo}, we made a Helmholtz decomposition of the eigenmodes, i.e we set $\bm{U}_{m} =\bm{U}_{m}^I + \bm{U}_{m}^S \equiv \nabla \phi_m + \nabla\times \bm{H}_{m}$, where superscripts $I$ and $S$ stand for the \textit{irrotational} and \textit{solenoidal} parts respectively and defined the quality factors for losses in the bulk
\begin{equation}
    Q_m^v=\frac{1}{\omega_m}\left(\frac{3\lambda + 6 \mu}{4\eta+3\zeta}+\frac{\mu}{\eta}\right)\,.
\end{equation}

The remaining surface integral in Eq. \eqref{eq:iowfeIEJWF} shows that viscous losses also occur at the boundary. By assuming that the damping terms are only significant when $q_m$ is enhanced by a resonance and thus $\delta \dot{\bm{x}}\approx \dot{q}_m\bm{U}_m$, we can characterize the surface losses through a quality factor as well 
\begin{equation}
    Q_m^a=\frac{\omega_mM\dot{q}_m}{\int da\,n_j\psi^{ij}_m\delta\dot{x}_i}\approx\frac{\omega_mM}{\int da\,n_j\psi^{ij}_mU_m^i}\,.
\end{equation}
Thus, we arrive at the final equations of motion Eq. \eqref{eq:mech_EOM_q} and the solution \eqref{eq:delta_x_no_y}, where the total quality factor is given by the sum of bulk and boundary losses
\begin{equation}
    \frac{1}{Q_m}=\frac{1}{Q_m^v}+\frac{1}{Q_m^a}\,.
\end{equation}
Similar dissipation mechanisms have already been considered in the past literature, see e.g. \cite{ballantini_microwave_2005}.

As we show now explicitly, our solution for $\delta \bm{x}$ is a solution to the equations of elasticity \eqref{eq:mech_elasticity_equation} and $\bm{y}$ is only constrained at the boundary and arbitrary within the bulk.
We start with Eq.~\eqref{eq:mech_elasticity_equation}, insert the expansion for $\delta\bm{x}$ in Eq. \eqref{eq:delta_x_no_y} and take an overlap integral with a mode $\bm{U}_m$. Neglecting damping for simplicity, we find
\begin{align}
    \ddot{q}_m+\omega_m^2(q_m-y_m)-\frac1M\int dv\,U_m^i\partial_j\sigma_y^{ij}=\frac1M f_m^\text{bulk}\,.
\end{align}
Using our equation of motion for the $q_m$ coefficients Eq.~\eqref{eq:mech_EOM_q}, this is equivalent to the equation for $y$
\begin{equation}
    \omega_m^2y_m+\frac1M\int dv\,U_m^i\partial_j\sigma_y^{ij}=\frac1M f_m^\text{bdy}\,.
\end{equation}
After integrating by parts, using the boundary condition $n_j\sigma_y^{ij}=-n_j\sigma_h^{ij}$ and $n_j\sigma_m^{ij}=0$, we can verify $\int dv\,U_m^i\partial_j\sigma_y^{ij}=f_m^\text{bdy}-M\omega_m^2 y_m$, i.e. our decomposition solves the equations of motion.
Since we only used the boundary condition for $\bm{y}$, we have confirmed that the choice of $\bm{y}$ in the bulk of the solid is not constrained by the equation of motion \eqref{eq:mech_elasticity_equation}. However, including $\bm{y}$ is necessary to correctly account for the boundary force $f_m^\text{bdy}$.

\section{Electromagnetic cavity perturbation}\label{ap:EM}

In this appendix, we explicitly derive the equations of motion for the cavity eigenmodes in the presence of perturbed boundary conditions, assuming our decomposition Eq.~\eqref{eq:eigenmode_expansion}. We will then show that the fields $\bm{F}, \bm{G}$ are not additionally constrained by Maxwell's equations. 

\subsection{Equations of motion with damping}

Starting from Eq.~\eqref{eq:eigenmode_expansion} and following \cite{meidlinger} by multiplying both sides of Maxwell's equations \eqref{eq:GWMaxwell} with an eigenmode field and integrating by parts, the equations of motion of the $s$th solenoidal and $i$th irrotational mode coefficients become
\begin{subequations}\label{eq:EM_mode_coefficients_EOM_first_order}
\begin{align}
&\dot{b}_s+\frac{\omega_s}{Q_s}b_s=i\omega_se_s-j^\text{bdy}_s\,,\\
&\dot{e}_s=i\omega_sb_s-j^\text{bulk}_s\, ,\\
&\dot{e}_i=-j_i^\text{bulk}\,,
\end{align}
\end{subequations}
where the bulk and boundary source terms are defined in Eqs. \eqref{eq:bulk_current} and \eqref{eq:boundary_current} and where we have used the definition of the internal quality factor
\begin{equation}
    Q_s\coloneqq\frac{\omega_s\int dV\bm{B}_s^*\cdot\delta\bm{B}}{\int_{\partial V} dA\,Z_s\bm{B}_s^*\cdot\delta\bm{B}}=\frac{\omega_s V}{\int_{\partial V} dA\,Z_s\left|\bm{B}_s\right|^2 }\,, 
\end{equation}
where we used that $\int dA \,\bm{B}^*_s\cdot \bm{B}_i=0$ because the cavity is a closed surface, $\int dA \,\bm{B}^*_s\cdot\bm{G}=0$, because $\bm{G}$ is orthogonal to all the magnetic modes at the surface and we assume $b_s \int dA|\bm{B}_s|^2 \gg \sum_{k\in \mathbf{S},k\neq s}b_k\int dA\bm{B}_s^*\cdot\bm{B}_k$. Then, we recover the familiar expression for the quality factor typically obtained from simulation or measurement \cite{Padamsee98}.

Note that the surface losses and the resulting quality factor are affected if antennas are actively removing power in addition to the unavoidable dissipation around the entire surface. However, the boundary condition \eqref{eq:boundary_condition_perturbation_B} does not describe antennas well and the resulting power loss is best added using equation \eqref{eq:Qext} and modifying $Q_s\to(1/Q_s+1/Q^\text{ext}_s)^{-1}$ which is called the \emph{loaded} quality factor. 

Combining these equations, we can separate the electric and magnetic coefficients and find Eqs. \eqref{eq:perturbation_coupling_eom} and the solution in Fourier space in Eqs. \eqref{eq:EM_mode_coeff_no_F}.

As derived in the next section of this appendix, the choice of the vector fields $\bm{F}$ and $\bm{G}$ for constructing $\delta\bm{E}$ and $\delta\bm{B}$ is not additionally constrained by Maxwell's equations. As we show below, this means our formalism is actually independent of the choice of $\bm{F},\bm{G}$ within the cavity volume. Focusing on the electric field, from Eqs.~\eqref{eq:eigenmode_expansion} and \eqref{eq:EM_mode_coeff_no_F}, the dependence of $\delta \bm{E}$ on $\bm{F}$ is simply
\begin{align}
\delta\tilde{\bm{E}}\supset\tilde{\bm{F}}-\frac{1}{V}\sum_n\bm{E}_n\int \, dV \, \bm{E}^{*}_n \cdot \tilde{\bm{F}} \, ,
\end{align}
i.e in regions where an expansion of $\bm{F}$ in eigenmodes converge to $\bm{F}$, $\delta \bm{E}$ becomes independent of $\bm{F}$. While this statement is clearly not true at the boundary, where $\bm{F}_\parallel \neq 0$, it becomes valid at an arbitrarily small distance away from the boundary, as the eigenmodes form a complete basis for any vector field living \textit{within} the cavity. In other words, the only practical importance of $\bm{F}$ is on the boundary such that we find Eqs. \eqref{eq:delta_E_no_F}. 

\subsection{Consistency with Maxwell's equations}

We now check the consistency of our decomposition with Maxwell's equations (without damping and background current for simplicity) and confirm that they do not impose further constraints on the lifting functions $\bm{F}$ and $\bm{G}$. We will focus on $\bm{F}$ but our findings are extended to $\bm{G}$. In the following, we decompose the vector field $\bm{F}$ as $\bm{F}= \bm{F}^I+\bm{F}^S$, where the former is divergence free, and the latter curl free. We start with the modified Gauss law, which reads $\nabla\cdot\delta\bm{E}=J^0_\text{eff}$ in the presence of a GW. Using Eq.~\eqref{eq:perturbation_coupling_eom} and the conservation of the effective current $\partial_\mu J^\mu_\mathrm{eff}=0$, Gauss's law is equivalent to 
\begin{subequations}
\begin{align}
    -\sum_n \frac{1}{V}\left[\int \, dV\, \bm{E}_i^{*}\cdot(\bm{J}_\mathrm{eff}+\dot{\bm{F}}^I)\right]\nabla\cdot\bm{E}_i+\nabla\cdot\dot{\bm{F}}^I &=-\nabla\cdot\bm{J}_\mathrm{eff} \, ,
\end{align}
which, after taking the overlap with the potential of an irrotational mode $\int dV\phi^*_j$ on both sides and integrating by parts, is 
\begin{align}
   \int_{\partial V} d\bm{A} \cdot \left[\dot{\bm{F}}^I+\bm{J}_\mathrm{eff}-\sum_n\left(\int \, dV\,\bm{E}^{*}_i \cdot (\bm{J}_\mathrm{eff}+\dot{\bm{F}}^I)\right)\bm{E}_i\right]\phi^*_j &= 0 \, , 
\end{align}
which is true when $\phi_j\big|_{\partial V} =0$ as we chose it in chapter \ref{sec:EM_eigenmode_decomposition}\footnote{Note that while the irrotational eigenmodes can describe any irrotational vector field within the cavity, this statement is not true anymore at the surface.}.
\end{subequations}

The wave equation reads $\delta\ddot{\bm{E}}-\nabla^2\delta\bm{E}=-\nabla J^0_\mathrm{eff}-\dot{\bm{J}}_\mathrm{eff}$. Taking the overlap with an irrotational mode $\bm{E}^{*}_i$ and using the modified Gauss law and Eq.~\eqref{eq:perturbation_coupling_eom}, this leads to 
\begin{align}
    \int \, dV \, \bm{E}^{*}_i \cdot \nabla^2 \bm{F}^S &= 0 \, , 
\end{align}
which is satisfied using $\phi_i\big|_{\partial V} =0$ and that the cavity walls form a closed surface. Taking now the overlap of the wave equation with a solenoidal mode, we find
\begin{equation}\label{eq:check_eom_solution}
    i\omega_s\int_{\partial V} d\bm{A}\cdot\left(\bm{B}_s^{*}\times\bm{\mathcal{V}}\right)=\int dV\bm{E}_s^{*}\cdot\left(\omega_s^2\bm{F}+\nabla^2\bm{F}^S\right) \,.
\end{equation}
Manipulating the LHS, we find
\begin{subequations}
\begin{align}
    &i\omega_s\int_{\partial V} d\bm{A}\cdot\left(\bm{B}_s^{*}\times\bm{\mathcal{V}}\right)=i\omega_s\int_{\partial V} d\bm{A}\cdot\left(\bm{B}_s^{*}\times\bm{F}\right)=\int dV\left(\omega_s^2\bm{E}_s^{*}\cdot\bm{F}-i\omega_s(\nabla\times\bm{F}^S)\cdot\bm{B}_s^{*}\right)\\
    &=\int dV\left(\omega_s^2\bm{E}_s^{*}\cdot\bm{F}-(\nabla\times\bm{F}^S)\cdot(\nabla\times\bm{E}_s^{*})\right)\\
    &=\int dV\left(\bm{E}_s^{*}\cdot\left(\omega_s^2\bm{F}+\nabla^2\bm{F}^S\right)-\nabla\cdot[\bm{E}_s^{*}\times(\nabla\times \bm{F}^S)]\right)\\
    &=\int dV\bm{E}_s^{*}\cdot\left(\omega_s^2\bm{F}+\nabla^2\bm{F}^S\right)-\int_{\partial V} d\bm{A}\cdot[\bm{E}_s^{*}\times(\nabla\times \bm{F}^S)]\\
    &=\int dV\bm{E}_s^{*}\cdot\left(\omega_n^2\bm{F}+\nabla^2\bm{F}^S\right)\,,
\end{align}
\end{subequations}
where we use $\nabla\cdot\bm{E}_s^*=0$ and $\bar{\bm{N}}\times \bm{E}_s^*=0$. Then, Eq.~\eqref{eq:check_eom_solution} is automatically satisfied.

This means that the choice of $\bm{F}$ in the cavity is not constrained by Maxwell's equations, i.e apart from satisfying Eq.~\eqref{eq:F_V_eq}, it is free.

\subsection{Transformation of the mode coefficients}

In this section, we show explicitly how the electric mode coefficients transform under the $\mathcal{O}(h)$ coordinate transformation $x^\mu \rightarrow x^\mu + \xi^\mu$, in the case of a magnetostatic background. The effective current transforms as shown in Eq.~\eqref{eq:gauge_transfo_j} and $\bm{\mathcal{V}}'=\bm{\mathcal{V}}+\bm{\bar{B}}\times\bm{\dot{\xi}}$.
Using these transformations, for a monochromatic GW, we can find the transformed expansion coefficients from the solution of Eq.~\eqref{eq:perturbation_coupling_eom} \textit{inside} the cavity as\footnote{Note that one must include the terms that are suppressed by $1/Q_s$ in order to find the correct transformation.}
\begin{subequations}
\begin{align}\label{eq:e_n_transformation}
     e'_s &= e_s+\frac{i\omega_s\int_{\partial V} d\bm{A}\cdot\left[\bm{B}^{*}_s\times\left(\bm{\bar{B}}\times\dot{\bm{\xi}}\right)\right]-\int dV\bm{E}^{*}_s\cdot\left(\omega^2\bm{\bar{B}}\times\dot{\bm{\xi}}-\nabla\times\left[\nabla\times\left(\bm{\bar{B}}\times\dot{\bm{\xi}}\right)\right]\right)}{\left(\omega_s^2-\omega^2+\frac{i\omega \omega_s}{Q_s}\right)V} \, \nonumber \\
    &+\frac{\omega_s}{Q_s}\frac{\int dV\bm{E}^{*}_s\cdot\left(i\omega\bm{\bar{B}}\times\dot{\bm{\xi}}-\frac{i}{\omega}\nabla\times\left[\nabla\times\left(\bm{\bar{B}}\times\dot{\bm{\xi}}\right)\right]\right)}{\left(\omega_s^2-\omega^2+\frac{i\omega \omega_s}{Q_s}\right)V}\, \\
    &=  e_s+\frac{\int dV\bm{E}^*_s\cdot\left(\bm{\bar{B}}\times\dot{\bm{\xi}}\right)}{V}\,,
\end{align}
for the solenoidal eigenmodes coefficients and
\begin{align}
     e'_i&= e_i+\frac{\int dV\bm{E}^{*}_i\cdot\left(\omega^2\bm{\bar{B}}\times\dot{\bm{\xi}}-\nabla\times\left[\nabla\times\left(\bm{\bar{B}}\times\dot{\bm{\xi}}\right)\right]\right)}{\omega^2V} = e_i+\frac{\int dV\bm{E}^{*}_i\cdot\left(\bm{\bar{B}}\times\dot{\bm{\xi}}\right)}{V}\,,
\end{align}
\end{subequations}
for the irrotational mode coefficients. In both cases we have used the vector identity 
\begin{equation}
    \bm{A}\cdot\left[\nabla\times(\nabla\times \bm{B})\right]-\bm{B}\cdot\nabla(\nabla\cdot \bm{A})=\nabla\cdot\left((\nabla\times \bm{B})\times \bm{A}-(\nabla\times \bm{A})\times \bm{B}\right)-\bm{B}\cdot\nabla^2\bm{A}\,,
\end{equation}
and the eigenvalue equations and boundary conditions of the eigenmodes \eqref{eq:eigenmode_maxwell}. 

\section{Numerical implementation}
\subsection{General considerations}\label{sec:numerical_considerations}
In the main text we have shown that all coordinate frames are equally valid when calculating the electromagnetic response of detectors to GWs. However, the same can not be said for practical implementations of our formulas. In particular, the mechanical and electromagnetic response \emph{off resonance} in most frames relies on using a \emph{complete} set of eigenmodes and not just a selection, and further requires delicate cancellations, which can easily be spoiled by numerical precision. In this small section, we will first point out those problems and discuss how to address them. In the following, we neglect the boundary condition functions $\bm{F}$ and $\bm{y}$ as their relevance cancels in the relevant formulas and does not need to be included. 

First, the difference between the motion of an elastic solid in two frames is given by
\begin{equation}
    \delta\bm{x}'-\delta\bm{x}=\frac1M\sum_m\bm{U}_m\int dv\rho \,\bm{U}_m\cdot\bm{\xi}=\bm{\xi}\,.
\end{equation}
However, the last equation is only valid when all mechanical eigenmodes with significant overlap with $\bm{\xi}$ are considered. In practice, this can be a prohibitively large amount of modes, especially at high frequencies when Eq.~\eqref{eq:free_falling_limit} is fulfilled, and the detector is expected to be in free fall $\delta \bm{x}^\text{TT}\approx 0$, $\delta\bm{x}^\text{PD}\approx\bm{\xi}_\text{PD}^\text{TT}$. Therefore, when using only a subset $\mathbb{m}\subset\mathbb{N}$ of mechanical eigenmodes, we recommend approximating
\begin{align}\label{eq:delta_x_PD_numerical}
    \delta\bm{x}^\text{PD}&=\bm{\xi}_\text{PD}^\text{TT}+\sum_{m\in\mathbb{N}}\bm{U}_m\left(q_m^\text{PD}-\frac1M\int dv\rho \,\bm{U}_m\cdot\bm{\xi}_\text{PD}^\text{TT}\right)\\\nonumber&\approx \bm{\xi}_\text{PD}^\text{TT}+\sum_{m\in \mathbb{m}}\bm{U}_m\left(q_m^\text{PD}-\frac1M\int dv\rho \,\bm{U}_m\cdot\bm{\xi}_\text{PD}^\text{TT}\right)\,,
\end{align}
while in freely falling coordinates we can simply limit the sum
\begin{equation}\label{eq:delta_x_TT_numerical}
    \delta\bm{x}^\text{TT}\approx \sum_{m\in \mathbb{m}}\bm{U}_mq_m^\text{TT}\,.
\end{equation}
This ensures the correct freely falling limit when using only a subset of mechanical eigenmodes, without compromising the accuracy of the calculation near the resonances $\omega_{m\in\mathbb{m}}$. 
We use this method to calculate the frequency-dependent mechanical response in TT and PD in Fig. \ref{fig:delta_x}. Note that the result is clearly not correct near mechanical resonances excluded from $\mathbb{m}$ if they have non-vanishing coupling to the GW.

At low frequencies, within the rigid limit $\delta \bm{x}^\text{PD}\approx 0$, $\delta\bm{x}^\text{TT}\approx-\bm{\xi}_\text{PD}^\text{TT}$, the inverse problem occurs, since an incomplete set of eigenmodes can not fully expand the correct $\delta\bm{x}^\text{TT}$. In that regime we can use 
\begin{align}\label{eq:delta_x_TT_numerical_rigid}
    \delta\bm{x}^\text{TT}&\approx -\bm{\xi}_\text{PD}^\text{TT}+\sum_{m\in \mathbb{m}}\bm{U}_m\left(q_m^\text{TT}+\frac1M\int dv\rho \,\bm{U}_m\cdot\bm{\xi}_\text{PD}^\text{TT}\right)\,,
\end{align}
and
\begin{equation}\label{eq:delta_x_PD_rigid}
    \delta\bm{x}^\text{PD}\approx \sum_{m\in \mathbb{m}}\bm{U}_mq_m^\text{PD}\,
\end{equation}
for higher numerical accuracy. Consequently, for practical applications we recommend using TT coordinates to obtain the displacement field of detectors in the freely falling regime and PD coordinates for detectors in the rigid regime.

A similar problem occurs when evaluating the solenoidal electromagnetic mode coefficient \eqref{eq:EM_mode_coefficient_e_S} at a frequency below its resonance $\omega\ll\omega_n$ but still in the freely falling regime in PD coordinates
\begin{equation}\label{eq:e_PD_low_f_approx}
    e_s^\text{PD}\approx\frac{i}{\omega_s}\int_{\partial V} d\bm{A}\cdot(\bm{B}_s^*\times\bm{\mathcal{V}^\text{PD}})=\int dV\bm{E}_s^*\cdot\bm{\mathcal{V}}^\text{PD}-\frac{i}{\omega_s}\int dV\bm{B}_s^*\cdot(\nabla\times\bm{\mathcal{V}}^\text{PD})\,.
\end{equation}
When constructing $\delta \bm{E}^\text{tetrad}$, the first term cancels since $\delta \bm{E}_\text{tetrad}\supset \sum_s \bm{E}_s\int dV\bm{E}_s^*\cdot\bm{\mathcal{V}}^\text{PD}/V-\bm{\mathcal{V}}^\text{PD}=0$, as it should, since a calculation in TT coordinates shows that the electric field should drop a factor $\omega_g/\omega_n$ faster than the first term in \eqref{eq:e_PD_low_f_approx} would allow. However, when $\bm{\mathcal{V}}^\text{PD}$ is not fully expanded in solenoidal \emph{and} irrotational modes, the cancellation does not work and the electric field can be overestimated by a factor $\omega_n/\omega_g\gg1$. The situation can be mitigated using the same trick as before
\begin{equation}\label{eq:delt_E_numerical_help}
    \delta\bm{E}\approx \bm{\mathcal{V}}+\sum_{n\in\mathbb{n}}\left(e_n-\frac{1}{V}\int dV\bm{E}_n^*\cdot\bm{\mathcal{V}}\right)\bm{E}_n \, ,
\end{equation}
to ensure the correct transformation between coordinate frames. However, note that this requires the overlap with $\bm{\mathcal{V}}$ from Eq. \eqref{eq:delt_E_numerical_help} and from the surface integral in Eq. \eqref{eq:e_PD_low_f_approx} to cancel to a relative accuracy $\omega_g/\omega_n$ which can require prohibitive computational resources for some frequencies. 

For this reason, we recommend to always use TT coordinates for freely falling detectors, where $j^\text{bdy}_n\approx0$ and the problem can be avoided.

\subsection{Details on the examples}\label{sec:details_on_examples}

In order to construct an analytic toy example to demonstrate our treatment of elasticity theory for a cylindrical shell with inner radius $R_i$, outer radius $R_o$ and constant mass density $\rho$, we consider an irrotational mode $\nabla\times \bm{U}_m=0$ with only radial dependence
\begin{equation}\label{eq:minimal_mech_mode}
    \bm{U}_m(r, \phi, z)= \frac {1}{\langle F_m\rangle}F_m(r)\hat{\bm{r}}\,,
\end{equation}
which is a solution to the eigenmode equations of elasticity \eqref{eq:mech_eigenmode_equation} for the combination of Bessel functions
\begin{equation}
    F_m(r)=J'_0(k_m r)-\frac{J_0''(k_m R_i)}{Y_0''(k_mR_i)} Y'_0(k_mr)\,.
\end{equation}
The free boundary condition \eqref{eq:mech_eigenmode_boundary} under the approximation of the material parameters $\lambda\ll\mu$ is fulfilled if  $k_m$ is chosen to be the $m$th smallest solution of the equation
\begin{equation}
    J_0''(k_m R_i)Y_0''(k_mR_o)=J_0''(k_m R_o)Y_0''(k_mR_i)\,,
\end{equation}
and the normalization is found from $\langle F_m\rangle^2=\frac{1}{R_o^2-R_i^2}\int_{R_i}^{R_o}  \,rF_m(r)^2dr$. 
The resonant frequencies are then found from $\omega_m=2\pi f_m=v_s\,k_m$, where $v_s = \sqrt{2\mu/\rho}$ is the speed of sound in the material.

In order to construct $\delta\bm{x}$ for the figures in the main text, we use equations \eqref{eq:delta_x_TT_numerical} and \eqref{eq:delta_x_PD_numerical} to construct the movement of the cylindrical shell from our minimal eigenmode example \eqref{eq:minimal_mech_mode}. For the ends of the cylinder, we assume a non-elastic material with $\mu=\lambda=0$ which is freely falling for all frequencies. For the quality factor of the mechanical modes we assume $Q_m=10^4\frac{k_0}{k_m}$. Furthermore, we assume $R_o-R_i=0.1\,\text{m}$ and the material parameters $\mu=37.5\,\text{GPa}$, $\rho= 8570\,\text{kg/m}^3$, which are typical for Niobium. However, our simplification $\lambda\ll\mu$ is not consistent with a physical metal like Niobium. 

\section{Off-resonant signals, magnetostatic case }\label{ap:signals}

In this appendix, we discuss the expected signals of magnetostatic cavity experiments off EM resonance i.e. $|\omega^2-\omega_s^2    |\gg\omega_s^2/Q_s$. Key differences to the sensitivity near resonances are that several modes can contribute to the expansion simultaneously and irrotational modes can not be neglected in general. 

\subsection{Free falling limit}

In the (elastic) free-falling limit, the electric mode coefficients are
\begin{subequations}
\begin{align}
    e_s^\text{TT} &\approx \frac{-i\omega \,(j_s^\text{bulk})^\text{TT}}{\omega_s^2-\omega^2}\, , \\
    e_i^\text{TT}&\approx \frac{i}{\omega}(j_i^\text{bulk})^\text{TT} \, .
\end{align}
\end{subequations}
Assuming that we are measuring the fields in the interior of the cavity with a freely falling antenna, we have 
\begin{align}
     \delta \bm{E}^\mathrm{obs} = \delta \bm{E} &= \frac{i}{\omega}\left(\sum_i(j_i^\text{bulk})^\text{TT} \bm{E}_i-\sum_s\frac{\omega^2 (j_s^\text{bulk})^\text{TT}}{\left(\omega_s^2-\omega^2\right)} \bm{E}_s \right) \, . 
\end{align}
Then, the signal power is 
\begin{subequations}
\begin{align}
    P^\mathrm{off. \: res.}_\mathrm{ant} &= \frac{1}{\omega^2 Z_\mathrm{eff}}\left|-\sum_s\frac{\omega^2(j_s^\text{bulk})^\text{TT}}{\omega^2_s - \omega^2}\int d \bar{\bm{\ell}}\cdot\bm{E}_s +\sum_i(j_i^\text{bulk})^\text{TT} \int d \bar{\bm{\ell}}\cdot \bm{E}_i \right|^2 \, \\
    &=\frac{V}{2\omega^2}\left|-\sum_s\frac{\omega^2(j_s^\text{bulk})^\text{TT}}{\omega^2_s - \omega^2}\sqrt{\frac{\omega_s}{Q_{s}^\mathrm{ext}}} + \sum_i\sqrt{\frac{\omega_i}{Q_{i}^\mathrm{ext}}}(j_i^\text{bulk})^\text{TT} \right|^2 \, \\
    &=\frac{h^2 \bar{B}^2 V}{2}\left|-\sum_s\frac{\omega^2}{\omega^2_s - \omega^2}\sqrt{\frac{\omega_s}{Q_{s}^\mathrm{ext}}}(\eta^\mathrm{bulk}_{s})^\text{TT} +\sum_i \sqrt{\frac{\omega_i}{Q_{i}^\mathrm{ext}}}(\eta^\mathrm{bulk}_{i})^\text{TT} \right|^2 \, ,
\end{align}
\end{subequations}
where $(\eta^\mathrm{bulk}_{s,i})^\text{TT}$ are dimensionless couplings given in Eq.~\eqref{eq:eta_bulk_TT_magneto}.
Note that the second to last line can be obtained by considering a choice of phase where the electric eigenmode fields are purely real. As expected, the power out of resonance is suppressed $\propto1/Q^2_n$ compared to the power on resonance in Eq.~\eqref{eq:static_B_power_res}.

\subsection{Elastic limit}

In the regime where $\omega \ll \omega_s$, the electric mode coefficients simplify to
\begin{subequations}
    \begin{align}
        e_s &\approx \frac{-i\,j_s^\text{bdy}}{\omega_s}\, , \\
        e_i&\approx \frac{i\,j_i^\text{bulk}}{\omega} \, .
    \end{align}
\end{subequations}
For simplicity, we assume that a monochromatic GW excites the $m^{th}$ mechanical resonance, i.e $\omega_g=\omega_{m}$. In this case, the boundary current is enhanced by the associated mechanical quality factor and therefore dominates over the bulk current in any frame. Specifically, using Eqs.~\eqref{eq:EM_couplings_static} and \eqref{eq:eta_n_bdy}, it is given by
\begin{align}
    j^\text{bdy}_s &=  \frac{\omega_{m}\bar{B}h}{2}Q_{m}\eta^{m}_s(\Gamma^\text{bdy}_{m}+\Gamma^\text{bulk}_{m}) \, .
\end{align}
In the elastic regime, we can work with PD coordinates which are more convenient in which case, $\Gamma^\text{bdy}_{m} \sim 0$. Then, the observed electric field is  
\begin{align}
    \delta \bm{E}^\mathrm{obs} &= -\frac{\omega_{m} V^{1/3}h}{2}Q_{m}\Gamma^\text{bulk}_{m}\left(i\bar{B}\sum_s \frac{\eta^{m}_s}{\omega_s V^{1/3}}\bm{E}_s + \bar{\bm{B}}\times \bm{U}_{m}\right) \, .
\end{align}
To simplify the calculation, we assume a cylindrical cavity where $\bm{\bar{B}}= \bar{B} \hat{\bm{z}}$ and that the probe is oriented in the same direction, i.e, $\bar{\bm{\ell}}= \bar{\ell} \hat{\bm{z}}$ in Eq.~\eqref{eq:current_antenna}. In this case, the second term does not deliver any power to the antenna, such that the power in PD coordinates reads
\begin{subequations}
\begin{align}
P^{m}_\mathrm{sig} &= \frac{\omega^2_{m} h^2 \bar{B}^2}{4 Z_\mathrm{eff}}\left| Q_{m}\Gamma^\text{bulk}_{m}\sum_s \frac{\eta^{m}_s}{\omega_s} \int \, d\bm{\bar{\ell}}\cdot\bm{E}_s\right|^2 \, \\
&= \frac{\omega^2_{m} h^2 \bar{B}^2 V}{8}\left| Q_{m}\Gamma^\text{bulk}_{m}\sum_s \frac{\eta^{m}_s}{\sqrt{\omega_sQ_s^\text{ext}}}\right|^2 \, .
\end{align}
\end{subequations}
Assuming that only one EM and mechanical mode contribute significantly to the signal power, the ratio of this power to the one at EM resonance Eq.~\eqref{eq:static_B_power_res} is
 \begin{align}
     \frac{P^{m}_\mathrm{sig}}{P^n_\mathrm{sig}} &\sim \left(\frac{Q_{m}}{Q_n}\right)^2\left(\frac{\omega_{m}}{\omega_n}\right)^2 \, ,
 \end{align}
up to dimensionless couplings and $\mathcal{O}(1)$ factors. This is $\ll 1$ in general and is consistent with what is shown in Fig.~\ref{fig:delta_E_static}.

\section{One-dimensional toy example}\label{sec:toy_example}
The key aspects of our eigenmode perturbation formalism can be understood in a simple one-dimensional (1D) wave equation with a dynamical boundary condition. There, we can compare our eigenmode expansion with a boundary condition lifting function against an \emph{exact} solution. This allows us to demonstrate the importance of the lifting functions and understand the convergence of our solution off-resonance. 

As a physical example, let us first consider a 1D elasticity equation with stress tensor $\sigma=\rho v_s^2\delta x'$ and no sources (as is the case in TT coordinates)
\begin{equation}\label{eq:toy_EOM}
    \delta \ddot{x}-v_s^2\delta x''=0 \, ,
\end{equation}
with dynamic Neumann boundary conditions\footnote{A similar toy example can be constructed for Dirichlet boundary conditions for $\delta x$, which is analogous to the EM case discussed in Sec. \ref{sec:covariant_EM}.} on the spatial derivatives of $\delta x$
\begin{subequations}
\begin{align}\label{eq:toy_bdy_condition}
    \delta x'(t, 0)&= h e^{i\omega t}\,,\\
    \delta x'(t, L) &=h e^{i\omega (t-L)}\,.
\end{align}
\end{subequations}
An exact solution is given by
\begin{equation}\label{eq:toy_exact_solution}
    \delta x(t, x)=h e^{i\omega t}\frac{v_s}{\omega}\left(\frac{\cos\frac{\omega L}{v_s}-e^{-i\omega L}}{\sin{\frac{\omega L}{v_s}}}\cos\frac{\omega}{v_s}x+\sin{\frac{\omega}{v_s}x}\right)\,.
\end{equation}
We can also use our eigenmode formalism to solve the problem using the modes
\begin{align}
    U_m=a_m\cos k_m x\,,
\end{align}
with $k_m=m\pi/L$, so that they fulfill the unperturbed boundary condition $U_m'(0)=U_m'(L)=0$ and the normalization $a_0=1$, $a_{m>0}=\sqrt{2}$.
Then, we can decompose
\begin{subequations}
\begin{align}
    \delta x=\sum_m(q_m-y_m)U_m+y
\end{align}
with 
\begin{align}
    q_m&=\frac1L\int_0^Ldx\, \delta x U_m\,,\\
    y_m&=\frac1L\int_0^Ldx\,y U_m\,,
\end{align}
\end{subequations}
and we can choose the lifting function e.g. 
\begin{equation}
    y=\frac{i}{\omega}h e^{i\omega (t-x)}\,,
\end{equation} 
so that it enforces the boundary condition. Multiplying \eqref{eq:toy_EOM} with $U_m$ and integrating by parts, we find the equation of motion
\begin{equation}
    \ddot q_m+\omega_m^2q_m=\frac{a_mv_s^2}{L}he^{i\omega (t-L)}((-1)^{m}-e^{i\omega L}) \, ,
\end{equation}
where we have defined $\omega_m=v_sk_m$. On the right-hand side, we have found a boundary source term just like $f_m^\text{bdy}$ in three dimensions. The equation can be solved by a monochromatic ansatz $q_m(t)\propto e^{i\omega t}$
\begin{equation}
    q_m(t)=a_m\frac{v_s^2}{L}he^{i\omega (t-L)}\frac{(-1)^{m}-e^{i\omega L}}{\omega_m^2-\omega^2}\,,
\end{equation}
and the overlap coefficients for the boundary lifting function are
\begin{equation}
    y_m(t)=a_m\frac{v_s^2}{L}h e^{i\omega (t-L)}\frac{(-1)^{m}-e^{i\omega L}}{\omega_m^2-(v_s\omega)^2}\,.
\end{equation}
This allows us to write down the full expansion
\begin{equation}\label{eq:toy_example_sum}
    \delta x(t, x)=he^{i\omega t}\left[\frac{(v_s\omega)^2}{L}e^{-i\omega L}\sum_m\frac{a_m^2(1-v_s^2)\left((-1)^{m}-e^{i\omega L}\right)}{(\omega_m^2-\omega^2)(\omega_m^2-(v_s\omega)^2)}\cos k_m x\,+e^{-i\omega x}\frac{i}{\omega}\right]\,.
\end{equation}

\begin{figure}[ht!]
\centering
\begin{subfigure}{0.54\textwidth}
  \includegraphics[width=\textwidth]{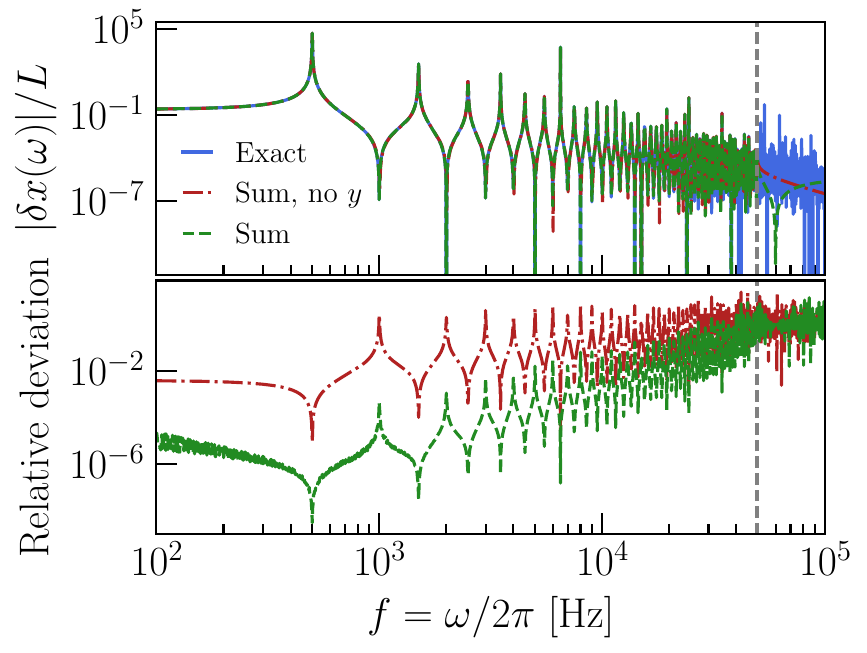}
\end{subfigure}
~ 
\begin{subfigure}{0.435\textwidth}
\includegraphics[width=\textwidth]{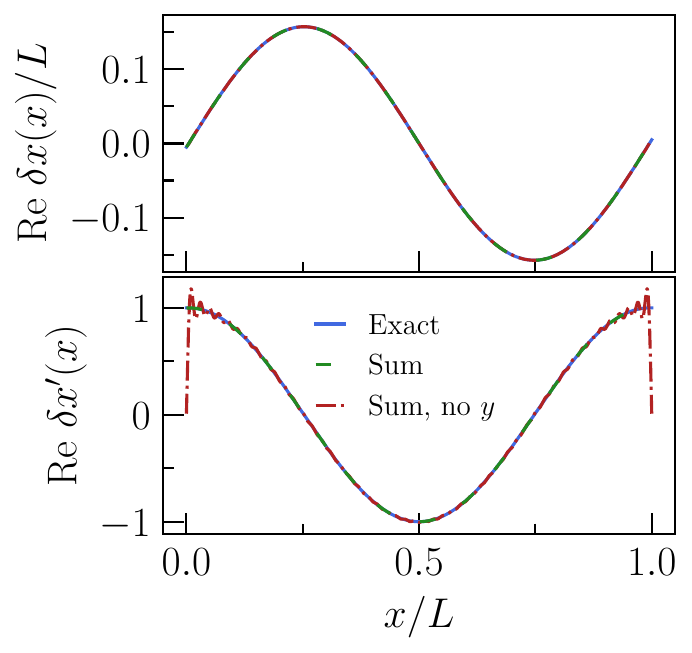}
\end{subfigure}
\caption{A comparison of the solutions to the toy model equations. The \emph{Exact} curve is obtained from Eq. \eqref{eq:toy_exact_solution}, the \emph{Sum} curve from Eq. \eqref{eq:toy_example_sum} and \emph{Sum, no $y$} is showing only $\sum_mq_m(\omega)U_m(x)$. The left plot is showing $\delta x(x=0, \omega)$ along with the relative deviation from the exact result $|\delta x-\delta x^\text{exact}|/|\delta x^\text{exact}|$. The largest eigenmode frequency used in the sum is shown as grey dashed line. The right plot is showing $\delta x(x)$ and its spatial derivative $\delta x'(x)$ at a non-resonant frequency $\omega/2\pi=1.01\,\text{kHz}$. Furthermore, we used $v_s=1\,\text{km}/\text{s}$ and $L=1\,\text{m}$.}\label{fig:toy_example_neumann}
\end{figure}

As shown in Fig. \ref{fig:toy_example_neumann}, the solution in Eq. \eqref{eq:toy_example_sum} converges to the exact solution Eq. \eqref{eq:toy_exact_solution}. Furthermore, a curve of the sum without the lifting function $y$ demonstrates its relevance near the cavity boundary, where the eigenmodes can not fulfil the dynamic boundary condition. Away from the boundary, we see that the sum converges to the exact solution without $y$ as well. We can also see that the lifting function does not improve the convergence of $\delta x$ as much as $\delta x'$, for which the boundary condition Eq. \eqref{eq:toy_bdy_condition} was imposed. This would not be the case when considering Dirichlet-type boundary conditions as in electromagnetism. Furthermore, we see that divergences due to undamped resonances exist at arbitrarily high frequencies and no free-falling limit is obtained. This confirms the argument from the main text that the free fall condition Eq. \eqref{eq:free_falling_limit} must depend on the quality factor and does not exist as $Q_m\to\infty$. From the exact solution, we see that the displacement off resonance drops with frequency $\delta x\propto \omega^{-1}$, however its derivative does not as demanded by the boundary condition \eqref{eq:toy_bdy_condition}. This supports the conclusion in the main text that elastic solids can enter free fall as $\omega_g\to\infty$ in the sense that $\delta \bm{x}^\text{TT}\to0$ but \emph{not} in the sense that $\nabla\delta \bm{x}^\text{TT}\to0$. A difference of the toy model to real elastic solids is given by the fact that $\delta x'$ never converges to zero in the bulk as $\omega\to\infty$, as can be seen from Eq. \eqref{eq:toy_exact_solution}. However, this is due to the unbounded resonances existing at arbitrarily high frequencies, and disappears when considering a damping term in the equations \eqref{eq:toy_EOM} using $\sigma\to\sigma+\eta\delta\dot{x}'$, which modifies the solution in Eq. \eqref{eq:toy_exact_solution} as $v_s\to v_s\sqrt{1+i\omega \eta/\mu}$. The additional frequency scaling of the damping term then ensures the free-falling limit of $\nabla\delta\bm{x}$ in the bulk as claimed in section \ref{sec:free_fall_limit}.


\bibliographystyle{JHEP}
\bibliography{Bibliography}

@article{Hudelist_2023,
doi = {10.1088/1361-6382/acc230},
url = {https://dx.doi.org/10.1088/1361-6382/acc230},
year = {2023},
month = {mar},
publisher = {IOP Publishing},
volume = {40},
number = {8},
pages = {085007},
author = {Hudelist, Mario and Mieling, Thomas B and Palenta, Stefan},
title = {Relativistic theory of elastic bodies in the presence of gravitational waves},
journal = {Classical and Quantum Gravity}
}

@inproceedings{meidlinger,
    author = {D. Meidlinger},
    title = {{A General Perturbation Theory for Cavity Mode Field Patterns}},
    booktitle = {Proc. SRF'09},
    pages = {523--527},
    paper = {THPPO005},
    venue = {Berlin, Germany, Sep. 2009},
    year = {2009},
    publisher = {JACoW Publishing, Geneva, Switzerland},
    url = {https://jacow.org/SRF2009/papers/THPPO005.pdf},
    language = {english}
}

@Article{Rawson-Harris1972,
author={Rawson-Harris, D.},
title={Covariant derivation of the electromagnetic boundary conditions in General Relativity},
journal={International Journal of Theoretical Physics},
year={1972},
month={Dec},
day={01},
volume={6},
number={5},
pages={339-346},
issn={1572-9575},
doi={10.1007/BF01258727},
url={https://doi.org/10.1007/BF01258727}
}

@article{Ahn2024,
  title = {Electromagnetic field in a cavity induced by gravitational waves},
  author = {Ahn, Danho and Bae, Yeong-Bok and Im, Sang Hui and Park, Chan},
  journal = {Phys. Rev. D},
  volume = {110},
  issue = {6},
  pages = {064061},
  numpages = {12},
  year = {2024},
  month = {Sep},
  publisher = {American Physical Society},
  doi = {10.1103/PhysRevD.110.064061},
  url = {https://link.aps.org/doi/10.1103/PhysRevD.110.064061}
}

@Article{Ratzinger2024,
author={Ratzinger, Wolfram
and Schenk, Sebastian
and Schwaller, Pedro},
title={A coordinate-independent formalism for detecting high-frequency gravitational waves},
journal={Journal of High Energy Physics},
year={2024},
month={Aug},
day={22},
volume={2024},
number={8},
pages={195},
issn={1029-8479},
doi={10.1007/JHEP08(2024)195},
url={https://doi.org/10.1007/JHEP08(2024)195}
}

@article{Belgacem24,
doi = {10.1088/1475-7516/2024/07/028},
url = {https://doi.org/10.1088/1475-7516/2024/07/028},
year = {2024},
month = {jul},
publisher = {IOP Publishing},
volume = {2024},
number = {07},
pages = {028},
author = {Belgacem, Enis and Maggiore, Michele and Moreau, Thomas},
title = {Coupling elastic media to gravitational waves: an effective field theory approach},
journal = {Journal of Cosmology and Astroparticle Physics},

}

@book{Pozar12, title={Microwave Engineering}, ISBN={978-1-118-21363-6}, url={https://books.google.fr/books?id=JegbAAAAQBAJ}, publisher={Wiley}, author={Pozar, D.M.}, year={2012} }

@article{Berlin22, title={Detecting high-frequency gravitational waves with microwave cavities}, volume={105}, ISSN={2470-0010, 2470-0029}, DOI={10.1103/PhysRevD.105.116011}, number={11}, journal={Physical Review D}, author={Berlin, Asher and Blas, Diego and D’Agnolo, Raffaele Tito and Ellis, Sebastian A. R. and Harnik, Roni and Kahn, Yonatan and Schütte-Engel, Jan}, year={2022}, month=jun, pages={116011}, language={en} }

@book{Misner1973,
    author = "Misner, Charles W. and Thorne, K. S. and Wheeler, J. A.",
    title = "{Gravitation}",
    isbn = "978-0-7167-0344-0, 978-0-691-17779-3",
    publisher = "W. H. Freeman",
    address = "San Francisco",
    year = "1973"
}

@article{ADMX:2020ote,
    author = "Khatiwada, R. and others",
    collaboration = "ADMX",
    title = "{Axion Dark Matter Experiment: Detailed design~and operations}",
    eprint = "2010.00169",
    archivePrefix = "arXiv",
    primaryClass = "astro-ph.IM",
    reportNumber = "FERMILAB-PUB-20-331-AD-E-QIS",
    doi = "10.1063/5.0037857",
    journal = "Rev. Sci. Instrum.",
    volume = "92",
    number = "12",
    pages = "124502",
    year = "2021"
}

@article{berlin2020axion,
  title={Axion dark matter detection by superconducting resonant frequency conversion},
  author={Berlin, Asher and D’Agnolo, Raffaele Tito and Ellis, Sebastian AR and Nantista, Christopher and Neilson, Jeffrey and Schuster, Philip and Tantawi, Sami and Toro, Natalia and Zhou, Kevin},
  journal={J. High Energ. Phys.},
  volume={2020},
  number={7},
  pages={1--42},
  year={2020},
  publisher={Springer},
  doi={10.1007/JHEP07(2020)088}
}

@article{LIGOScientific:2016aoc,
    author = "Abbott, B. P. and others",
    collaboration = "LIGO Scientific, Virgo",
    title = "{Observation of Gravitational Waves from a Binary Black Hole Merger}",
    eprint = "1602.03837",
    archivePrefix = "arXiv",
    primaryClass = "gr-qc",
    reportNumber = "LIGO-P150914",
    doi = "10.1103/PhysRevLett.116.061102",
    journal = "Phys. Rev. Lett.",
    volume = "116",
    number = "6",
    pages = "061102",
    year = "2016"
}

@article{Fischer_2025,
doi = {10.1088/1361-6382/add8da},
url = {https://doi.org/10.1088/1361-6382/add8da},
year = {2025},
month = {may},
publisher = {IOP Publishing},
volume = {42},
number = {11},
pages = {115015},
author = {Fischer, Lars and Giaccone, Bianca and Gonin, Ivan and Grassellino, Anna and Hillert, Wolfgang and Khabiboulline, Timergali and Krokotsch, Tom and Moortgat-Pick, Gudrid and Muhs, Andrea and Orlov, Yuriy and Paulsen, Michel and Peters, Krisztian and Posen, Sam and Pronitchev, Oleg and Wenskat, Marc},
title = {First characterisation of the MAGO cavity, a superconducting RF detector for kHz–MHz gravitational waves},
journal = {Classical and Quantum Gravity}
}

@article{Fischer_waveforms,
  title = {Laboratory frame representation of time-dependent gravitational waveforms},
  author = {Fischer, Lars and Krokotsch, Tom and Moortgat-Pick, Gudrid},
  journal = {Phys. Rev. D},
  volume = {112},
  issue = {12},
  pages = {124085},
  numpages = {9},
  year = {2025},
  month = {Dec},
  publisher = {American Physical Society},
  doi = {10.1103/h3nh-3hxg},
  url = {https://link.aps.org/doi/10.1103/h3nh-3hxg}
}

@misc{ballantini_microwave_2005,
	title = {Microwave apparatus for gravitational waves observation},
	url = {http://arxiv.org/abs/gr-qc/0502054},

	urldate = {2024-01-25},
	publisher = {arXiv},
	author = {Ballantini, R. and Bernard, Ph and Calatroni, S. and Chiaveri, E. and Chincarini, A. and others},
	month = feb,
	year = {2005},
	note = {arXiv:gr-qc/0502054},
}

@article{berlin_mago20_2023,
  title = {Electromagnetic cavities as mechanical bars for gravitational waves},
  author = {Berlin, Asher and Blas, Diego and D'Agnolo, Raffaele Tito and Ellis, Sebastian A. R. and Harnik, Roni and Kahn, Yonatan and Sch\"utte-Engel, Jan and Wentzel, Michael},
  journal = {Phys. Rev. D},
  volume = {108},
  issue = {8},
  pages = {084058},
  numpages = {28},
  year = {2023},
  month = {Oct},
  publisher = {American Physical Society},
  doi = {10.1103/PhysRevD.108.084058},
  url = {https://link.aps.org/doi/10.1103/PhysRevD.108.084058}
}

@book{jackson_classical_1999,
  address = {New York, {NY}},
  author = {Jackson, John David},
  edition = {3rd ed.},
  isbn = {9780471309321},
  lccn = {538.3537.8},
  publisher = {Wiley},
  title = {Classical electrodynamics},
  url = {http://cdsweb.cern.ch/record/490457},
  year = 1999
}

@Article{Löwenberg2023,
author={L{\"o}wenberg, Robin
and Moortgat-Pick, Gudrid},
title={Lorentz force detuning in heterodyne gravitational wave experiments},
journal={The European Physical Journal C},
year={2023},
month={Dec},
day={19},
volume={83},
number={12},
pages={1153},
issn={1434-6052},
doi={10.1140/epjc/s10052-023-12304-1},
url={https://doi.org/10.1140/epjc/s10052-023-12304-1}
}

@book{lan84,
  address = {New York},
  author = {Landau, L. D. and Lifshitz, E. M.},
  comment = {Mezzi Continui},
  publisher = {Pergamon},
  title = {Electrodynamics of Continuous Media},
  year = 1984
}

@article{Kahn:2023mrj,
    author = {Kahn, Yonatan and Sch{\"u}tte-Engel, Jan and Trickle, Tanner},
    title = "{Searching for high-frequency gravitational waves with phonons}",
    eprint = "2311.17147",
    archivePrefix = "arXiv",
    primaryClass = "hep-ph",
    reportNumber = "FERMILAB-PUB-23-668-T, RIKEN-iTHEMS-Report-23",
    doi = "10.1103/PhysRevD.109.096023",
    journal = "Phys. Rev. D",
    volume = "109",
    number = "9",
    pages = "096023",
    year = "2024"
}

@article{Valero:2024ncz,
    author = "Valero, Jos{\'e} Reina and Madrid, Jose R. Navarro and Blas, Diego and Morcillo, Alejandro D{\'\i}az and Irastorza, Igor Garc{\'\i}a and Gimeno, Benito and Cabrera, Juan Monz{\'o}",
    title = "{High-frequency gravitational waves detection with the BabyIAXO haloscopes}",
    eprint = "2407.20482",
    archivePrefix = "arXiv",
    primaryClass = "hep-ex",
    doi = "10.1103/PhysRevD.111.043024",
    journal = "Phys. Rev. D",
    volume = "111",
    number = "4",
    pages = "043024",
    year = "2025"
}

@article{Fermilab_heterodyne,
    author = "Giaccone, B. and others",
    title = "{Design of axion and axion dark matter searches based on ultra high Q SRF cavities}",
    eprint = "2207.11346",
    archivePrefix = "arXiv",
    primaryClass = "hep-ex",
    reportNumber = "FERMILAB-PUB-22-592-SQMS-T-TD",
    month = "7",
    year = "2022"
}

@article{Slac_heterodyne,
    author = "Li, Zenghai and Zhou, Kevin and Oriunno, Marco and Berlin, Asher and Calatroni, Sergio and Tito D'Agnolo, Raffaele and Ellis, Sebastian A. R. and Schuster, Philip and Tantawi, Sami G. and Toro, Natalia",
    title = "{A Prototype Hybrid Mode Cavity for Heterodyne Axion Detection}",
    eprint = "2507.07173",
    archivePrefix = "arXiv",
    primaryClass = "physics.ins-det",
    reportNumber = "FERMILAB-PUB-25-0465-PPD",
    month = "7",
    year = "2025"
}

@article{ADMX_SLIC,
    author = "Crisosto, N. and Sikivie, P. and Sullivan, N. S. and Tanner, D. B. and Yang, J. and Rybka, G.",
    title = "{ADMX SLIC: Results from a Superconducting $LC$ Circuit Investigating Cold Axions}",
    eprint = "1911.05772",
    archivePrefix = "arXiv",
    primaryClass = "astro-ph.CO",
    doi = "10.1103/PhysRevLett.124.241101",
    journal = "Phys. Rev. Lett.",
    volume = "124",
    number = "24",
    pages = "241101",
    year = "2020"
}

@article{SHANHE,
  title = {First Scan Search for Dark Photon Dark Matter with a Tunable Superconducting Radio-Frequency Cavity},
  author = {Tang, Zhenxing and Wang, Bo and Chen, Yifan and Zeng, Yanjie and Li, Chunlong and Yang, Yuting and Feng, Liwen and Sha, Peng and Mi, Zhenghui and Pan, Weimin and Zhang, Tianzong and Jin, Yirong and Hao, Jiankui and Lin, Lin and Wang, Fang and Xie, Huamu and Huang, Senlin and Shu, Jing},
  collaboration = {SHANHE},
  journal = {Phys. Rev. Lett.},
  volume = {133},
  issue = {2},
  pages = {021005},
  numpages = {7},
  year = {2024},
  month = {Jul},
  publisher = {American Physical Society},
  doi = {10.1103/PhysRevLett.133.021005},
  url = {https://link.aps.org/doi/10.1103/PhysRevLett.133.021005}
}

@article{Atonga:2025utf,
    author = "Atonga, Eduard and Aboushelbaya, Ramy and Norreys, Peter A.",
    title = "{The gravito-optic effect}",
    eprint = "2504.21225",
    archivePrefix = "arXiv",
    primaryClass = "gr-qc",
    month = "4",
    year = "2025"
}

@article{plasma_haloscopes,
  title = {Gravitational wave detection with plasma haloscopes},
  author = {Capdevilla, Rodolfo and Gelmini, Graciela B. and Hyman, Jonah and Millar, Alexander J. and Vitagliano, Edoardo},
  journal = {Phys. Rev. D},
  volume = {112},
  issue = {5},
  pages = {055011},
  numpages = {16},
  year = {2025},
  month = {Sep},
  publisher = {American Physical Society},
  doi = {10.1103/zn2s-2jp6},
  url = {https://link.aps.org/doi/10.1103/zn2s-2jp6}
}

@article{Domcke_Dielectric,
  title = {Dielectric haloscopes as gravitational wave detectors},
  author = {Domcke, Valerie and Ellis, Sebastian A. R. and Kopp, Joachim},
  journal = {Phys. Rev. D},
  volume = {111},
  issue = {3},
  pages = {035031},
  numpages = {18},
  year = {2025},
  month = {Feb},
  publisher = {American Physical Society},
  doi = {10.1103/PhysRevD.111.035031},
  url = {https://link.aps.org/doi/10.1103/PhysRevD.111.035031}
}

@article{Ellis71,
    author = "Ellis, G. F. R.",
    title = "{Relativistic cosmology}",
    doi = "10.1007/s10714-009-0760-7",
    journal = "Proc. Int. Sch. Phys. Fermi",
    volume = "47",
    pages = "104--182",
    year = "1971"
}

@misc{Lobo98,
      title={Viscoelastic effects in a spherical Gravitational Wave antenna}, 
      author={J. A. Lobo and J. A. Ortega},
      year={1998},
      eprint={gr-qc/9802018},
      archivePrefix={arXiv},
      primaryClass={gr-qc},
      url={https://arxiv.org/abs/gr-qc/9802018}, 
}

@article{Domcke25,
  title = {Magnets are Weber Bar Gravitational Wave Detectors},
  author = {Domcke, Valerie and Ellis, Sebastian A. R. and Rodd, Nicholas L.},
  journal = {Phys. Rev. Lett.},
  volume = {134},
  issue = {23},
  pages = {231401},
  numpages = {7},
  year = {2025},
  month = {Jun},
  publisher = {American Physical Society},
  doi = {10.1103/966v-r5fm},
  url = {https://link.aps.org/doi/10.1103/966v-r5fm}
}

@article{Brouwer22,
   title={Proposal for a definitive search for GUT-scale QCD axions},
   volume={106},
   ISSN={2470-0029},
   url={http://dx.doi.org/10.1103/PhysRevD.106.112003},
   DOI={10.1103/physrevd.106.112003},
   number={11},
   journal={Physical Review D},
   publisher={American Physical Society (APS)},
   author={Brouwer, L. and Chaudhuri, S. and Cho, H.-M. and Corbin, J. and Dawson, C. S. and Droster, A. and Foster, J. W. and Fry, J. T. and Graham, P. W. and Henning, R. and Irwin, K. D. and Kadribasic, F. and Kahn, Y. and Keller, A. and Kolevatov, R. and Kuenstner, S. and Leder, A. F. and Li, D. and Ouellet, J. L. and Pappas, K. M. W. and Phipps, A. and Rapidis, N. M. and Safdi, B. R. and Salemi, C. P. and Simanovskaia, M. and Singh, J. and van Assendelft, E. C. and van Bibber, K. and Wells, K. and Winslow, L. and Wisniewski, W. J. and Young, B. A.},
   year={2022},
   month=dec }

@article{Salemi21,
  title = {Search for Low-Mass Axion Dark Matter with ABRACADABRA-10 cm},
  author = {Salemi, Chiara P. and Foster, Joshua W. and Ouellet, Jonathan L. and Gavin, Andrew and Pappas, Kaliro\"e M. W. and Cheng, Sabrina and Richardson, Kate A. and Henning, Reyco and Kahn, Yonatan and Nguyen, Rachel and Rodd, Nicholas L. and Safdi, Benjamin R. and Winslow, Lindley},
  journal = {Phys. Rev. Lett.},
  volume = {127},
  issue = {8},
  pages = {081801},
  numpages = {7},
  year = {2021},
  month = {Aug},
  publisher = {American Physical Society},
  doi = {10.1103/PhysRevLett.127.081801},
  url = {https://link.aps.org/doi/10.1103/PhysRevLett.127.081801}
}

@Article{Aggarwal25,
author={Aggarwal, Nancy
and Aguiar, Odylio D.
and Blas, Diego
and Bauswein, Andreas
and Cella, Giancarlo
and Clesse, Sebastian
and Cruise, Adrian Michael
and Domcke, Valerie
and Ellis, Sebastian
and Figueroa, Daniel G.
and Franciolini, Gabriele
and Garc{\'i}a-Cely, Camilo
and Geraci, Andrew
and Goryachev, Maxim
and Grote, Hartmut
and Hindmarsh, Mark
and Ito, Asuka
and Kopp, Joachim
and Lee, Sung Mook
and Martineau, Killian
and McDonald, Jamie
and Muia, Francesco
and Mukund, Nikhil
and Ottaway, David
and Peloso, Marco
and Peters, Krisztian
and Quevedo, Fernando
and Ricciardone, Angelo
and Ringwald, Andreas
and Steinlechner, Jessica
and Steinlechner, Sebastian
and Sun, Sichun
and Tamarit, Carlos
and Tobar, Michael E.
and Torrenti, Francisco
and {\"U}nal, Caner
and White, Graham},
title={Challenges and opportunities of gravitational-wave searches above 10 kHz},
journal={Living Reviews in Relativity},
year={2025},
month={Nov},
day={03},
volume={28},
number={1},
pages={10},
issn={1433-8351},
doi={10.1007/s41114-025-00060-5},
url={https://doi.org/10.1007/s41114-025-00060-5}
}

@article{Foster25,
    author = "Foster, Joshua W. and Blas, Diego and Bourgoin, Adrien and Hees, Aurelien and Herrero-Valea, M{\'\i}riam and Jenkins, Alexander C. and Xue, Xiao",
    title = "{Discovering $\mu$Hz gravitational waves and ultra-light dark matter with binary resonances}",
    eprint = "2504.15334",
    archivePrefix = "arXiv",
    primaryClass = "astro-ph.CO",
    reportNumber = "FERMILAB-PUB-25-0091-T",
    month = "4",
    year = "2025"
}

@misc{Colpi24,
      title={LISA Definition Study Report}, 
      author={Monica Colpi and Karsten Danzmann and Martin Hewitson and Kelly Holley-Bockelmann and Philippe Jetzer and Gijs Nelemans and Antoine Petiteau and David Shoemaker and Carlos Sopuerta and Robin Stebbins and Nial Tanvir and Henry Ward and William Joseph Weber and Ira Thorpe and Anna Daurskikh and Atul Deep and Ignacio Fernández Núñez and César García Marirrodriga and Martin Gehler and Jean-Philippe Halain and Oliver Jennrich and Uwe Lammers and Jonan Larrañaga and Maike Lieser and Nora Lützgendorf and Waldemar Martens and Linda Mondin and Ana Piris Niño and Pau Amaro-Seoane and Manuel Arca Sedda and Pierre Auclair and Stanislav Babak and Quentin Baghi and Vishal Baibhav and Tessa Baker and Jean-Baptiste Bayle and Christopher Berry and Emanuele Berti and Guillaume Boileau and Matteo Bonetti and Richard Brito and Riccardo Buscicchio and Gianluca Calcagni and Pedro R. Capelo and Chiara Caprini and Andrea Caputo and Eleonora Castelli and Hsin-Yu Chen and Xian Chen and Alvin Chua and Gareth Davies and Andrea Derdzinski and Valerie Fiona Domcke and Daniela Doneva and Irna Dvorkin and Jose María Ezquiaga and Jonathan Gair and Zoltan Haiman and Ian Harry and Olaf Hartwig and Aurelien Hees and Anna Heffernan and Sascha Husa and David Izquierdo and Nikolaos Karnesis and Antoine Klein and Valeriya Korol and Natalia Korsakova and Thomas Kupfer and Danny Laghi and Astrid Lamberts and Shane Larson and Maude Le Jeune and Marek Lewicki and Tyson Littenberg and Eric Madge and Alberto Mangiagli and Sylvain Marsat and Ivan Martin Vilchez and Andrea Maselli and Josh Mathews and Maarten van de Meent and Martina Muratore and Germano Nardini and Paolo Pani and Marco Peloso and Mauro Pieroni and Adam Pound and Hippolyte Quelquejay-Leclere and Angelo Ricciardone and Elena Maria Rossi and Andrea Sartirana and Etienne Savalle and Laura Sberna and Alberto Sesana and Deirdre Shoemaker and Jacob Slutsky and Thomas Sotiriou and Lorenzo Speri and Martin Staab and Danièle Steer and Nicola Tamanini and Gianmassimo Tasinato and Jesus Torrado and Alejandro Torres-Orjuela and Alexandre Toubiana and Michele Vallisneri and Alberto Vecchio and Marta Volonteri and Kent Yagi and Lorenz Zwick},
      year={2024},
      eprint={2402.07571},
      archivePrefix={arXiv},
      primaryClass={astro-ph.CO},
      url={https://arxiv.org/abs/2402.07571}, 
}

@misc{Li24,
      title={Gravitational Wave Astronomy With TianQin}, 
      author={En-Kun Li and Shuai Liu and Alejandro Torres-Orjuela and Xian Chen and Kohei Inayoshi and Long Wang and Yi-Ming Hu and Pau Amaro-Seoane and Abbas Askar and Cosimo Bambi and Pedro R. Capelo and Hong-Yu Chen and Alvin J. K. Chua and Enrique Condés-Breña and Lixin Dai and Debtroy Das and Andrea Derdzinski and Hui-Min Fan and Michiko Fujii and Jie Gao and Mudit Garg and Hongwei Ge and Mirek Giersz and Shun-Jia Huang and Arkadiusz Hypki and Zheng-Cheng Liang and Bin Liu and Dongdong Liu and Miaoxin Liu and Yunqi Liu and Lucio Mayer and Nicola R. Napolitano and Peng Peng and Yong Shao and Swarnim Shashank and Rongfeng Shen and Hiromichi Tagawa and Ataru Tanikawa and Martina Toscani and Verónica Vázquez-Aceves and Hai-Tian Wang and Han Wang and Shu-Xu Yi and Jian-dong Zhang and Xue-Ting Zhang and Lianggui Zhu and Lorenz Zwick and Song Huang and Jianwei Mei and Yan Wang and Yi Xie and Jiajun Zhang and Jun Luo},
      year={2024},
      eprint={2409.19665},
      archivePrefix={arXiv},
      primaryClass={astro-ph.GA},
      url={https://arxiv.org/abs/2409.19665}, 
}

@article{Ruan20,
   title={Taiji program: Gravitational-wave sources},
   volume={35},
   ISSN={1793-656X},
   url={http://dx.doi.org/10.1142/S0217751X2050075X},
   DOI={10.1142/s0217751x2050075x},
   number={17},
   journal={International Journal of Modern Physics A},
   publisher={World Scientific Pub Co Pte Ltd},
   author={Ruan, Wen-Hong and Guo, Zong-Kuan and Cai, Rong-Gen and Zhang, Yuan-Zhong},
   year={2020},
   month=jun, pages={2050075} }

@article{Badurina19,
    author = "Badurina, L. and others",
    title = "{AION: An Atom Interferometer Observatory and Network}",
    eprint = "1911.11755",
    archivePrefix = "arXiv",
    primaryClass = "astro-ph.CO",
    reportNumber = "AION-2019-001, CERN-TH-2019-199",
    doi = "10.1088/1475-7516/2020/05/011",
    journal = "JCAP",
    volume = "05",
    pages = "011",
    year = "2020"
}

@article{Abe21,
   title={Matter-wave Atomic Gradiometer Interferometric Sensor (MAGIS-100)},
   volume={6},
   ISSN={2058-9565},
   url={http://dx.doi.org/10.1088/2058-9565/abf719},
   DOI={10.1088/2058-9565/abf719},
   number={4},
   journal={Quantum Science and Technology},
   publisher={IOP Publishing},
   author={Abe, Mahiro and Adamson, Philip and Borcean, Marcel and Bortoletto, Daniela and Bridges, Kieran and Carman, Samuel P and Chattopadhyay, Swapan and Coleman, Jonathon and Curfman, Noah M and DeRose, Kenneth and Deshpande, Tejas and Dimopoulos, Savas and Foot, Christopher J and Frisch, Josef C and Garber, Benjamin E and Geer, Steve and Gibson, Valerie and Glick, Jonah and Graham, Peter W and Hahn, Steve R and Harnik, Roni and Hawkins, Leonie and Hindley, Sam and Hogan, Jason M and Jiang (姜一君), Yijun and Kasevich, Mark A and Kellett, Ronald J and Kiburg, Mandy and Kovachy, Tim and Lykken, Joseph D and March-Russell, John and Mitchell, Jeremiah and Murphy, Martin and Nantel, Megan and Nobrega, Lucy E and Plunkett, Robert K and Rajendran, Surjeet and Rudolph, Jan and Sachdeva, Natasha and Safdari, Murtaza and Santucci, James K and Schwartzman, Ariel G and Shipsey, Ian and Swan, Hunter and Valerio, Linda R and Vasonis, Arvydas and Wang, Yiping and Wilkason, Thomas},
   year={2021},
   month=jul, pages={044003} }

@article{Abbott23,
   title={Open Data from the Third Observing Run of LIGO, Virgo, KAGRA, and GEO},
   volume={267},
   ISSN={1538-4365},
   url={http://dx.doi.org/10.3847/1538-4365/acdc9f},
   DOI={10.3847/1538-4365/acdc9f},
   number={2},
   journal={The Astrophysical Journal Supplement Series},
   publisher={American Astronomical Society},
   author={Abbott, R. and Abe, H. and Acernese, F. and Ackley, K. and Adhicary, S. and Adhikari, N. and Adhikari, R. X. and Adkins, V. K. and Adya, V. B. and Affeldt, C. and Agarwal, D. and Agathos, M. and Aguiar, O. D. and Aiello, L. and Ain, A. and Ajith, P. and Akutsu, T. and Albanesi, S. and Alfaidi, R. A. and Al-Jodah, A. and Alléné, C. and Allocca, A. and Almualla, M. and Altin, P. A. and Amato, A. and Amez-Droz, L. and Amorosi, A. and Anand, S. and Ananyeva, A. and Andersen, R. and Anderson, S. B. and Anderson, W. G. and Andia, M. and Ando, M. and Andrade, T. and Andres, N. and Andrés-Carcasona, M. and Andrić, T. and Ansoldi, S. and Antelis, J. M. and Antier, S. and Aoumi, M. and Apostolatos, T. and Appavuravther, E. Z. and Appert, S. and Apple, S. K. and Arai, K. and Araya, A. and Araya, M. C. and Areeda, J. S. and Arène, M. and Aritomi, N. and Arnaud, N. and Arogeti, M. and Aronson, S. M. and Arun, K. G. and Asada, H. and Ashton, G. and Aso, Y. and Assiduo, M. and Assis de Souza Melo, S. and Aston, S. M. and Astone, P. and Aubin, F. and AultONeal, K. and Babak, S. and Badalyan, A. and Badaracco, F. and Badger, C. and Bae, S. and Bagnasco, S. and Bai, Y. and Baier, J. G. and Baiotti, L. and Baird, J. and Bajpai, R. and Baka, T. and Ball, M. and Ballardin, G. and Ballmer, S. W. and Baltus, G. and Banagiri, S. and Banerjee, B. and Bankar, D. and Baral, P. and Barayoga, J. C. and Barber, J. and Barish, B. C. and Barker, D. and Barneo, P. and Barone, F. and Barr, B. and Barsotti, L. and Barsuglia, M. and Barta, D. and Barthelmy, S. D. and Barton, M. A. and Bartos, I. and Basak, S. and Basalaev, A. and Bassiri, R. and Basti, A. and Bawaj, M. and Bayley, J. C. and Baylor, A. C. and Bazzan, M. and Bécsy, B. and Bedakihale, V. M. and Beirnaert, F. and Bejger, M. and Bell, A. S. and Benedetto, V. and Beniwal, D. and Benoit, W. and Bentley, J. D. and Yaala, M. Ben and Bera, S. and Berbel, M. and Bergamin, F. and Berger, B. K. and Bernuzzi, S. and Beroiz, M. and Berry, C. P. L. and Bersanetti, D. and Bertolini, A. and Betzwieser, J. and Beveridge, D. and Bevins, N. and Bhandare, R. and Bhandari, A. V. and Bhardwaj, U. and Bhatt, R. and Bhattacharjee, D. and Bhaumik, S. and Bianchi, A. and Bilenko, I. A. and Bilicki, M. and Billingsley, G. and Bini, S. and Birnholtz, O. and Biscans, S. and Bischi, M. and Biscoveanu, S. and Bisht, A. and Biswas, B. and Bitossi, M. and Bizouard, M.-A. and Blackburn, J. K. and Blair, C. D. and Blair, D. G. and Blair, R. M. and Bobba, F. and Bode, N. and Boër, M. and Bogaert, G. and Boileau, G. and Boldrini, M. and Bolingbroke, G. N. and Bonavena, L. D. and Bondarescu, R. and Bondu, F. and Bonilla, E. and Bonilla, G. S. and Bonnand, R. and Booker, P. and Bork, R. and Boschi, V. and Bose, N. and Bose, S. and Bossilkov, V. and Boudart, V. and Bouffanais, Y. and Bozzi, A. and Bradaschia, C. and Brady, P. R. and Braglia, M. and Branch, A. and Branchesi, M. and Brau, J. E. and Breschi, M. and Briant, T. and Brillet, A. and Brinkmann, M. and Brockill, P. and Brooks, A. F. and Brooks, J. and Brown, D. D. and Brunett, S. and Bruno, G. and Bruntz, R. and Bryant, J. and Bucci, F. and Buchanan, J. and Bulashenko, O. and Bulik, T. and Bulten, H. J. and Buonanno, A. and Burtnyk, K. and Buscicchio, R. and Buskulic, D. and Buy, C. and Byer, R. L. and Cabourn Davies, G. S. and Cabras, G. and Cabrita, R. and Cadonati, L. and Caesar, S. and Cagnoli, G. and Cahillane, C. and Calderón Bustillo, J. and Callaghan, J. D. and Callister, T. A. and Calloni, E. and Camp, J. B. and Canepa, M. and Santoro, G. Caneva and Cannavacciuolo, M. and Cannon, K. C. and Cao, H. and Cao, Z. and Capistran, L. A. and Capocasa, E. and Capote, E. and Carapella, G. and Carbognani, F. and Carlassara, M. and Carlin, J. B. and Carpinelli, M. and Carter, J. J. and Carullo, G. and Casanueva Diaz, J. and Casentini, C. and Castaldi, G. and Castro-Lucas, S. Y. and Caudill, S. and Cavaglià, M. and Cavalieri, R. and Cella, G. and Cerdá-Durán, P. and Cesarini, E. and Chaibi, W. and Chakalis, W. and Chalathadka Subrahmanya, S. and Champion, E. and Chan, C. and Chan, C. L. and Chandra, K. and Chang, I. P. and Chang, W. and Chanial, P. and Chao, S. and Chapman-Bird, C. and Charlton, E. L. and Charlton, P. and Chassande-Mottin, E. and Chastain, L. and Chatterjee, C. and Chatterjee, Debarati and Chatterjee, Deep and Chaturvedi, M. and Chaty, S. and Chatziioannou, K. and Chen, D. and Chen, H. and Chen, H. Y. and Chen, J. and Chen, K. H. and Chen, X. and Chen, Y.-R. and Chen, Y. and Cheng, H. and Chessa, P. and Cheung, H. Y. and Chia, H. Y. and Chiadini, F. and Chiang, C-I. and Chiang, C. and Chiarini, G. and Chiba, A. and Chiba, R. and Chierici, R. and Chincarini, A. and Chiofalo, M. L. and Chiummo, A. and Choudhary, S. and Christensen, N. and Chua, S. S. Y. and Chung, K. W. and Ciani, G. and Ciecielag, P. and Cieślar, M. and Cifaldi, M. and Ciobanu, A. A. and Ciolfi, R. and Clara, F. and Clark, J. A. and Clarke, T. A. and Clearwater, P. and Clesse, S. and Cleva, F. and Coccia, E. and Codazzo, E. and Cohadon, P.-F. and Colleoni, M. and Collette, C. G. and Colombo, A. and Colpi, M. and Compton, C. M. and Conti, L. and Cooper, S. J. and Corban, P. and Corbitt, T. R. and Cordero-Carrión, I. and Corezzi, S. and Cornish, N. J. and Corsi, A. and Cortese, S. and Coschizza, A. C. and Cottingham, R. and Coughlin, M. W. and Coulon, J.-P. and Countryman, S. T. and Coupechoux, J.-F. and Cousins, B. and Couvares, P. and Coward, D. M. and Cowart, M. J. and Cowburn, B. D. and Coyne, D. C. and Coyne, R. and Craig, K. and Creighton, J. D. E. and Creighton, T. D. and Criswell, A. W. and Crockett-Gray, J. C. G. and Croquette, M. and Crowder, S. G. and Cudell, J. R. and Cullen, T. J. and Cumming, A. and Cummings, R. and Cuoco, E. and Curyło, M. and Dabadie, P. and Dal Canton, T. and Dall’Osso, S. and Dálya, G. and D’Angelo, B. and Danilishin, S. and D’Antonio, S. and Danzmann, K. and Darroch, K. E. and Darsow-Fromm, C. and Dasgupta, A. and Datrier, L. E. H. and Datta, Sayantani and Dattilo, V. and Dave, I. and Davenport, A. and Davier, M. and Davis, D. and Davis, M. C. and Daw, E. J. and Dax, M. and DeBra, D. and Deenadayalan, M. and Degallaix, J. and De Laurentis, M. and Deléglise, S. and Del Favero, V. and De Lillo, F. and De Lillo, N. and Dell’Aquila, D. and Del Pozzo, W. and De Matteis, F. and D’Emilio, V. and Demos, N. and Dent, T. and Depasse, A. and De Pietri, R. and De Rosa, R. and De Rossi, C. and DeSalvo, R. and De Simone, R. and Dhurandhar, S. and Diab, R. and Diamond, P. Z. and Díaz, M. C. and Didio, N. A. and Dietrich, T. and Di Fiore, L. and Di Fronzo, C. and Di Giorgio, C. and Di Giovanni, F. and Di Giovanni, M. and Di Girolamo, T. and Diksha, D. and Di Lieto, A. and Di Michele, A. and Di Pace, S. and Di Palma, I. and Di Renzo, F. and Divyajyoti and Dmitriev, A. and Doctor, Z. and Dohmen, E. and Doleva, P. P. and Donahue, L. and D’Onofrio, L. and Donovan, F. and Dooley, K. L. and Dooney, T. and Doravari, S. and Dorosh, O. and Drago, M. and Driggers, J. C. and Drori, Y. and Ducoin, J.-G. and Dunn, L. and Dupletsa, U. and Durante, O. and D’Urso, D. and Duverne, P.-A. and Dwyer, S. E. and Eassa, C. and Easter, P. J. and Ebersold, M. and Eckhardt, T. and Eddolls, G. and Edelman, B. and Edo, T. B. and Edy, O. and Effler, A. and Eichholz, J. and Eisenmann, M. and Eisenstein, R. A. and Ejlli, A. and Engelby, E. and Engl, A. J. and Errico, L. and Essick, R. C. and Estellés, H. and Estevez, D. and Etzel, T. and Evans, C. and Evans, M. and Evans, T. M. and Evstafyeva, T. and Ewing, B. E. and Fabrizi, F. and Faedi, F. and Fafone, V. and Fair, H. and Fairhurst, S. and Fan, P. C. and Fan, X. and Farah, A. M. and Farr, B. and Farr, W. M. and Fauchon-Jones, E. J. and Favaro, G. and Favata, M. and Fays, M. and Feicht, J. and Fejer, M. M. and Fenyvesi, E. and Ferguson, D. L. and Fernandez-Galiana, A. and Ferrante, I. and Ferreira, T. A. and Fidecaro, F. and Figura, P. and Fiori, A. and Fiori, I. and Fishbach, M. and Fisher, R. P. and Fittipaldi, R. and Fiumara, V. and Flaminio, R. and Fleischer, S. M. and Fleming, L. S. and Floden, E. and Fong, H. K. and Font, J. A. and Fornal, B. and Forsyth, P. W. F. and Franke, A. and Frasca, S. and Frasconi, F. and Freed, J. P. and Frei, Z. and Freise, A. and Freitas, O. and Frey, R. and Fritschel, P. and Frolov, V. V. and Fronzé, G. G. and Fujimoto, Y. and Fukunaga, I. and Fulda, P. and Fyffe, M. and Gabbard, H. A. and Gabella, W. E. and Gadre, B. U. and Gaglani, K. and Gair, J. R. and Gais, J. and Galaudage, S. and Gallardo, S. and Gamba, R. and Ganapathy, D. and Ganguly, A. and Gao, D. and Gaonkar, S. G. and Garaventa, B. and Garcia-Bellido, J. and García-Núñez, C. and García-Quirós, C. and Gardner, K. A. and Gargiulo, J. and Garufi, F. and Gasbarra, C. and Gateley, B. and Gayathri, V. and Gemme, G. and Gennai, A. and George, J. and Gerberding, O. and Gergely, L. and Ghonge, S. and Ghosh, Abhirup and Ghosh, Archisman and Ghosh, Shaon and Ghosh, Shrobana and Ghosh, T. and Giacoppo, L. and Giaime, J. A. and Giardina, K. D. and Gibson, D. R. and Gier, C. and Giri, P. and Gissi, F. and Gkaitatzis, S. and Glanzer, J. and Gleckl, A. E. and Glotin, F. and Godfrey, J. and Godwin, P. and Goetz, E. and Goetz, R. and Golomb, J. and Goncharov, B. and González, G. and Gosselin, M. and Gouaty, R. and Gould, D. W. and Goyal, S. and Grace, B. and Grado, A. and Graham, V. and Granata, M. and Granata, V. and Gras, S. and Grassia, P. and Gray, C. and Gray, R. and Greco, G. and Green, A. C. and Green, R. and Green, S. and Green, S. R. and Gretarsson, A. M. and Gretarsson, E. M. and Griffith, D. and Griffiths, W. L. and Griggs, H. L. and Grignani, G. and Grimaldi, A. and Grote, H. and Gruson, A. S. and Guerra, D. and Guetta, D. and Guidi, G. M. and Guimaraes, A. R. and Gulati, H. K. and Gulminelli, F. and Gunny, A. M. and Guo, H. and Guo, Y. and Gupta, Anchal and Gupta, Anuradha and Gupta, Ish and Gupta, N. C. and Gupta, P. and Gupta, S. K. and Gurs, J. and Gushima, Y. and Gustafson, E. K. and Gutierrez, N. and Guzman, F. and Haegel, L. and Hain, G. and Haino, S. and Halim, O. and Hall, E. D. and Hamilton, E. Z. and Hammond, G. and Han, W.-B. and Haney, M. and Hanks, J. and Hanna, C. and Hannam, M. D. and Hannuksela, O. A. and Hansen, H. and Hanson, J. and Harada, R. and Harder, T. and Haris, K. and Harmark, T. and Harms, J. and Harry, G. M. and Harry, I. W. and Hartwig, D. and Haskell, B. and Haster, C.-J. and Hathaway, J. S. and Haughian, K. and Hayakawa, H. and Hayama, K. and Hayes, F. J. and Healy, J. and Heffernan, A. and Heidmann, A. and Heintze, M. C. and Heinze, J. and Heinzel, J. and Heitmann, H. and Hellman, F. and Hello, P. and Helmling-Cornell, A. F. and Hemming, G. and Hendry, M. and Heng, I. S. and Hennes, E. and Hennig, J.-S. and Hennig, M. and Henshaw, C. and Hernandez Vivanco, F. and Heurs, M. and Hewitt, A. L. and Higginbotham, S. and Hild, S. and Hill, P. and Himemoto, Y. and Hines, A. S. and Hirata, N. and Hirose, C. and Ho, J. and Hochheim, S. and Hofman, D. and Hohmann, J. N. and Holcomb, D. G. and Holland, N. A. and Holley-Bockelmann, K. and Hollows, I. J. and Holmes, Z. J. and Holt, K. and Holz, D. E. and Hong, Q. and Hornung, J. and Hoshino, S. and Hough, J. and Hourihane, S. and Howell, D. and Howell, E. J. and Hoy, C. G. and Hoyland, D. and Hsieh, B.-H. and Hsieh, H.-F. and Hsiung, C. and Hsu, H. and Hu, P. and Hu, Q. and Huang, H.-Y. and Huang, Y.-J. and Huang, Y. and Huang, Y. T. and Hübner, M. T. and Huddart, A. D. and Hughey, B. and Hui, D. C. Y. and Hui, V. and Husa, S. and Huttner, S. H. and Huxford, R. and Huynh-Dinh, T. and Hyland, J. and Iakovlev, A. and Iandolo, G. A. and Idzkowski, B. and Iess, A. and Inayoshi, K. and Inoue, Y. and Iorio, G. and Iosif, P. and Irwin, J. and Isi, M. and Ismail, M. A. and Itoh, Y. and Iyer, B. R. and JaberianHamedan, V. and Jacqmin, T. and Jacquet, P.-E. and Jadhav, S. J. and Jadhav, S. P. and Jain, D. and Jain, T. and James, A. L. and Jan, A. Z. and Jani, K. and Janiurek, L. and Janquart, J. and Janssens, K. and Janthalur, N. N. and Jaraba, S. and Jaranowski, P. and Jarov, S. and Jasal, P. and Jaume, R. and Javed, W. and Jenkins, A. C. and Jenner, K. and Jennings, A. and Jia, W. and Jiang, J. and Liu, Jian and Jin, H.-B. and Johansmeyer, K. and Johns, G. R. and Johnson, N. A. and Johnston, R. and Johny, N. and Jones, A. W. and Jones, D. H. and Jones, D. I. and Jones, P. and Jones, R. and Joshi, P. and Ju, L. and Jung, K. and Junker, J. and Juste, V. and Kajita, T. and Kalaghatgi, C. and Kalogera, V. and Kamai, B. and Kamiizumi, M. and Kanda, N. and Kandhasamy, S. and Kang, G. and Kanner, J. B. and Kapadia, S. J. and Kapasi, D. P. and Karat, S. and Karathanasis, C. and Karki, S. and Kasamatsu, D. and Kas-danouche, Y. A. and Kashyap, R. and Kasprzack, M. and Kastaun, W. and Kato, J. and Katsanevas, S. and Katsavounidis, E. and Katsuren, J. K. and Katzman, W. and Kaur, T. and Kawabe, K. and Kawazoe, K. and Kéfélian, F. and Keitel, D. and Kellard, I. and Kelley-Derzon, J. and Kennington, J. and Key, J. S. and Khadka, S. and Khalili, F. Y. and Khan, S. and Khanam, T. and Khazanov, E. A. and Khursheed, M. and Kijbunchoo, N. and Kim, C. and Kim, J. C. and Kim, K. and Kim, M. H. and Kim, P. and Kim, S. and Kim, W. S. and Kim, Y.-M. and Kimball, C. and Kimura, N. and Kinley-Hanlon, M. and Kirchhoff, R. and Kissel, J. S. and Kiyota, T. and Klimenko, S. and Klinger, T. and Knee, A. M. and Knust, N. and Kobayashi, Y. and Koch, P. and Koehlenbeck, S. M. and Koekoek, G. and Kohri, K. and Kokeyama, K. and Koley, S. and Koliadko, N. D. and Kolitsidou, P. and Kolstein, M. and Kondrashov, V. and Kong, A. K. H. and Kontos, A. and Korobko, M. and Kossak, R. V. and Kouvatsos, N. and Kovalam, M. and Koyama, N. and Kozak, D. B. and Kranzhoff, L. and Kranzhoff, S. L. and Kringel, V. and Krishnendu, N. V. and Królak, A. and Kuehn, G. and Kuijer, P. and Kukihara, M. and Kulkarni, S. and Kumar, A. and Kumar, Praveen and Kumar, Prayush and Kumar, Rahul and Kumar, Rakesh and Kume, J. and Kuns, K. and Kuroyanagi, S. and Kuwahara, S. and Kwak, K. and Lacaille, G. and Lagabbe, P. and Laghi, D. and Lakkis, M. H. and Lalande, E. and Lalleman, M. and Lamberts, A. and Landry, M. and Lane, B. B. and Lang, R. N. and Lange, J. and Lantz, B. and La Rana, A. and La Rosa, I. and Lartaux-Vollard, A. and Lasky, P. D. and Lawrence, J. and Laxen, M. and Lazzarini, A. and Lazzaro, C. and Leaci, P. and Leavey, S. and LeBohec, S. and Lecoeuche, Y. K. and Lee, E. and Lee, H. M. and Lee, H. W. and Lee, K. and Lee, R.-L. and Lee, R. and Lee, S. and Legred, I. N. and Lehmann, J. and Lehner, L. and Lemaître, A. and Lenti, M. and Leonardi, M. and Leonova, E. and Leroy, N. and Letendre, N. and Lethuillier, M. and Levesque, C. and Levin, Y. and Leyde, K. and Li, A. K. Y. and Li, K. L. and Li, T. G. F. and Li, X. and Lin, C.-Y. and Lin, E. T. and Lin, F-K. and Lin, F-L. and Lin, F. and Lin, H. L. and Lin, H. and Lin, L. C.-C. and Linde, F. and Linker, S. D. and Littenberg, T. B. and Liu, A. and Liu, G. C. and Llamas, F. and Lo, R. K. L. and Lo, T. and London, L. T. and Longo, A. and Lopez, D. and Lopez Portilla, M. and Lorenzini, M. and Loriette, V. and Lormand, M. and Losurdo, G. and Lott, T. P. and Lough, J. D. and Loughlin, H. A. and Lousto, C. O. and Lovelace, G. and Lowry, M. J. and Lück, H. and Lumaca, D. and Lundgren, A. P. and Lung, Y. and Lussier, A. W. and Lynam, J. E. and Ma, L. and Ma, S. and Ma’arif, M. and Macas, R. and MacInnis, M. and Macleod, D. M. and MacMillan, I. A. O. and Macquet, A. and Magaña Hernandez, I. and Magazzù, C. and Magee, R. M. and Maggiore, R. and Magnozzi, M. and Mahesh, M. and Mahesh, S. and Maini, M. and Majorana, E. and Makarem, C. N. and Maliakal, S. and Malik, A. and Man, N. and Mandic, V. and Mangano, V. and Mannix, B. and Mansell, G. L. and Mansingh, G. and Manske, M. and Mantovani, M. and Mapelli, M. and Marchesoni, F. and Pina, D. Marín and Marion, F. and Márka, S. and Márka, Z. and Markakis, C. and Markosyan, A. S. and Markowitz, A. and Maros, E. and Marquina, A. and Marsat, S. and Martelli, F. and Martin, I. W. and Martin, R. M. and Martinez, B. B. and Martinez, M. and Martinez, V. A. and Martinez, V. and Martinovic, K. and Martynov, D. V. and Marx, E. J. and Masalehdan, H. and Mason, K. and Masserot, A. and Reid, M. Masso and Mastrodicasa, M. and Mastrogiovanni, S. and Mateu-Lucena, M. and Matiushechkina, M. and Matsunaga, K. and Mavalvala, N. and McCarthy, R. and McClelland, D. E. and McClincy, P. K. and McCormick, S. and McCuller, L. and McGhee, G. I. and McGinn, J. and McIsaac, C. and McIver, J. and McLeod, A. and McRae, T. and McWilliams, S. T. and Meacher, D. and Mehmet, M. and Mehta, A. K. and Meijer, Q. and Melatos, A. and Mendell, G. and Menendez-Vazquez, A. and Menoni, C. S. and Mercer, R. A. and Mereni, L. and Merfeld, K. and Merilh, E. L. and Merritt, J. D. and Merzougui, M. and Messenger, C. and Messick, C. and Meyers, P. M. and Meylahn, F. and Mhaske, A. and Miani, A. and Miao, H. and Michaloliakos, I. and Michel, C. and Michimura, Y. and Middleton, H. and Mihaylov, D. P. and Miller, A. and Miller, A. L. and Miller, B. and Miller, S. and Millhouse, M. and Mills, J. C. and Milotti, E. and Minenkov, Y. and Mio, N. and Mir, Ll. M. and Miravet-Tenés, M. and Mishra, A. and Mishra, C. and Mishra, T. and Mistry, T. and Mitchell, A. L. and Mitra, S. and Mitrofanov, V. P. and Mitselmakher, G. and Mittleman, R. and Miyakawa, O. and Miyoki, S. and Mo, Geoffrey and Modafferi, L. M. and Moguel, E. and Mohapatra, S. R. P. and Mohite, S. R. and Molina-Ruiz, M. and Mondal, C. and Mondin, M. and Montani, M. and Moore, C. J. and Moragues, J. and Moraru, D. and Morawski, F. and More, A. and More, S. and Moreno, C. and Moreno, G. and Morisaki, S. and Moriwaki, Y. and Morras, G. and Moscatello, A. and Mours, B. and Mow-Lowry, C. M. and Mozzon, S. and Muciaccia, F. and Mukherjee, D. and Mukherjee, Soma and Mukherjee, Subroto and Mukherjee, Suvodip and Mukund, N. and Mullavey, A. and Munch, J. and Muñiz, E. A. and Murray, P. G. and Murray-Dean, J. and Muusse, S. and Nadji, S. L. and Nagar, A. and Nagar, T. and Nagarajan, N. and Nakamura, K. and Nakano, H. and Nakano, M. and Nakayama, Y. and Napolano, V. and Nardecchia, I. and Narikawa, T. and Narola, H. and Naticchioni, L. and Nayak, R. K. and Neil, B. F. and Neilson, J. and Nelson, A. and Nelson, T. J. N. and Nery, M. and Nesseris, S. and Neunzert, A. and Ng, K. Y. and Ng, S. W. S. and Nguyen, C. and Nguyen, P. and Nguyen, R. and Nguyen, T. and Nguyen Quynh, L. and Nichols, S. A. and Nieradka, G. and Nishino, Y. and Nishizawa, A. and Nissanke, S. and Nitoglia, E. and Niu, W. and Nocera, F. and Norman, M. and North, C. and Novak, J. and Nuño Siles, J. F. and Nurbek, G. and Nuttall, L. K. and Oberling, J. and O’Dell, J. and Oelker, E. and Oertel, M. and Oganesyan, G. and Oh, J. J. and Oh, K. and Oh, S. H. and O’Hanlon, T. and Ohashi, M. and Ohashi, T. and Ohkawa, M. and Ohme, F. and Ohta, H. and Oliveira, A. S. and Oliveri, R. and Oohara, K. and O’Reilly, B. and Ormiston, R. G. and Ormsby, N. D. and Orselli, M. and O’Shaughnessy, R. and O’Shea, E. and Oshima, Y. and Oshino, S. and Ossokine, S. and Osthelder, C. and Ottaway, D. J. and Overmier, H. and Pace, A. E. and Pagano, R. and Page, M. A. and Pai, A. and Pai, S. A. and Pal, S. and Palashov, O. and Pálfi, M. and Palomba, C. and Pan, K. C. and Panda, P. K. and Pang, P. T. H. and Pannarale, F. and Pant, B. C. and Panther, F. H. and Paoletti, F. and Paoli, A. and Paolone, A. and Papalexakis, E. E. and Pappas, G. and Parisi, A. and Park, J. and Parker, W. and Pascucci, D. and Pasqualetti, A. and Passaquieti, R. and Passuello, D. and Patel, M. and Pathak, M. and Patra, A. and Patricelli, B. and Patron, A. S. and Paul, S. and Payne, E. and Pearce, T. and Pedraza, M. and Pedurand, R. and Pegna, R. and Pegoraro, M. and Pele, A. and Arellano, F. E. Peña and Penn, S. and Perego, A. and Pereira, A. and Perez, C. J. and Périgois, C. and Perkins, C. C. and Perreca, A. and Perriès, S. and Perry, J. W. and Pesios, D. and Petermann, J. and Petrillo, C. and Pfeiffer, H. P. and Pham, H. and Pham, K. A. and Phukon, K. S. and Phurailatpam, H. and Piccinni, O. J. and Pichot, M. and Piendibene, M. and Piergiovanni, F. and Pierini, L. and Pierra, G. and Pierro, V. and Pillant, G. and Pillas, M. and Pilo, F. and Pinard, L. and Pineda-Bosque, C. and Pinto, I. M. and Piotrzkowski, B. J. and Piotrzkowski, K. and Pirello, M. and Pitkin, M. D. and Placidi, A. and Placidi, E. and Planas, M. L. and Plastino, W. and Poggiani, R. and Polini, E. and Pompili, L. and Pong, D. Y. T. and Ponrathnam, S. and Porcelli, E. and Portell, J. and Porter, E. K. and Posnansky, C. and Poulton, R. and Powell, Jade and Powell, Jonathan and Pracchia, M. and Pradier, T. and Prajapati, A. K. and Prasai, K. and Prasanna, R. and Pratten, G. and Principe, M. and Prodi, G. A. and Prokhorov, L. and Prosposito, P. and Prudenzi, L. and Puecher, A. and Pullin, J. and Punturo, M. and Puosi, F. and Puppo, P. and Pürrer, M. and Qi, H. and Quetschke, V. and Quinonez, P. J. and Quitzow-James, R. and Raab, F. J. and Raaijmakers, G. and Radulesco, N. and Raffai, P. and Rail, S. X. and Raja, S. and Rajan, C. and Ramirez, K. E. and Ramirez, T. D. and Ramos-Buades, A. and Rana, D. and Rana, J. and Randel, E. and Rangnekar, P. R. and Rapagnani, P. and Ray, A. and Raymond, V. and Raza, N. and Razzano, M. and Read, J. and Regimbau, T. and Rei, L. and Reid, S. and Reid, S. W. and Reitze, D. H. and Relton, P. and Renzini, A. and Rettegno, P. and Revenu, B. and Reza, A. and Rezac, M. and Rezaei, A. S. and Ricci, F. and Richards, D. and Richardson, J. W. and Rijal, A. and Riles, K. and Riley, H. K. and Rinaldi, S. and Robertson, C. and Robertson, N. A. and Robinet, F. and Rocchi, A. and Rodriguez, S. and Rolland, L. and Rollins, J. G. and Romanelli, M. and Romano, R. and Romel, C. L. and Romero, A. and Romero-Shaw, I. M. and Romie, J. H. and Ronchini, S. and Roocke, T. J. and Rosa, L. and Rosauer, T. J. and Rose, C. A. and Rosińska, D. and Ross, M. P. and Rossello, M. and Roussel, A. and Rowan, S. and Rowlinson, S. J. and Roy, S. and Royzman, A. and Rozza, D. and Ruggi, P. and Ruiz Morales, E. and Ruiz-Rocha, K. and Ryan, K. and Sachdev, S. and Sadecki, T. and Sadiq, J. and Saffarieh, P. and Saha, S. S. and Saha, S. and Saito, Y. and Sakai, K. and Sakellariadou, M. and Sako, T. and Sakon, S. and Salafia, O. S. and Salces-Carcoba, F. and Salconi, L. and Saleem, M. and Salemi, F. and Sallé, M. and Samajdar, A. and Sanchez, E. J. and Sanchez, J. H. and Sanchez, L. E. and Sanchis-Gual, N. and Sanders, J. R. and Sanuy, A. and Saravanan, T. R. and Sarin, N. and Sasli, A. and Sassi, P. and Sassolas, B. and Satari, H. and Sauter, O. and Savage, R. L. and Savant, V. and Sawada, T. and Sawant, H. L. and Sayah, S. and Schaetzl, D. and Scheel, M. and Scherf, S. J. and Scheuer, J. and Schiworski, M. G. and Schmidt, P. and Schmidt, S. and Schmitz, S. J. and Schnabel, R. and Schneewind, M. and Schofield, R. M. S. and Schönbeck, A. and Schuler, H. and Schulte, B. W. and Schutz, B. F. and Schwartz, E. and Scott, J. and Scott, S. M. and Seetharamu, T. C. and Seglar-Arroyo, M. and Sekiguchi, Y. and Sellers, D. and Sengupta, A. S. and Sentenac, D. and Seo, E. G. and Sequino, V. and Sergeev, A. and Servignat, G. and Setyawati, Y. and Shaffer, T. and Shahriar, M. S. and Shaikh, M. A. and Shams, B. and Shao, L. and Sharma, P. and Chaudhary, S. Sharma and Shawhan, P. and Shcheblanov, N. S. and Sheela, A. and Shen, B. and Shepard, K. G. and Sheridan, E. and Shikano, Y. and Shikauchi, M. and Shimizu, H. and Shimode, K. and Shinkai, H. and Shoemaker, D. H. and Shoemaker, D. M. and ShyamSundar, S. and Sider, A. and Siegel, H. and Sieniawska, M. and Sigg, D. and Silenzi, L. and Singer, L. P. and Singh, D. and Singh, M. K. and Singh, N. and Singha, A. and Sintes, A. M. and Sipala, V. and Skliris, V. and Slagmolen, B. J. J. and Slaven-Blair, T. J. and Smetana, J. and Smith, J. R. and Smith, L. and Smith, R. J. E. and Soldateschi, J. and Somala, S. N. and Somiya, K. and Soni, K. and Soni, S. and Sordini, V. and Sorrentino, F. and Sorrentino, N. and Sotani, H. and Soulard, R. and Souradeep, T. and Sowell, E. and Spagnuolo, V. and Spencer, A. P. and Spera, M. and Spinicelli, P. and Srivastava, A. K. and Srivastava, V. and Stachie, C. and Stachurski, F. and Steer, D. A. and Steinlechner, J. and Steinlechner, S. and Stergioulas, N. and StPierre, M. and Strang, L. C. and Stratta, G. and Strong, M. D. and Strunk, A. and Sturani, R. and Stuver, A. L. and Suchenek, M. and Sudhagar, S. and Sueltmann, N. and Sugiyama, T. and Suh, H. G. and Sullivan, A. G. and Summerscales, T. Z. and Sun, L. and Sunil, S. and Sur, A. and Suresh, J. and Sutton, P. J. and Suzuki, Takamasa and Suzuki, Takanori and Swinkels, B. L. and Syx, A. and Szczepańczyk, M. J. and Szewczyk, P. and Tacca, M. and Tagoshi, H. and Tait, S. C. and Takahashi, H. and Takahashi, R. and Takamori, A. and Takano, S. and Takeda, H. and Takeda, M. and Talbot, C. J. and Talbot, C. and Tamaki, M. and Tamanini, N. and Tanabe, D. and Tanaka, K. and Tanaka, T. and Tanasijczuk, A. J. and Tanioka, S. and Tanner, D. B. and Tao, D. and Tao, L. and Tapia, R. D. and San Martín, E. N. Tapia and Tarafder, R. and Taranto, C. and Taruya, A. and Tasson, J. D. and Teloi, M. and Tenorio, R. and Terhune, J. E. S. and Terkowski, L. and Themann, H. and Thirugnanasambandam, M. P. and Thomas, L. M. and Thomas, M. and Thomas, P. and Thomas, S. and Thompson, J. E. and Thondapu, S. R. and Thorne, K. A. and Thrane, E. and Tiwari, Shubhanshu and Tiwari, Srishti and Tiwari, V. and Toivonen, A. M. and Tolley, A. E. and Tomaru, T. and Tomita, K. and Tomura, T. and Tonelli, M. and Torres-Forné, A. and Torrie, C. I. and e Melo, I. Tosta and Tournefier, E. and Trapananti, A. and Travasso, F. and Traylor, G. and Trenado, J. and Trevor, M. and Tringali, M. C. and Tripathee, A. and Troiano, L. and Trovato, A. and Trozzo, L. and Trudeau, R. J. and Tsang, K. W. and Tsang, T. and Tse, M. and Tso, R. and Tsuchida, S. and Tsukada, L. and Tsutsui, T. and Turbang, K. and Turconi, M. and Turski, C. and Tuyenbayev, D. and Ubach, H. and Ubhi, A. S. and Uchikata, N. and Uchiyama, T. and Udall, R. P. and Uehara, T. and Ueno, K. and Unnikrishnan, C. S. and Ushiba, T. and Utina, A. and Vahlbruch, H. and Vaidya, N. and Vajente, G. and Vajpeyi, A. and Valdes, G. and Valentini, M. and Vallero, S. and Valsan, V. and van Bakel, N. and van Beuzekom, M. and van Dael, M. and van den Brand, J. F. J. and Van Den Broeck, C. and Vander-Hyde, D. C. and van der Sluys, M. and Van de Walle, A. and van Dongen, J. and van Haevermaet, H. and van Heijningen, J. V. and Vanosky, J. and Putten, M. H. P. M. van and van Ranst, Z. and van Remortel, N. and Vardaro, M. and Vargas, A. F. and Varma, V. and Vasúth, M. and Vecchio, A. and Vedovato, G. and Veitch, J. and Veitch, P. J. and Venneberg, J. and Venugopalan, G. and Verdier, P. and Verkindt, D. and Verma, P. and Verma, Y. and Vermeulen, S. M. and Veske, D. and Vetrano, F. and Viceré, A. and Vidyant, S. and Viets, A. D. and Vijaykumar, A. and Villa-Ortega, V. and Vina, M. and Vincent, E. T. and Vinet, J.-Y. and Viret, S. and Virtuoso, A. and Vitale, S. and Vocca, H. and Voigt, D. and von Reis, E. R. G. and von Wrangel, J. S. A. and Vorvick, C. and Vyatchanin, S. P. and Wade, L. E. and Wade, M. and Wagner, K. J. and Walet, R. C. and Walker, M. and Wallace, G. S. and Wallace, L. and Wang, H. and Wang, J. Z. and Wang, W. H. and Ward, R. L. and Warner, J. and Was, M. and Washimi, T. and Washington, N. Y. and Watada, K. and Watarai, D. and Watchi, J. and Wayt, K. E. and Weaver, B. and Weaving, C. R. and Webster, S. A. and Weinert, M. and Weinstein, A. J. and Weiss, R. and Weller, C. M. and Weller, R. A. and Wellmann, F. and Wen, L. and Weßels, P. and Wette, K. and Whelan, J. T. and White, D. D. and Whiting, B. F. and Whittle, C. and Wilk, O. S. and Wilken, D. and Willetts, K. and Williams, D. and Williams, M. J. and Williamson, A. R. and Willis, J. L. and Willke, B. and Wipf, C. C. and Woan, G. and Woehler, J. and Wofford, J. K. and Wong, D. and Wong, H. T. and Wong, I. C. F. and Wright, M. and Wu, C. and Wu, D. S. and Wu, H. and Wysocki, D. M. and Xiao, L. and Xu, V. A. and Yadav, N. and Yamada, T. and Yamamoto, H. and Yamamoto, K. and Yamamoto, M. and Yamamoto, T. and Yamamoto, T. S. and Yamashita, K. and Yamazaki, R. and Yang, F. W. and Yang, K. Z. and Yang, Y.-C. and Yap, M. J. and Yeeles, D. W. and Yelikar, A. B. and Yeung, T. Y. and Yokoyama, J. and Yokozawa, T. and Yoo, J. and Yu, Hang and Yu, Haocun and Yuzurihara, H. and Zadrożny, A. and Zannelli, A. J. and Zanolin, M. and Zeeshan, M. and Zeidler, S. and Zelenova, T. and Zendri, J.-P. and Zevin, M. and Zhang, J. and Zhang, L. and Zhang, R. and Zhang, T. and Zhang, Y. and Zhao, C. and Zhao, Yue and Zhao, Yuhang and Zheng, Y. and Zhong, H. and Zhou, R. and Zhu, X. J. and Zhu, Z.-H. and Zimmerman, A. B. and Zucker, M. E. and Zweizig, J.},
   year={2023},
   month=jul, pages={29} }

@ARTICLE{Manchester13,
       author = {{Manchester}, R.~N. and {Hobbs}, G. and {Bailes}, M. and {Coles}, W.~A. and {van Straten}, W. and {Keith}, M.~J. and {Shannon}, R.~M. and {Bhat}, N.~D.~R. and {Brown}, A. and {Burke-Spolaor}, S.~G. and {Champion}, D.~J. and {Chaudhary}, A. and {Edwards}, R.~T. and {Hampson}, G. and {Hotan}, A.~W. and {Jameson}, A. and {Jenet}, F.~A. and {Kesteven}, M.~J. and {Khoo}, J. and {Kocz}, J. and {Maciesiak}, K. and {Oslowski}, S. and {Ravi}, V. and {Reynolds}, J.~R. and {Sarkissian}, J.~M. and {Verbiest}, J.~P.~W. and {Wen}, Z.~L. and {Wilson}, W.~E. and {Yardley}, D. and {Yan}, W.~M. and {You}, X.~P.},
        title = "{The Parkes Pulsar Timing Array Project}",
      journal={Publications of the Astronomical Society of Australia},
         year = 2013,
        month = jan,
       volume = {30},
          eid = {e017},
        pages = {e017},
          doi = {10.1017/pasa.2012.017},
archivePrefix = {arXiv},
       eprint = {1210.6130},
 primaryClass = {astro-ph.IM},
       adsurl = {https://ui.adsabs.harvard.edu/abs/2013PASA...30...17M}
}

@ARTICLE{Kramer13,
       author = {{Kramer}, Michael and {Champion}, David J.},
        title = "{The European Pulsar Timing Array and the Large European Array for Pulsars}",
      journal = {Classical and Quantum Gravity},
         year = 2013,
        month = nov,
       volume = {30},
       number = {22},
          eid = {224009},
        pages = {224009},
          doi = {10.1088/0264-9381/30/22/224009},
       adsurl = {https://ui.adsabs.harvard.edu/abs/2013CQGra..30v4009K}
}

@ARTICLE{Demorest13,
       author = {{Demorest}, P.~B. and {Ferdman}, R.~D. and {Gonzalez}, M.~E. and {Nice}, D. and {Ransom}, S. and {Stairs}, I.~H. and {Arzoumanian}, Z. and {Brazier}, A. and {Burke-Spolaor}, S. and {Chamberlin}, S.~J. and {Cordes}, J.~M. and {Ellis}, J. and {Finn}, L.~S. and {Freire}, P. and {Giampanis}, S. and {Jenet}, F. and {Kaspi}, V.~M. and {Lazio}, J. and {Lommen}, A.~N. and {McLaughlin}, M. and {Palliyaguru}, N. and {Perrodin}, D. and {Shannon}, R.~M. and {Siemens}, X. and {Stinebring}, D. and {Swiggum}, J. and {Zhu}, W.~W.},
        title = "{Limits on the Stochastic Gravitational Wave Background from the North American Nanohertz Observatory for Gravitational Waves}",
      journal = {The Astrophysical Journal},
         year = 2013,
        month = jan,
       volume = {762},
       number = {2},
          eid = {94},
        pages = {94},
          doi = {10.1088/0004-637X/762/2/94},
archivePrefix = {arXiv},
       eprint = {1201.6641},
 primaryClass = {astro-ph.CO},
       adsurl = {https://ui.adsabs.harvard.edu/abs/2013ApJ...762...94D}
}

@INPROCEEDINGS{Lee16,
       author = {{Lee}, K.~J.},
        title = "{Prospects of Gravitational Wave Detection Using Pulsar Timing Array for Chinese Future Telescopes}",
    booktitle = {Frontiers in Radio Astronomy and FAST Early Sciences Symposium 2015},
         year = 2016,
       editor = {{Qain}, L. and {Li}, D.},
       series = {Astronomical Society of the Pacific Conference Series},
       volume = {502},
        month = feb,
        pages = {19},
       adsurl = {https://ui.adsabs.harvard.edu/abs/2016ASPC..502...19L}
}

@ARTICLE{Joshi22,
       author = {{Joshi}, Bhal Chandra and {Gopakumar}, Achamveedu and {Pandian}, Arul and {Prabu}, Thiagaraj and {Dey}, Lankeswar and {Bagchi}, Manjari and {Desai}, Shantanu and {Tarafdar}, Pratik and {Rana}, Prerna and {Maan}, Yogesh and {Batra}, Neelam Dhanda and {Girgaonkar}, Raghav and {Agarwal}, Nikita and {Arumugam}, Paramasivan and {Basu}, Avishek and {Bathula}, Adarsh and {Dandapat}, Subhajit and {Gupta}, Yashwant and {Hisano}, Shinnosuke and {Kato}, Ryo and {Kharbanda}, Divyansh and {Kikunaga}, Tomonosuke and {Kolhe}, Neel and {Krishnakumar}, M.~A. and {Manoharan}, P.~K. and {Marmat}, Piyush and {Naidu}, Arun and {Banik}, Sarmistha and {Nobleson}, K. and {Paladi}, Avinash Kumar and {Pathak}, Dhruv and {Singha}, Jaikhomba and {Srivastava}, Aman and {Surnis}, Mayuresh and {Susarla}, Sai Chaitanya and {Susobhanan}, Abhimanyu and {Takahashi}, Keitaro},
        title = "{Nanohertz gravitational wave astronomy during SKA era: An InPTA perspective}",
      journal = {Journal of Astrophysics and Astronomy},
         year = 2022,
        month = dec,
       volume = {43},
       number = {2},
          eid = {98},
        pages = {98},
          doi = {10.1007/s12036-022-09869-w},
archivePrefix = {arXiv},
       eprint = {2207.06461},
 primaryClass = {astro-ph.HE},
       adsurl = {https://ui.adsabs.harvard.edu/abs/2022JApA...43...98J}
}

@ARTICLE{Miles23,
       author = {{Miles}, M.~T. and {Shannon}, R.~M. and {Bailes}, M. and {Reardon}, D.~J. and {Keith}, M.~J. and {Cameron}, A.~D. and {Parthasarathy}, A. and {Shamohammadi}, M. and {Spiewak}, R. and {van Straten}, W. and {Buchner}, S. and {Camilo}, F. and {Geyer}, M. and {Karastergiou}, A. and {Kramer}, M. and {Serylak}, M. and {Theureau}, G. and {Venkatraman Krishnan}, V.},
        title = "{The MeerKAT Pulsar Timing Array: first data release}",
      journal = {Monthly Notices of the Royal Astronomical Society},
         year = 2023,
        month = mar,
       volume = {519},
       number = {3},
        pages = {3976-3991},
          doi = {10.1093/mnras/stac3644},
archivePrefix = {arXiv},
       eprint = {2212.04648},
 primaryClass = {astro-ph.HE},
       adsurl = {https://ui.adsabs.harvard.edu/abs/2023MNRAS.519.3976M}
}

@book{Padamsee98,
      author        = "Padamsee, Hasan and Hays, Tom and Knobloch, Jens",
      title         = "{RF superconductivity for accelerators}",
      publisher     = "Wiley",
      address       = "New York, NY",
      series        = "Wiley series in beam physics and accelerator technology",
      year          = "1998",
      url           = "https://cds.cern.ch/record/366783",
}

@book{Balanis05,
  author    = {Constantine A. Balanis},
  title     = {Antenna Theory: Analysis and Design},
  edition   = {3rd},
  publisher = {John Wiley \& Sons},
  address   = {Hoboken, NJ},
  year      = {2005},
  isbn      = {047166782X}
}

\end{document}